\documentclass{emulateapj}
\usepackage{natbib}
\usepackage{graphicx}
\usepackage{amssymb}

\def\msun{{~M}_{\odot}}

\def\be{\begin{equation}}
\def\ee{\end{equation}}
\def\ergs{{\rm\,erg\,s^{-1}}}

\catcode`\@=11 
\def\@versim#1#2{\vcenter{\offinterlineskip
        \ialign{$\m@th#1\hfil##\hfil$\crcr#2\crcr\sim\crcr } }}
\def\mpy{M_{\sun} \ {\rm yr^{-1}}}

\begin{document}

\shorttitle{Nature, Origin, and Properties of Outflow}
\shortauthors{Yuan, Bu \& Wu}

\title{Numerical Simulation of Hot Accretion Flows (II): Nature, Origin, and Properties of Outflow and Their Possible Observational Applications}

\author{Feng Yuan, Defu Bu, and Maochun Wu}

\affil{Key Laboratory for Research in Galaxies and Cosmology, Shanghai Astronomical Observatory, Chinese Academy of Sciences, 80 Nandan Road, Shanghai 200030, China;
fyuan,dfbu,mcwu@shao.ac.cn}


\begin{abstract}
Previous hydrodynamical (HD) and magnetohydrodynamical (MHD) numerical simulations of hot accretion flows have indicated that the inflow (gas with inward radial velocity) accretion rate decreases with decreasing radius. Two models have been proposed to explain this result. In the adiabatic inflow-outflow solution (ADIOS), the inward decrease of accretion rate is because of the loss of gas in the outflow. In the alternative convection-dominated accretion flow (CDAF) model, the accretion flow is thought to be convectively unstable; the gas is then assumed to be locked in convective eddies, which results in the inward decrease of the accretion rate. In the present paper we investigate the nature of inward decrease of accretion rate using HD and MHD simulations. We calculate various properties of inflow and outflow, including the mass flux,  radial and rotational velocities, temperature, and the Bernoulli parameter ($Be$). Systematic and significant differences between inflow and outflow are found. For example, for HD flows, the temperature of outflow is significantly higher than inflow; while for MHD flows, the specific angular momentum of outflow is nearly Keplerian, which is significantly higher than inflow. These results suggest that the inflow and outflow are not dominated by convective turbulence, but they are systematic inward and outward motion. We have also analyzed the convective stability of MHD accretion flow and found that they are convectively stable. These results indicate that the inward decrease of inflow rate is because of the mass loss in outflow. The different properties of inflow and outflow also suggest that the mechanisms of producing outflow in HD and MHD flows are buoyancy associated with the convection and centrifugal force associated with the angular momentum transport mediated by the magnetic field, respectively. The latter mechanism is similar to the Blandford \& Payne mechanism but no large-scale open magnetic field is required; it is kind of ``micro-Blandford \& Payne'' mechanism. We also study the effect of initial conditions in the simulations. We find that the value of $Be$, whose sign determines whether the outflow can escape to infinity or not, is mainly determined by the value of $Be$ of the initial condition. We discuss some possible observational applications, including the Fermi bubble observed in the Galaxy center and winds widely observed in active galactic nuclei and black hole X-ray binaries.

\end{abstract}

\keywords{accretion, accretion discs -- hydrodynamics-- black hole physics}

\section{Introduction}

There are two series of black hole accretion solution, depending on the temperature of the accretion flow. The cold series include the standard thin disk and the slim disk, bounded by the Eddington accretion rate $\dot{M}_{\rm Edd}\equiv 10 L_{\rm Edd}/c^2$ (Sunkura \& Sunyaev 1973; Abramowicz et al. 1988). For the hot series, when the accretion rate is below $\sim \alpha^2 \dot{M}_{\rm Edd}$, where $\alpha$ is the viscous parameter, it is the advection-dominated accretion flow (ADAF; Narayan \& Yi 1994; 1995; Abramowicz et al. 1995; see reviews by Narayan, Mahadevan \& Quataert 1998, Kato, Fukue \& Mineshige 1998, and Yuan \& Narayan 2013). The main application of ADAFs is on the dim black hole sources, including the supermassive black hole in our Galactic center, low-luminosity AGNs, quiescent and hard states of black hole X-ray binaries (see reviews by Narayan 2005; Yuan 2007; Narayan \& McClintock 2008; Ho 2008). Above $\alpha^2\dot{M}_{\rm Edd}$, up to even close to $\dot{M}_{\rm Edd}$, the solution is described by the luminous hot accretion flow (LHAF; Yuan 2001; 2003). LHAFs correspond to higher accretion rates and radiative efficiency therefore the emitted luminosity is higher. This model has been applied to explain the origin of hard X-ray emission in relatively luminous sources such as luminous hard state of black hole X-ray binaries and some AGNs (e.g., Yuan \& Zdziarski 2004; Yuan et al. 2007).

In the early analytical studies of ADAFs, it was {\em assumed} that the mass accretion rate, or more precisely the inflow rate (inflow and outflow rates are defined as the mass flux of gas with a negative and positive radial velocity; refer to eqs. [\ref{inflowrate}] \& [\ref{outflowrate}] in the present paper for their definitions), is a constant of radius. Correspondingly, the density follows a power-law distribution: $\rho (r) \propto r^{-3/2}$. Later many numerical simulations have been performed to study the multi-dimensional dynamics of the hot accretion flow, including both hydrodynamical (HD) and magneto-hydrodynamical (MHD) ones (e.g., Stone, Pringle \& Begelman 1999, hereafter SPB99; Igumenshchev \& Abramowicz 1999, 2000; Stone \& Pringle 2001; Hawley, Balbus \& Stone 2001; Machida, Matsumoto \& Mineshige 2001; Hawley \& Balbus 2002; Igumenshchev, Narayan \& Abramowicz 2003; Pen, Matzener \& Wong 2003; De Villiers, Hawley \& Krolik 2003; De Villiers et al. 2005; Yuan \& Bu 2010; Pang et al. 2011; McKinney, Tchekhovskoy \& Blandford 2012; Narayan et al. 2012; Li, Ostriker \& Sunyaev 2012; Yuan, Wu \& Bu 2012, hereafter Paper I). One of the most important findings of these simulations is that the inflow accretion rate decreases with decreasing radius. The radial profile of the inflow rate can be described by a power-law function of radius, $\dot{M}(r)\propto r^{s}$ with $s\sim 0.5-1$. Correspondingly, the radial profile of density becomes flatter, $\rho(r)\propto r^{-p}$ with $p\approx 1.5-s\sim 1-0.5$. Such a result not only has theoretical significance, but also is of great importance to observations. This is because it determines the emitted spectrum of the accretion flow; and also the strength of ``mechanical AGNs feedback'', if such a profile is caused by mass outflow as we will illustrate in the present paper.

Because of its importance, we first need to critically examine the reliability of this  result. This is because, due to technical difficulties, all the above-mentioned numerical simulations have a rather small radial dynamical range, which usually spans two orders of magnitude. In this case, the results may suffer from the boundary conditions thus are suspectable. To overcome this problem, in Paper I we presented a ``two-zone'' approach, successfully extended the dynamical range to four orders of magnitude. We found that in this case, the profiles of accretion rate and density are almost remain unchanged compared to previous works. In addition, we combined previous numerical simulation works on hot accretion flows, including both HD and MHD ones, and found that the profiles of accretion rate and density in all these works are quite similar. Such a consistency is in agreement with the prediction of a recent work by Begelman (2012), although slopes are different. What is more exciting is that these theoretical results have been confirmed by the observations, e.g., to Sgr A* and NGC~3115 (Yuan, Quataert \& Narayan 2003; Wong et al. 2011; see Paper I for details).

The next question is then: what is the physical reason for the decrease of the accretion rate? Two main models have been proposed to answer this question. The first one is the adiabatic inflow-outflow solution (ADIOS; Blandford \& Begelman 1999; 2004; Begelman 2012; see also  Becker, Subramanian \& Kazanas 2001; Xue \& Wang 2005; and Jiao \& Wu 2011 for related works.). In this model there is inflowing and outflowing zones with almost equal but opposite mass fluxes. The net accretion rate is orders of magnitude smaller. The inward decrease of mass accretion rate is because of the mass loss in the outflow launched at every radius. In the early versions of the model (Blandford \& Begelman 1999; 2004), the origin of the outflow was assumed to be because of the positive sign of the Bernoulli parameter of the accretion flow. But in the latest version of the ADIOS model the mechanism of producing outflow is not specified but leave open (Begelman 2012).

The second model is the convection-dominated accretion flow (CDAF; Narayan, Igumenshchev, \& Abramowicz 2000; Quataert \& Gruzinov 2000). This model is based on the assumption that the hot accretion flow, both hydrodynamical (HD) and magnetohydrodynamical (MHD), is convectively unstable. In this model, when $\alpha$ is small, the inward angular momentum transport by convection and outward transport by viscous stress almost cancel each other. A convective envelope solution was then constructed which can reproduce the simulated flat density profile. In this scenario, the inward decrease of mass accretion rate is because that with accretion, more and more fluid is locked in convective eddies operating circular motion.

The first aim of the present work is to investigate which one of the above two scenarios is correct. For this purpose, we run some HD and MHD simulations, and calculate the respective properties of inflow and outflow, including their radial and rotational velocities, temperature, and Bernoulli parameter. If the CDAF scenario is correct, i.e, the motion of the flow is dominated by convective turbulence, and inflow and outflow rates are simply due to turbulent fluctuation, we should expect that the properties of inflow and outflow are similar. As we will see, however, we find that the properties of inflow and outflow are systematically and significantly different.

The second question we want to address is the convective stability of MHD accretion flow. The HD hot accretion flow was predicted to be convectively unstable (Narayan \& Yi 1994). Physically this is because the entropy increases with decreasing radius, which arises because of the entropy production by viscous dissipation and little entropy loss in the radiation. Such a prediction has been well confirmed by HD simulations (SPB99; Igumenshchev \& Abramowicz 1999, 2000). CDAF model further assumes that an MHD accretion flow is also convectively unstable and applies to an MHD flow as well. However, such an applicability was questioned from the beginning by some authors, because they think that the dynamics of MHD flow is controlled by magnetorotational instability (MRI; Balbus \& Hawley 1991, 1998) rather than convection (Stone \& Pringle 2001; Balbus \& Hawley 2002). Narayan et al. (2002) did a linear MHD stability analysis and argued that if the flow is unstable the long-wavelength modes of the instability are intrinsically ``convection'' so the CDAF model should be applicable to MHD accretion flows; moreover, the convective stability can be judged by the HD H{\o}iland criteria.  In the present work we use the simulation data to analyze the convective stability of MHD accretion flows. We find that they are convectively stable. Based on this result, combined with the fact that HD and MHD accretion flows have almost identical inward decrease of accretion rate, we conclude that the decrease of accretion rate for both HD and MHD flows is  not because of the convection, but systematic outflow.

Then two immediate questions arise. The question in the theoretical aspect is: what is the physical origin of these outflow? In the aspect of observational applications, the most important question that concerns us is the main properties of outflow such as their terminal velocity and kinetic power. We investigate these two questions again by systematically comparing the properties of inflow and outflow. One potential complication of the numerical simulation is the effect of initial condition adopted in the simulation. Many simulations choose a rotation torus as the initial condition; in some other works the gas is injected into the computation domain, i.e.,  ``injection-type'' initial condition. The effect of the initial condition according to our knowledge was never systematically studied. Our idea of the potential importance of initial condition is stimulated by the one-dimensional steady global solution of ADAFs. Since the accretion equations are a set of {\em differential} equations, and since the differential terms in ADAFs (i.e, advection) are very important, we expect that the outer boundary condition should be important in determining the global dynamics of ADAFs. In other words, for given parameters, the solutions of ADAFs are not unique, but comprise a series of solution corresponding to different outer boundary conditions. This expectation was fully confirmed by the detailed calculations of Yuan (1999). Following this spirit, the initial condition should also play some role in multidimensional numerical simulations, i.e., the final solution should preserve some memory to the initial condition, although we don't know to what extent before the detailed calculations are done. We expect that while the main properties of accretion flow should remain, some properties of outflow may depend on the initial conditions.

One particular example is the Bernoulli parameter of the accretion flow. This parameter is the sum of kinetic energy, enthalpy, and potential energy (eq. [{\ref{bernoulli}]).  As we will describe later, its value will determine whether the outflow can escape to infinity and what is the terminal velocity of the outflow. In the self-similar solution of ADAFs, it was found that $Be>0$ (Narayan \& Yi 1994). However, as later pointed out by several authors the positive sign of $Be$ is only a result of self-similar solution. In the global solution, its sign should depend on the the outer boundary condition (Nakamura 1998; Yuan 1999; Abramowicz, Lasota \& Igumenshchev 2000). This was confirmed by the numerical calculations of Yuan (1999).

We assign the systematic study of the effects of initial condition and boundary condition in numerical simulation of accretion flow in a separate paper (Bu, Yuan \& Wu 2012, in preparation). In this paper, we consider three different initial conditions in our HD simulations, but only one initial condition for the MHD simulation. The paper is structured as follows. The four models, one MHD and three HD ones with different initial conditions, will be described in \S2. The systematic comparisons of properties between inflow and outflow are presented in \S3. In \S4, we compare our simulation results with previous ones. We find that some puzzling results in previous works can be understood as due to different initial conditions. In \S\ref{adioscdaf}, we analyze the convective stability of MHD accretion flow, and argue that the ADIOS scenario is favored by our simulations. The origin of outflow in both HD and MHD accretion flows are discussed in \S\ref{outfloworigin}. In \S\ref{observation}, we discuss the observational application of our study.  The last section (\S\ref{summary}) devotes to a summary.

\section{Models}

\subsection{Equations}

We perform both HD and MHD calculation of two-dimensional axisymmetric accretion flows around black holes. The equations we use are exactly same with those in SPB99 and Stone \& Pringle (2001). For convenience of readers, we copy them here. The HD equations are:
\begin{equation}
\frac{d\rho}{dt}+\rho\nabla\cdot \mathbf{v}=0,\label{cont1}
\end{equation}
\begin{equation}
\rho\frac{d\mathbf{v}}{dt}=-\nabla p-\rho\nabla
\psi+\nabla\cdot\mathbf{T}, \label{rmon1}
\end{equation}
\begin{equation}
\rho\frac{d(e/\rho)}{dt}=-p\nabla\cdot\mathbf{v}+\mathbf{T}^2/\mu.
\label{energy1}
\end{equation}
The MHD equations are:
\begin{equation}
\frac{d\rho}{dt}+\rho\nabla\cdot \mathbf{v}=0,\label{cont}
\end{equation}
\begin{equation}
\rho\frac{d\mathbf{v}}{dt}=-\nabla p-\rho\nabla
\psi+\frac{1}{4\pi}(\nabla \times \mathbf{B})\times \mathbf{B}, \label{rmon}
\end{equation}
\begin{equation}
\rho\frac{d(e/\rho)}{dt}=-P\nabla\cdot\mathbf{v}+\eta\mathbf{J}^2,
\label{energy}
\end{equation}
\be \frac{\partial\mathbf{B}}{\partial t}=\nabla \times (v\times \mathbf{B}-\eta \mathbf{J}).\label{mag}
\ee

In the above equations, $\rho$, $P$, $v$, $e$, $\mathbf{B}$, $\mathbf{J}=(c/4\pi)\nabla\times \mathbf{B}$ are the mass density, pressure, velocity,  internal energy, magnetic filed, and the current density, respectively. We adopt an adiabatic equation of state $P=(\gamma
-1)e$ with $\gamma =5/3$. $\psi$ is the gravitational potential. We adopt the Paczy\'nsky \& Wiita potential $\psi=-GM/(r-r_s)$, where
$M$ is the center black hole mass, $G$ is the gravitational
constant, and $r_s\equiv 2GM/c^2$. The self gravity of the accretion flow is neglected. We use the spherical coordinate $(r, \theta, \phi )$ to solve these equations. We choose units such that $c=M=G=1$.

In the HD equations, $\mathbf{T}$ is anomalous stress tensor. The final two terms in eqs. (\ref{rmon1}) and (\ref{energy1}) represent anomalous stress and its corresponding heating, respectively. Since we don't have magnetic field, we add the two terms to transfer the angular momentum and convert energy, so that to mimic the real MHD case (see below). The viscosity coefficient
$\mu=\nu\rho$ and $\nu$ is the kinematic viscosity coefficient. We
adopt the form $\nu=\alpha {r^{1/2}}$ because it corresponds to the
usual ``$\alpha$'' description (refer to ``Run K'' in SPB99). We set
$\alpha$=0.01. Following
SPB99, to better mimic the MHD case, we assume that the only
non-zero components of $\mathbf{T}$ are the azimuthal components (refer to SPB99 for details). In the MHD equations, the final term in eqs. (\ref{energy}) and (\ref{mag}) are the magnetic heating and dissipation rate mediated by a finite resistivity $\eta$. Since the energy equation here is actually internal energy equation, numerical reconnection inevitably results in loss of energy from the system. By adding the anomalous resistivity $\eta$, the energy loss can be captured in the form of heating in the current sheet (Stone \& Pringle 2001).

MHD simulation is of course more realistic than HD one. Then what are the motivations of doing the HD simulation? The reasons are threefold. The first reason is mainly academic interest. We hope to compare with and understand previous HD works, both analytical and numerical. The second one is to study the effects of initial conditions. Although we can use MHD simulation to accomplish this, it will obviously be much more expensive. More importantly, we believe that in terms of the effect of the initial condition, the results for HD and MHD simulations should be intrinsically same. Lastly, by comparing the HD and MHD simulations we hope to be able to deepen our understanding to the production and properties of outflow.

\subsection{Model Description}

We investigate four models. Three models (Models A, B, and C) are HD and one (Model D) is MHD. The former three models have different initial conditions. The initial condition of Model D is identical to that of Model A, except that magnetic field is included. Our spirit is that we hope the models are as simple and clean as possible. Based on this consideration, except the two models that are devoted to studying the effect of initial conditions (i.e, Models B and C), the other two models (Models A and D) are the most ``popular'' ones in the literature. The details are as follows.

{\bf Mode A} This model is almost identical to ``Run K'' in SPB99. The initial condition is a rotating torus with constant angular momentum, which is identical to that adopted in Stone, Pringle \& Begelman (1999). The readers are referred to SPB99 for the detailed description of the torus. Different from SPB99 where Newtonian potential is adopted, we adopt the Paczy\'nsky \& Wiita (1980) potential. This introduces the length scale, so we put the center of the density maximum of the torus at $\sim 100 r_s$. We set the maximum density of the torus $\rho_{\rm max}=1.0$, the density of the surrounding medium $\rho_0=10^{-4}$, and pressure $ p_0=\rho_0/r$. The torus is a bound system, therefore the Bernoulli parameter $Be<0$ (refer to the dotted line in Fig. \ref{Fig:radialbernoulli} for its value at the equatorial plane).

{\bf Model B} A rotating torus has no radial velocity while a realistic accretion flow has. In this sense, we think that a perhaps better initial condition is that the gas  has both radial and rotational velocities. This is one of our motivations for the initial conditions in Models B \& C. In Model B, to get the initial condition, we first get the one-dimensional global solution of two-temperature ADAFs. This corresponds to solving the one-dimensional version of eqs. \ref{cont1}-\ref{energy1}, although the equations are for one-temperature accretion flow. We use the calculation results for ions (such as density and temperature) as the initial conditions in our simulation. We choose to solve the two-temperature rather than one-temperature simply because we think the former is more realistic; but a one-temperature solution should not make any notable differences.  The solution extends from the inner boundary to the outer boundary, and it has both rotation and radial velocities. The details of the calculations are described in Yuan (1999). Specifically, we choose the outer boundary condition so that $Be<0$ in the solution, but larger than that in Model A. We do not choose $Be>0$ in Model B in order to discriminate this model from Model C where $Be>0$. We then expand this one-dimensional solution to two-dimension by assuming that the vertical density profile follows an exponential distribution but all other quantities such as temperature and velocity remain their original one-dimensional values. The thickness of the accretion flow is determined by the scale-height $H=c_s/\Omega_{\rm k}$.

{\bf Model C} We adopt ``injection-type'' initial condition in this model. The gas is injected at the outer boundary. The properties of the injected gas, such as temperature, rotational and radial velocities, are completely determined by the self-similar solution of Narayan \& Yi (1994). In this case, $Be>0$ (Narayan \& Yi 1994).

{\bf Model D} This is an MHD model. The model is fully identical to ``Run F'' in Stone \& Pringle (2001)\footnote{The data we use in the present paper for Model D was kindly sent to us by Jim Stone. We also did simulation ourselves, using higher resolution and smaller inner boundary. The results have no notable differences.}. Namely,  the initial condition is a rotating torus with the density maximum located at $r=100r_s$, exactly same with Model A. So the initial $Be$ is identical in the two models. The difference is that an initial magnetic field is added. The field is poloidal, confined to the interior of the torus. The loops are parallel to the density loops. The readers are referred to Stone \& Pringle (2001) for details. One reason we adopt this model is because that, as Hawley \& Krolik (2002) argued, compared to the initial toroidal field configuration, the poloidal configuration is more realistic.

We will see that while the main results remain similar for the different initial conditions, some results do differ. Especially, as we will shown in \S3.3, the sign of $Be$ is largely dependent on the value of $Be$ in the initial condition. Then a question is what initial condition is more realistic. This is a difficult question, the answer may depend on different environments. These environments include, among other things, the accretion flow formed in Roche Lobe overflow or stellar wind in the case of X-ray binaries, the center of galaxies in the case of various type of AGNs, molecular cloud in the case of star formation, or even a rotation disk in the case of planetary formation. In terms of the Bernoulli parameter of the initial gas, if the radiative energy loss can be neglected during their accretion, we expect $Be\ga 0$ since they come from infinity. This may be the case of most elliptical galaxies in which hot interstellar medium (ISM) fuels the black hole such as M~87, and some spiral galaxies such as Sgr A*. In this case, Models B \& C are likely more realistic than Models A \& D. But if radiation is important, as perhaps in the case of most spiral galaxies and black hole X-ray binaries where the gas accreted is cool and dense, usually we think that at large radius the accretion flow is described by a standard thin disk. This disk may be truncated and replaced by a hot accretion flow within a transition radius, if the accretion rate is not too large (see \S7.1 for more details). It is unclear in this case the sign of $Be$ of the inner hot accretion flow, although it is certain that $Be<0$ for the outer truncated thin disk. This is because that the physics of the transition from the outer thin disk to the inner hot accretion flow is still an unsolved problem.  The proposed two models include the evaporation and radial turbulent transport. There must be energy transfer either vertically (in the former model) or radially (in the latter model). Therefore the value of $Be$ must be increased after the transition. Similar uncertainty exists for the hot corona sandwiched the thin disk.

\subsection{Approach}

We use the ZEUS code (Stone \& Norman 1992a, 1992b) in spherical geometry to solve the equations. The two modifications to the code are to include the shear stress
terms (in the HD case) and the implementation of the Paczy\'nsky \& Wiita (1980) potential. The computational grids extend from an inner boundary at $r=1.2r_s$ to $r=500r_s$. The inner boundary is small enough to make sure that our solution is transonic. The standard outflow boundary condition is adopted at both the inner and outer radial boundaries. In the angular direction, the boundary conditions are set by symmetry at the poles. To adequately resolve the flow, we adopt the same
non-uniform grid in both the radial and angular directions as in
SPB99 and Stone \& Pringle (2001).

\section{Results: Different Properties of Inflow and Outflow}

\subsection{General Results}

For the aim of the present paper we are only interested in the steady state of the solution. We judge the radial range within which the steady solution is achieved by examining whether the net accretion rate (eq. [\ref{netrate}]) is a constant of radius. The general descriptions to the simulation results of Models A and D can be found in SPB99 and Stone \& Pringle (2001). For Model B, at the beginning of simulation, we found a shock is formed at the inner boundary and it propagates outwards. The flow achieves a steady state after 0.7 orbital times at the outer boundary. In model C, initially, the whole computational domain is filled with very low density interstellar medium. Gas is steadily injected into the calculation domain from the equatorial plane at the outer boundary. The injected gas quickly moves inwards and forms an accretion flow. After about 9 orbital times at the outer boundary, the accretion flow achieves a quasi-steady state.

When we evaluate the radial distribution of physical quantities, such as $Be$, velocity, and temperature, we need to calculate the angle-integrated average. We adopt the mass flux-weighted value, which is defined as (for quantity $q$): \be
<q(r)>=\frac{2\pi r^2\int^\pi_0\rho q~{\rm max}(v_r,0){\rm
sin}{\theta}d\theta}{2\pi r^2\int^\pi_0\rho~ {\rm max}(v_r,0){\rm
sin}{\theta}d\theta},\label{fluxweight} \ee
for outflow; and
\be <q(r)>=\frac{2\pi r^2\int^\pi_0\rho q~{\rm
min}(v_r,0){\rm sin}{\theta}d\theta}{2\pi r^2\int^\pi_0\rho~ {\rm
min}(v_r,0){\rm sin}{\theta}d\theta}, \label{fluxweightin} \ee
for inflow. We have also tried the density-weighted values, which are defined as
\be <q_{d}(r)>=\frac{2\pi r^2\int^\pi_0\rho q~{\rm max}(v_r,0)/v_r{\rm
sin}{\theta}d\theta}{2\pi r^2\int^\pi_0\rho~{\rm max}(v_r,0)/v_r{\rm
sin}{\theta}d\theta},\label{densityweight} \ee and \be <q_{d}(r)>=\frac{2\pi r^2\int^\pi_0\rho q~{\rm min}(v_r,0)/v_r{\rm
sin}{\theta}d\theta}{2\pi r^2\int^\pi_0\rho~{\rm min}(v_r,0)/v_r{\rm
sin}{\theta}d\theta},\label{densityweightin} \ee for outflow and inflow, respectively. We found the results are very similar. In the following we compare the various properties of inflow and outflow.

\subsection{Inflow and Outflow Rates}

\begin{figure*}
\epsscale{0.45} \plotone{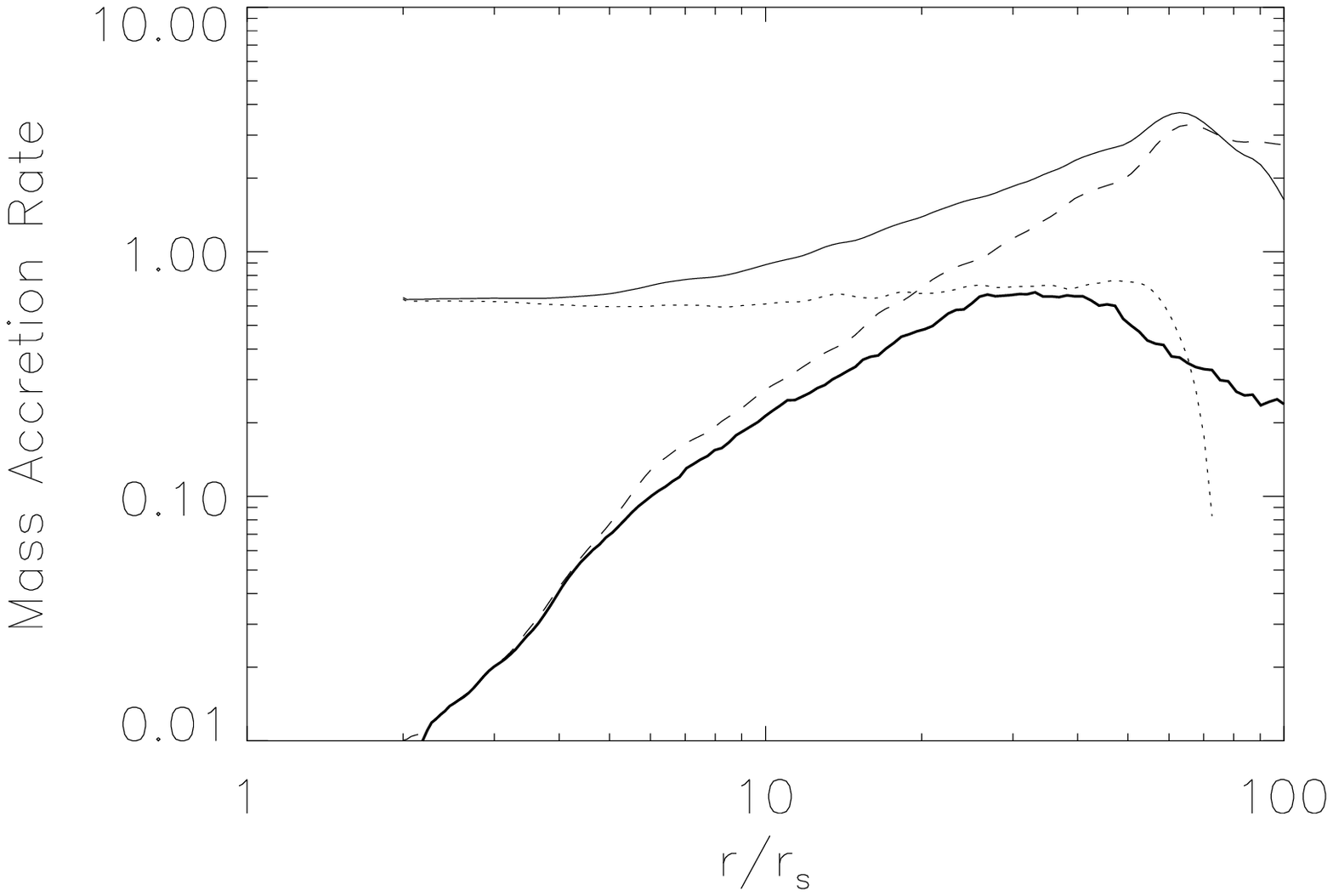}\hspace{1.cm} \epsscale{0.45}
\plotone{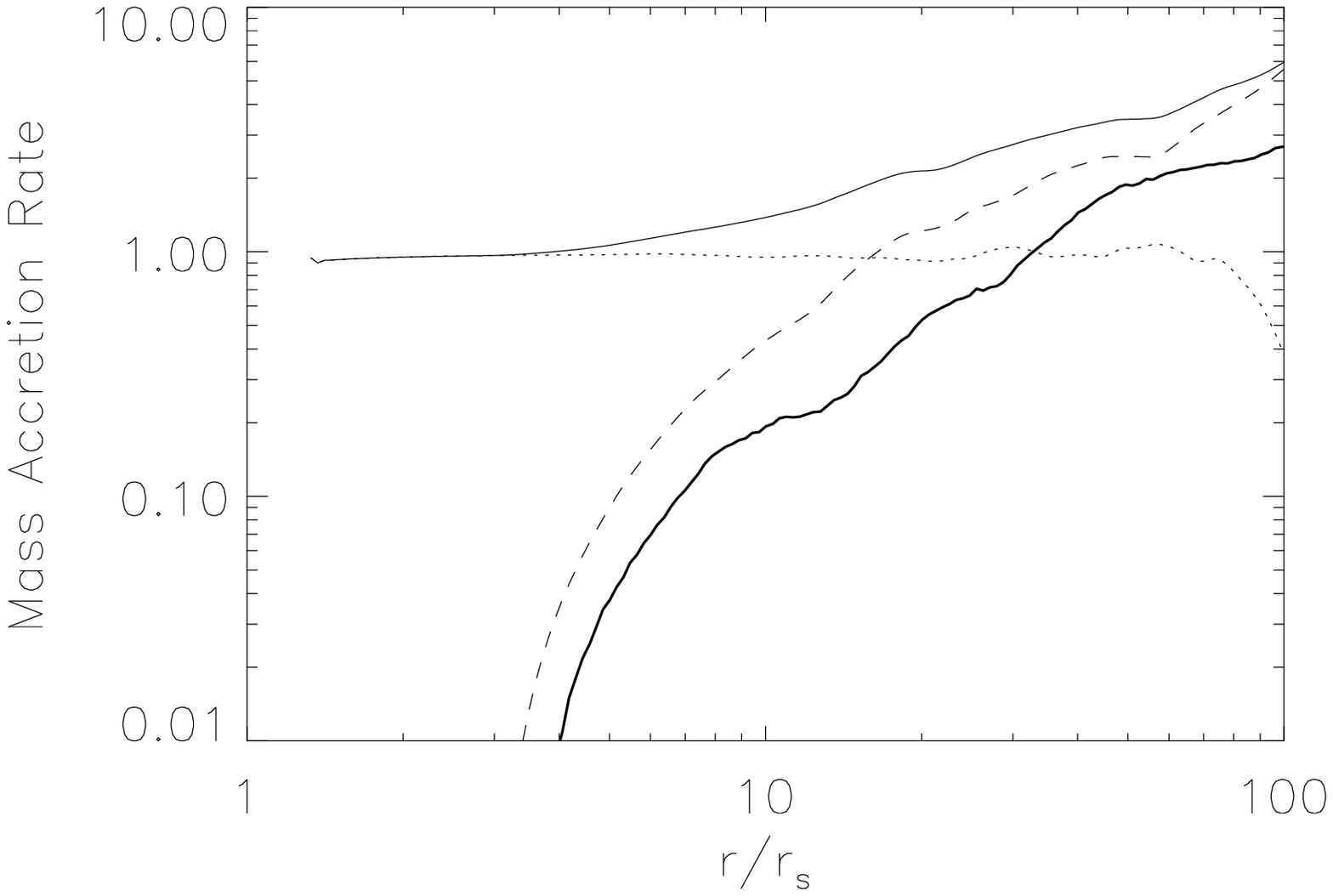} \vspace{0.2in}\epsscale{0.45} \plotone{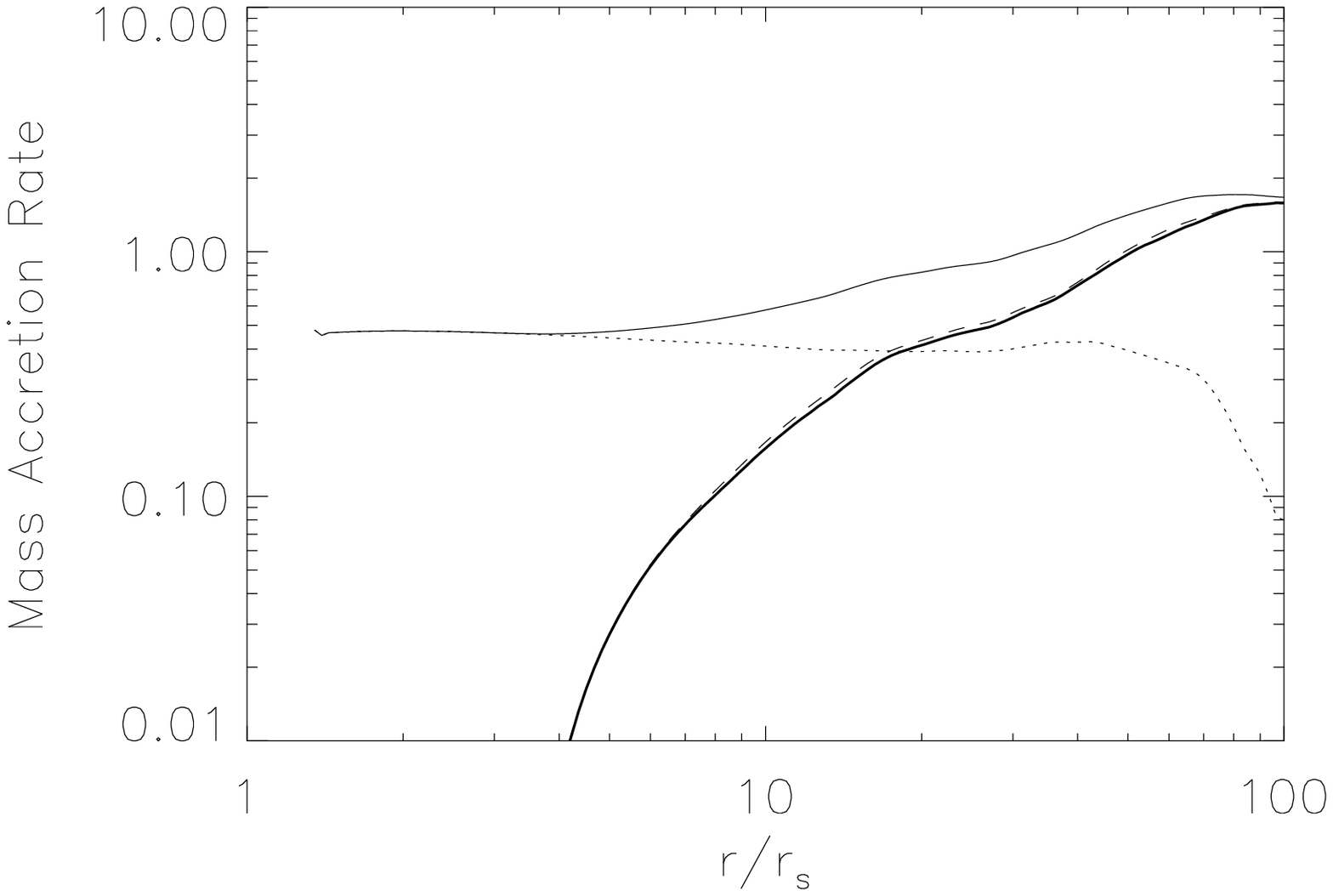}
\hspace{1cm} \epsscale{0.45} \plotone{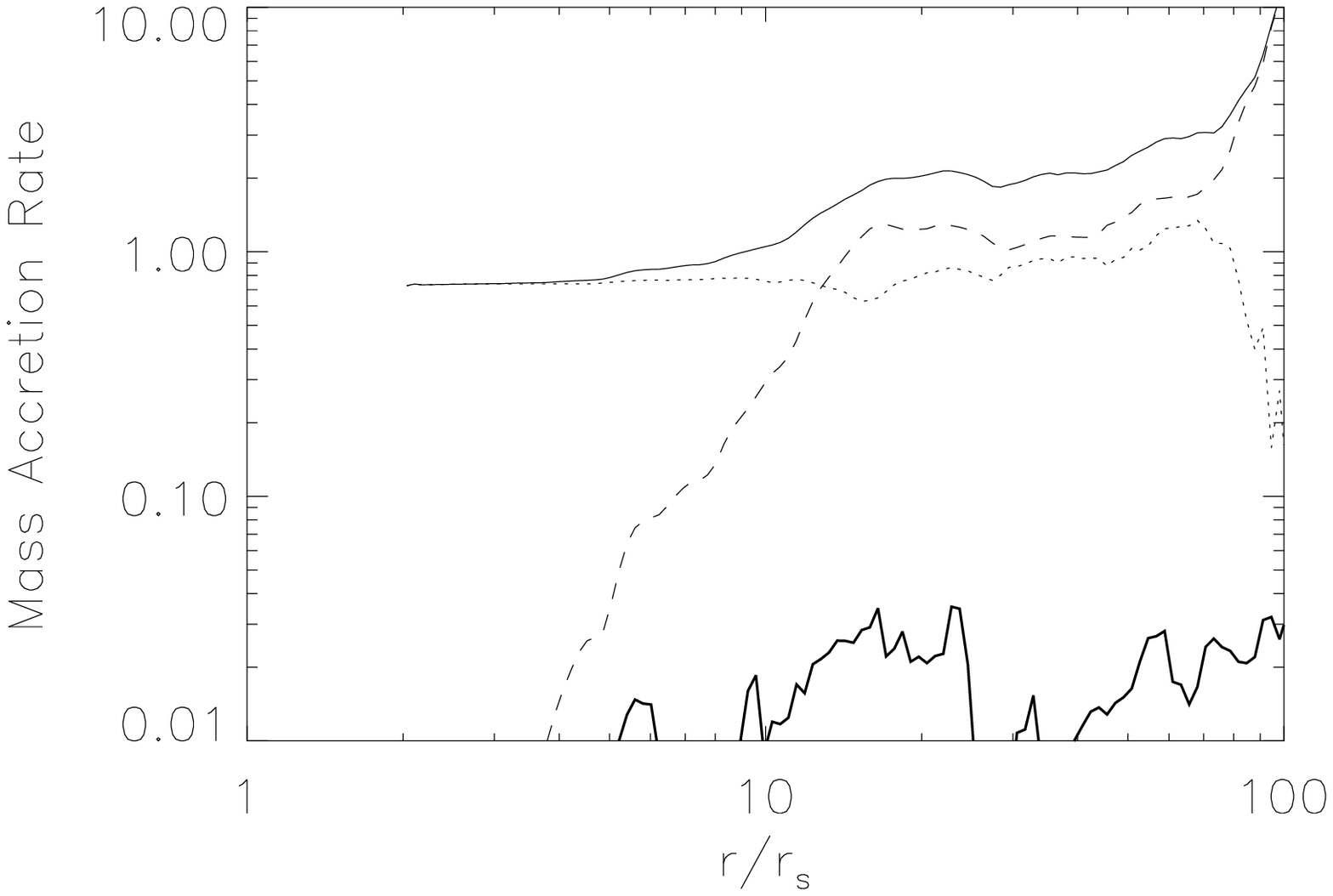}\vspace{0.3in}
\caption{The radial distribution of inflow (thin solid), outflow (dashed) and net rates (dotted). The thick solid lines denotes the rate of outflow with a positive $Be$. The top-left, top-right, bottom-left, and bottom-right plots are for Models A, B, C, and D, respectively.}
\label{Fig:mdot}
\end{figure*}

Following SPB99, the mass inflow and outflow rates, $\dot {M}_{\rm in}$ and $\dot {M}_{\rm out}$, are defined as the following time-averaged and angle-integrated quantities,
\begin{equation}
 \dot{M}_{\rm in}(r) = 2\pi r^{2} \int_{0}^{\pi} \rho \min(v_{r},0)
   \sin \theta d\theta,
\label{inflowrate}
\end{equation}
\begin{equation}
 \dot{M}_{\rm out}(r) = 2\pi r^{2} \int_{0}^{\pi} \rho \max(v_{r},0)
    \sin \theta d\theta.
\label{outflowrate}
\end{equation}
The net mass accretion rate is,
\begin{equation}
\dot{M}_{\rm net}(r)=\dot{M}_{\rm in}(r)-\dot{M}_{\rm out}(r).\label{netrate}
\end{equation}
The net rate is the accretion rate that finally falls onto the black hole. Note that the above rates are obtained by time-averaging integral rather than integrating the time-averages.

Fig. \ref{Fig:mdot} shows the radial distribution of the inflow and outflow rates and net rate of the four models. The radial profile of the inflow rate in all four models can be described by $\dot{M}_{\rm in}(r)\propto r^{s}$. Regarding the value of $s$, Paper I presents a more precise measurement since the dynamical range of simulations presented there is much larger than in the present work. The radial profile of density is also measured, and it was found that $\rho(r)\propto r^{-p}$. Moreover, the relationship of $s=1.5-p$ is well satisfied, which indicate that the flattening of density is caused by the inward decrease of accretion rate. Another notable result is that the accretion rate profiles for all the four models are roughly similar. In Paper I, we have discussed this consistency in a much more comprehensive and detailed way by combining all published HD and MHD simulation works. Also shown in the figure by the thick solid lines are the outflow rate with a positive $Be$. We will discuss this issue later.

\begin{figure*}
\epsscale{0.45} \plotone{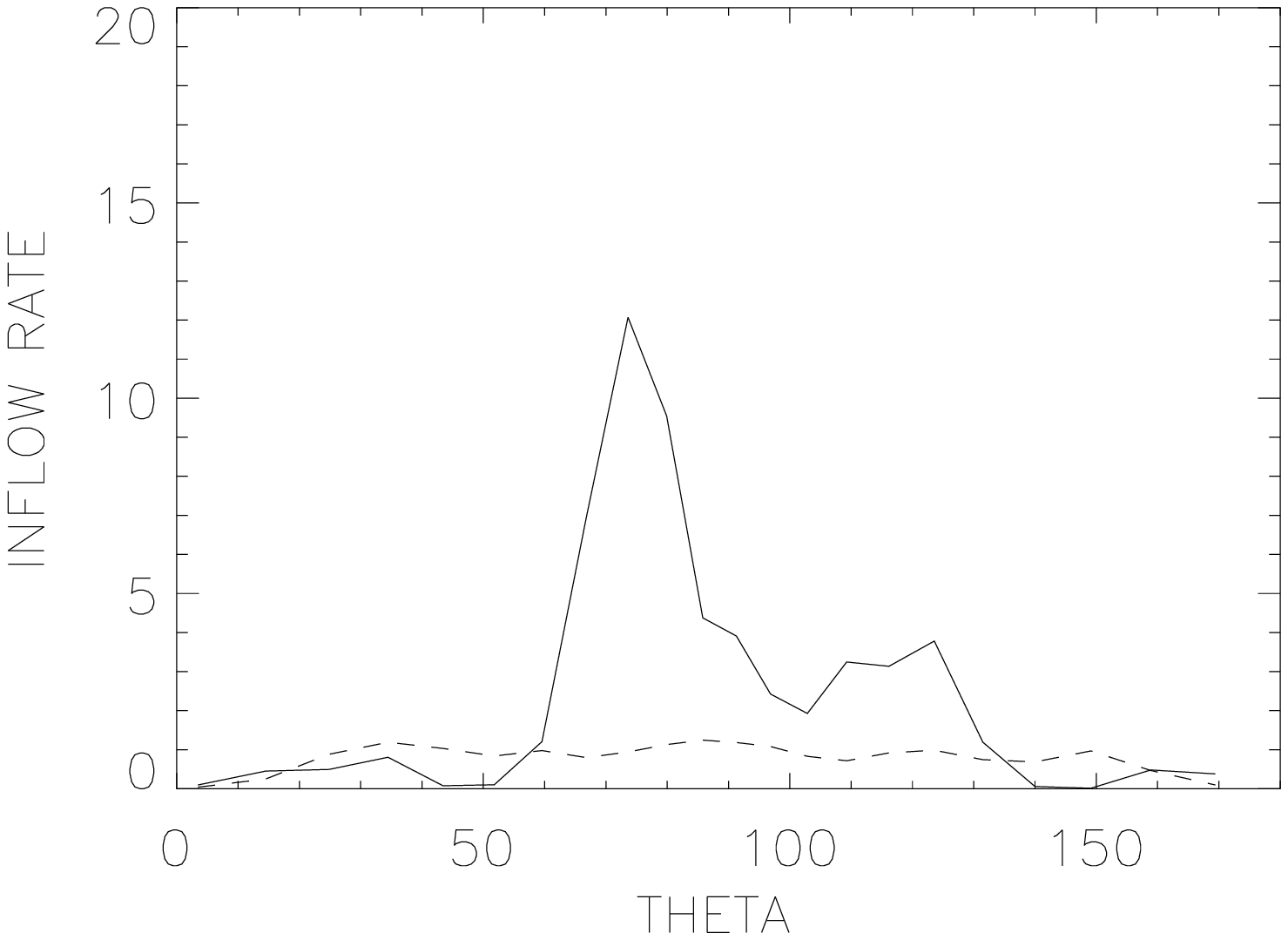}\hspace{1.cm} \epsscale{0.45}
\plotone{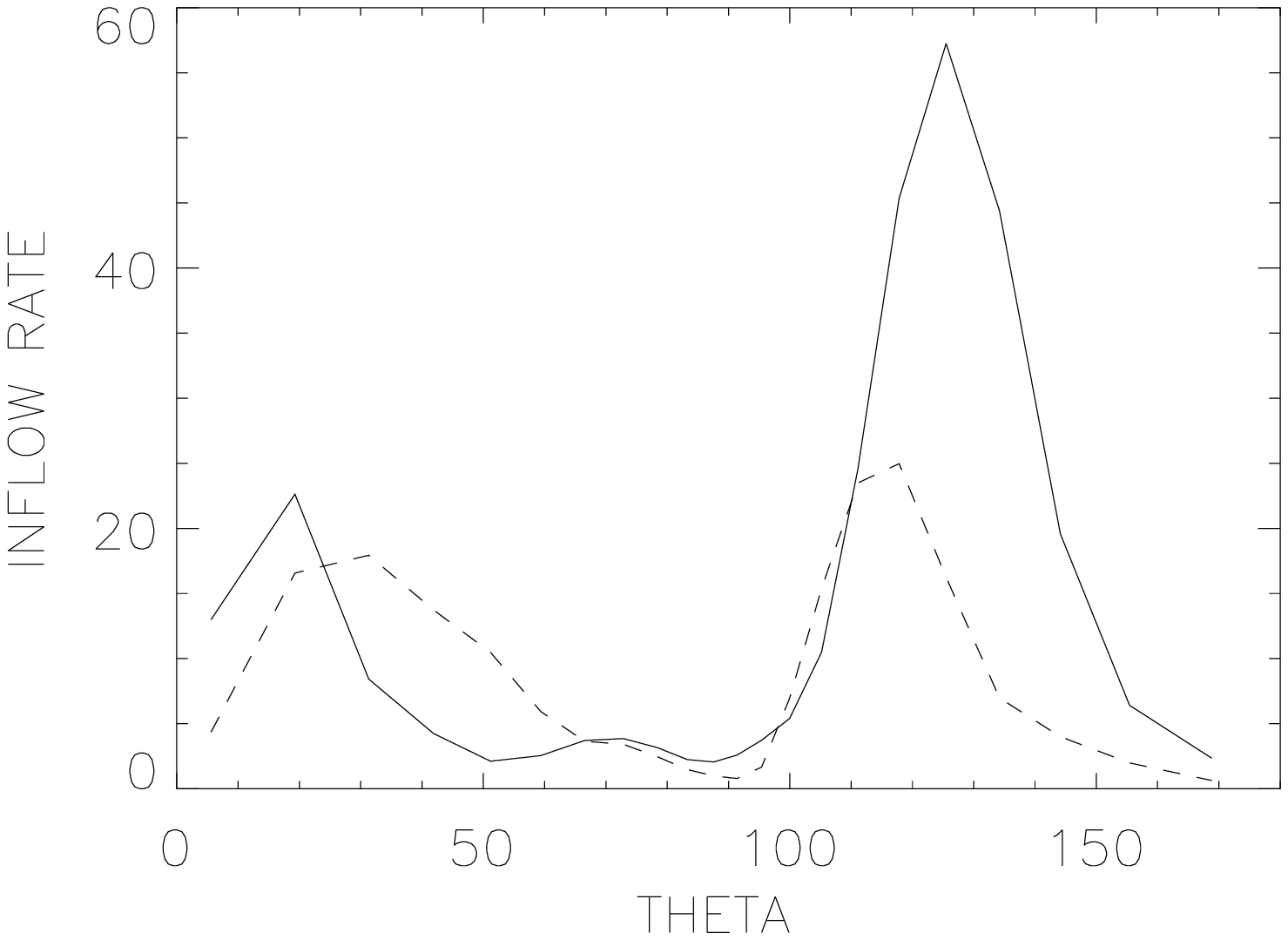} \vspace{0.2in}\epsscale{0.45} \plotone{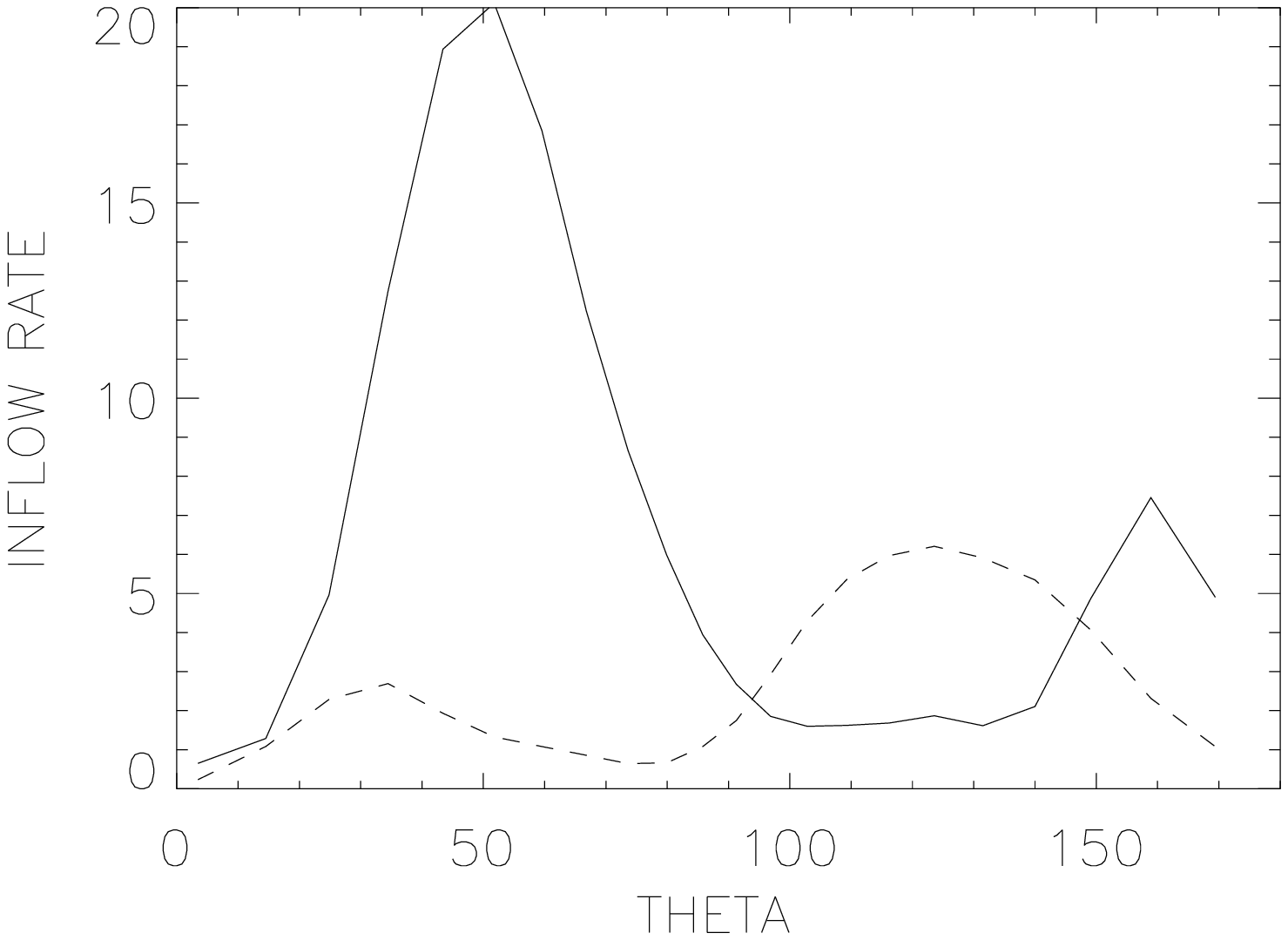}
\hspace{1cm} \epsscale{0.45} \plotone{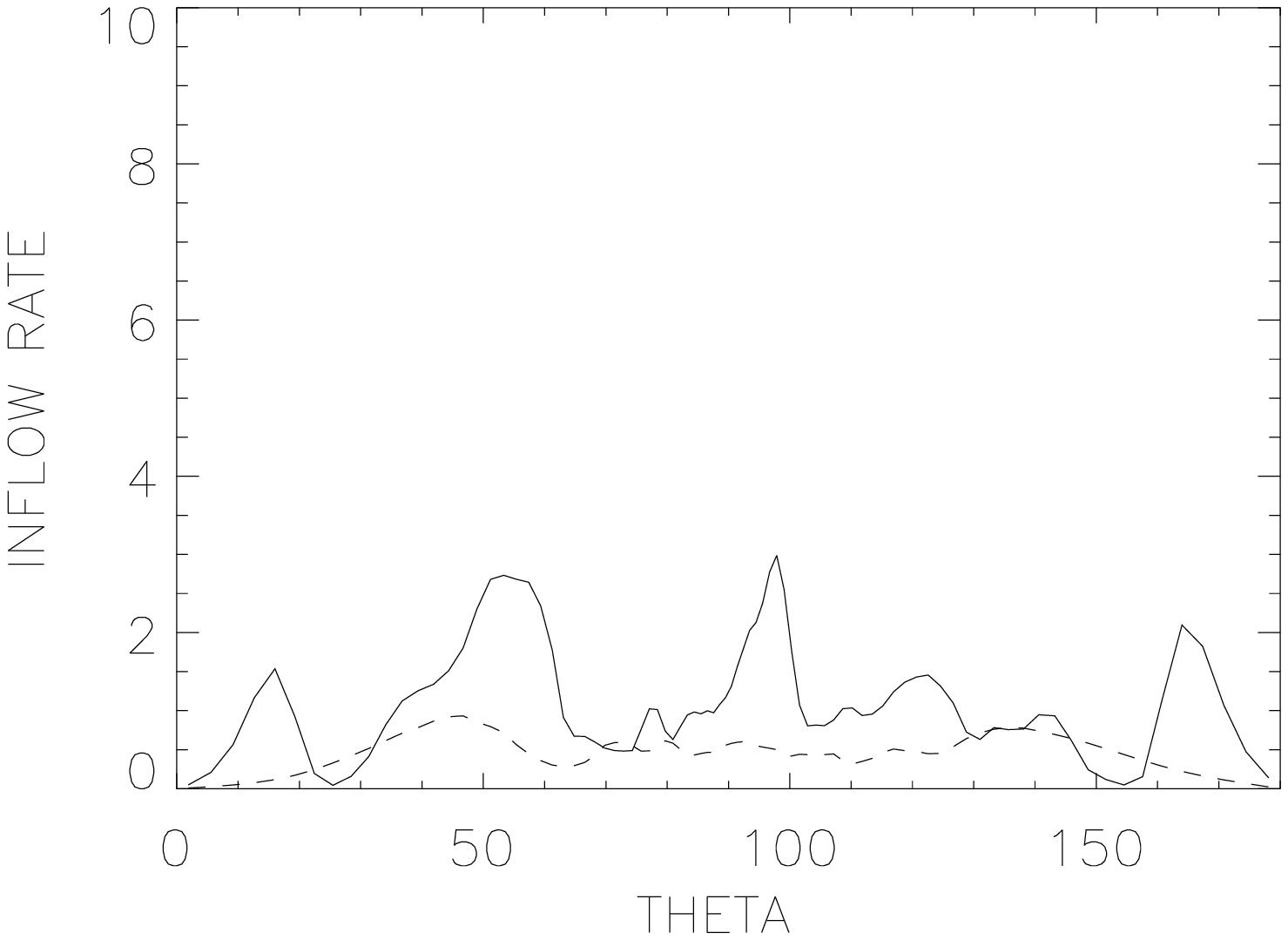}\vspace{0.3in}
\caption{The angular distribution of inflow rate. The solid and dashed lines are for $r=50r_s$ and $10r_s$, respectively. The top-left, top-right, bottom-left, and bottom-right plots are for Models A, B, C, and D, respectively.}
\label{Fig:mdotinflow}
\end{figure*}

\begin{figure*}
\epsscale{0.45} \plotone{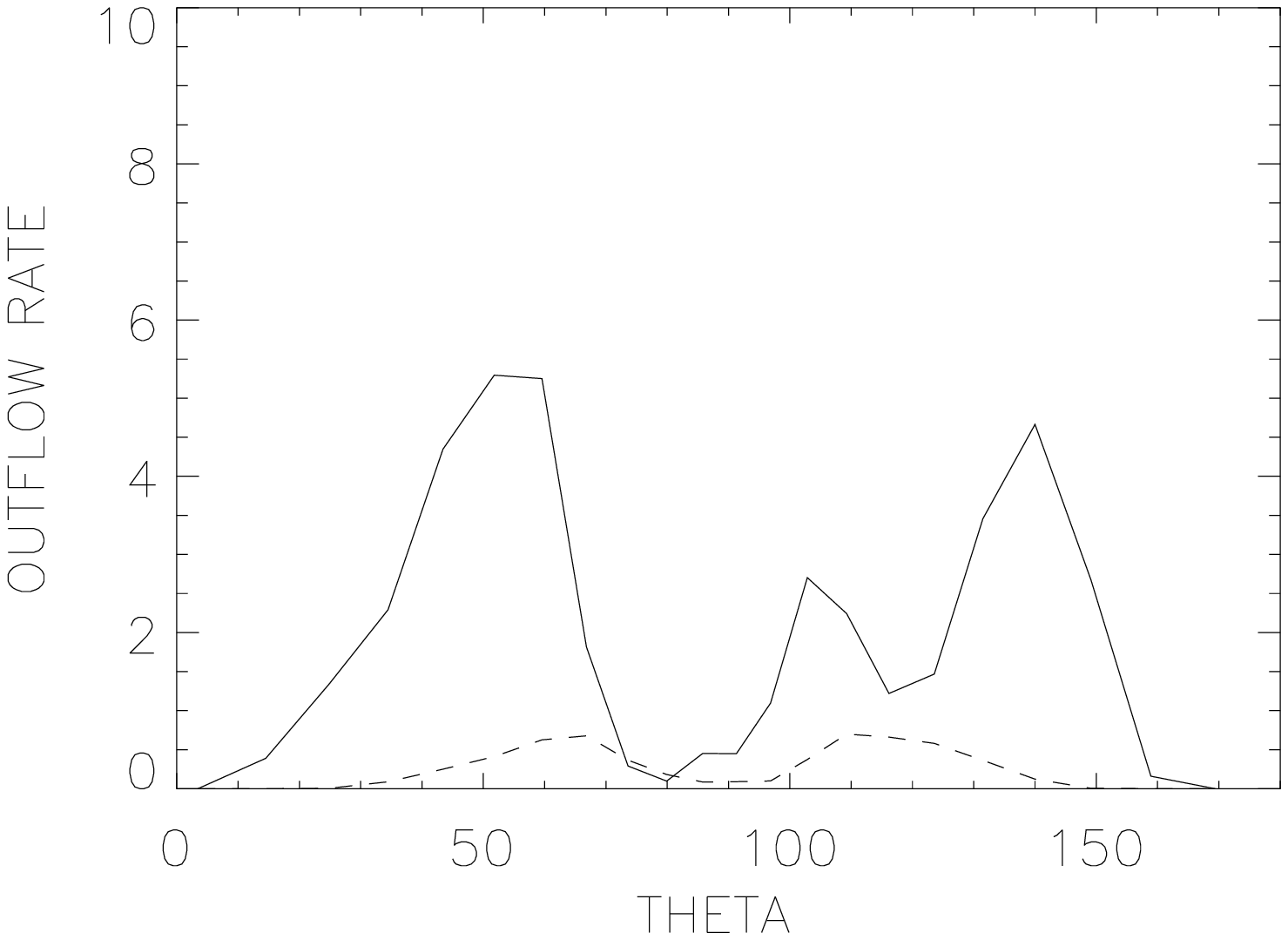}\hspace{1.cm} \epsscale{0.45}
\plotone{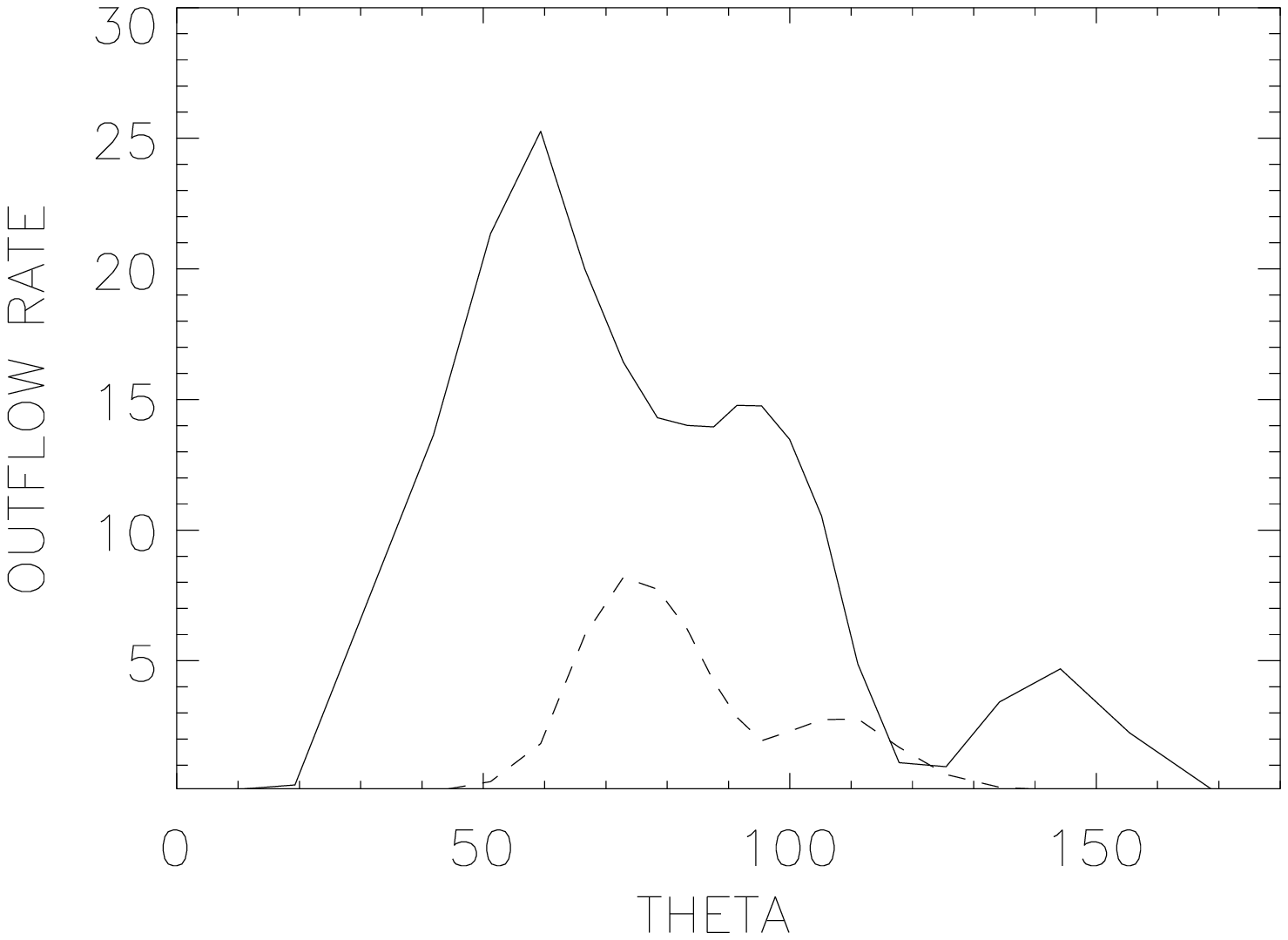} \vspace{0.2in}\epsscale{0.45} \plotone{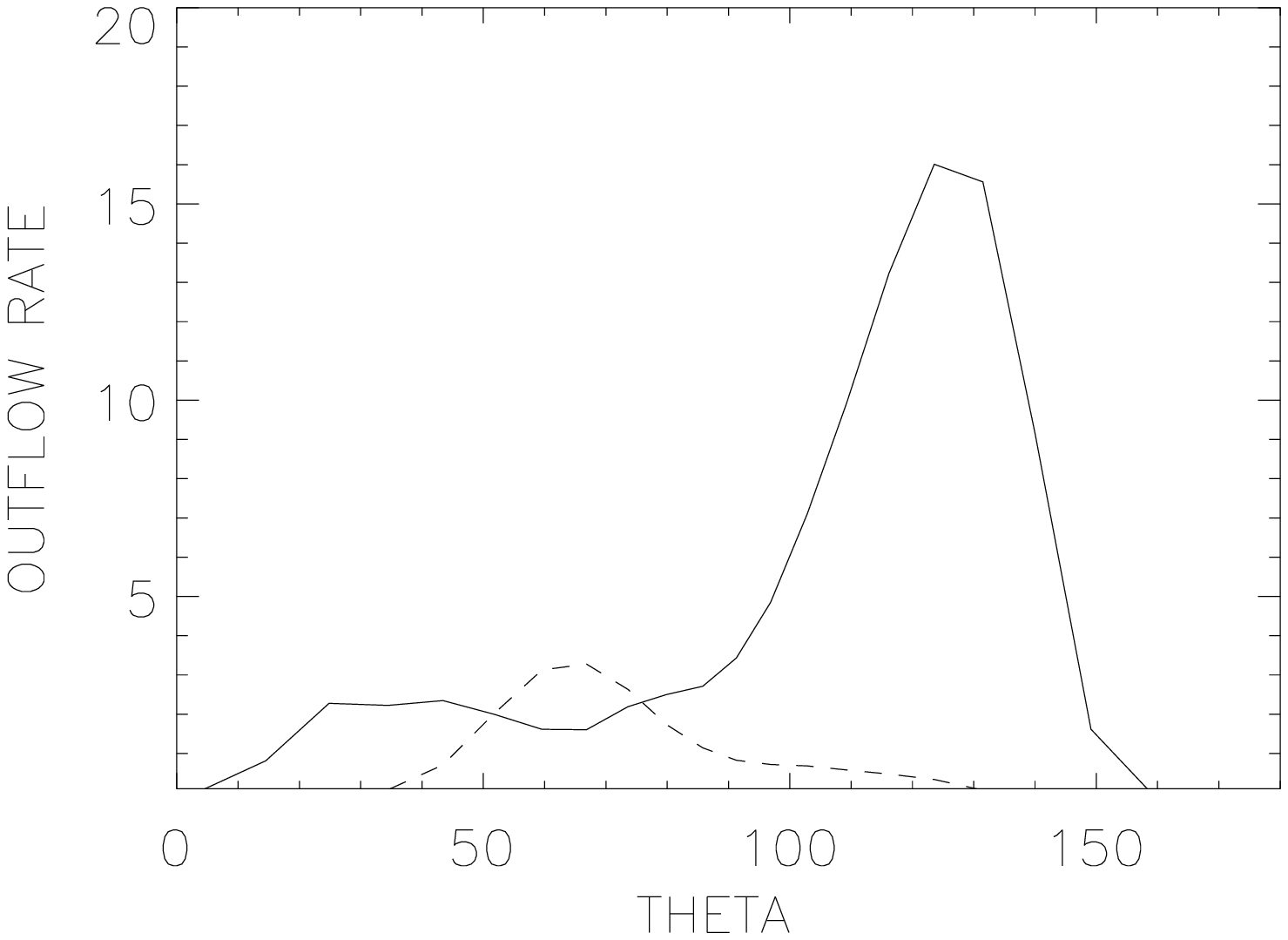}
\hspace{1cm} \epsscale{0.45} \plotone{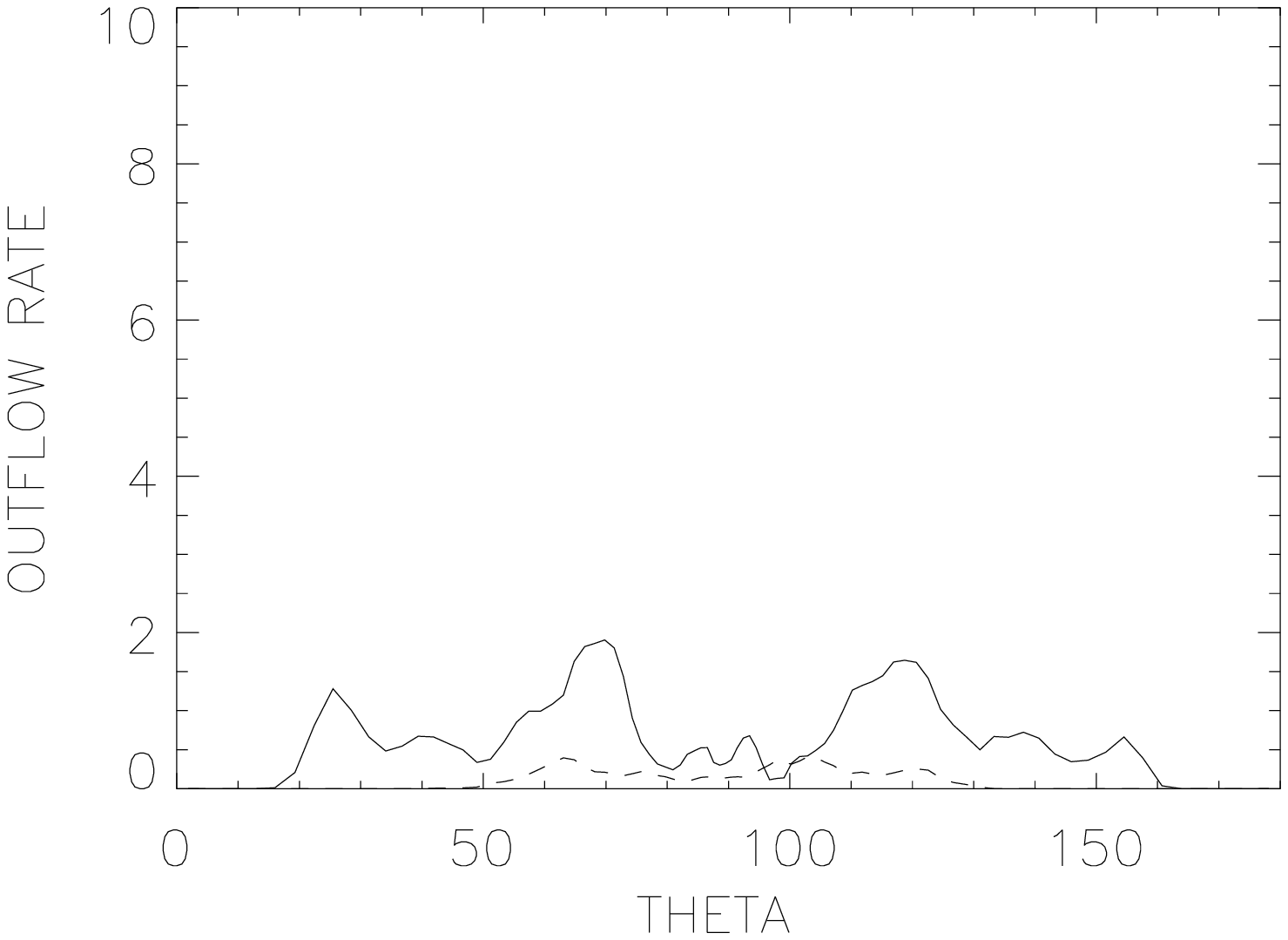}\vspace{0.3in}
\caption{The angular distribution of outflow rate. The solid and dashed lines are for $r=50r_s$ and $10r_s$, respectively. The top-left, top-right, bottom-left, and bottom-right plots are for Models A, B, C, and D, respectively.}
\label{Fig:mdotoutflow}
\end{figure*}

We study the angular distribution of the mass rate by time-averaging the following
outflow and inflow rates as a function of $\theta$: \be \dot{M}_{\rm
out}(\theta)=2\pi r^2 \rho ~{\rm max}(v_r,0){\rm sin}\theta \Delta
\theta.\label{angulardis}\ee \be \dot{M}_{\rm in}(\theta)=2\pi r^2 \rho ~{\rm min}(v_r,0){\rm
sin}\theta \Delta \theta.\label{angulardis2}\ee  Figs. \ref{Fig:mdotinflow} \& \ref{Fig:mdotoutflow} show the angular distribution of inflow rate and  outflow rate respectively. We can see that their angular distributions are in general quite broad; there is no angle where there is only inflow or outflow in the time-average sense. The distributions of Models A and D are similar, they are nearly symmetric to the equatorial plane. The distribution of Models B \& C are similar, peaked at a $\theta$ angle away from the equatorial plane. The discrepancy between these two groups is obviously because of the initial conditions. Models A \& D have the same initial condition, while Models B \& C are similar since in both models the gas has an initial radial velocity. Comparing these two figures, we can see that for all the four models, the angular distributions of inflow and outflow roughly match well respectively, in the sense that in the region where inflow is strong the outflow is weak.  For example, for Model A, most of the inflow occurs within $\theta=70^{\circ}-90^{\circ}$; while in this region the outflow is very weak. Mots of the outflow occurs at the surface of the accretion flow, peaking at $\theta=50^{\circ}$ and $140^{\circ}$. Model D is similar to Model A.

For Models B \& C, the distributions of both inflow and outflow rates are not symmetric to the equatorial plane and there is one peak, which locates at $\theta\sim 50^{\circ}$ and $130^{\circ}$ depending on radius and models. But again, outflow is strong in the region where inflow is weak. For example, for Model B, at $r=50$, most of the inflow occurs below the equatorial plane, around $\theta\sim (110-140)^{\circ}$; while most of the outflow occurs above the equatorial plane, around $\theta\sim (40-70)^{\circ}$. This suggests that the inflow and outflow rates are dominated by large-scale bulk motion rather than fluctuations, and the bulk of the inflowing gas is turn around and becomes outflow. As we will discuss later, the outflowing motion is caused by the buoyant force. We speculate that the reason for the difference of the angular distribution between Models B\&C and Models A\&D may be because of the difference of the value of $Be$ of the initial condition. For Models B\&C, the Bernoulli parameter of the initial gas is significantly higher, which means that the gas is more ``energetic''. This results in  apparently more ``violent'' motion.

\subsection{Bernoulli Parameter}

Bernoulli parameter is the sum of kinetic energy, enthalpy, and
potential energy. \be Be=\frac{v^2}{2}+\frac{\gamma
P}{(\gamma-1)\rho}-\frac{GM}{r-r_s}.\label{bernoulli}\ee
Here the magnetic field is not included. In the case of inviscid flow, the Bernoulli parameter of fluid element along a stream line is conserved.  A positive $Be$ means that the accretion flow has enough energy to overcome gravitational energy and doing work to its surroundings to escape to infinity.

\begin{figure}
\epsscale{0.45} \plotone{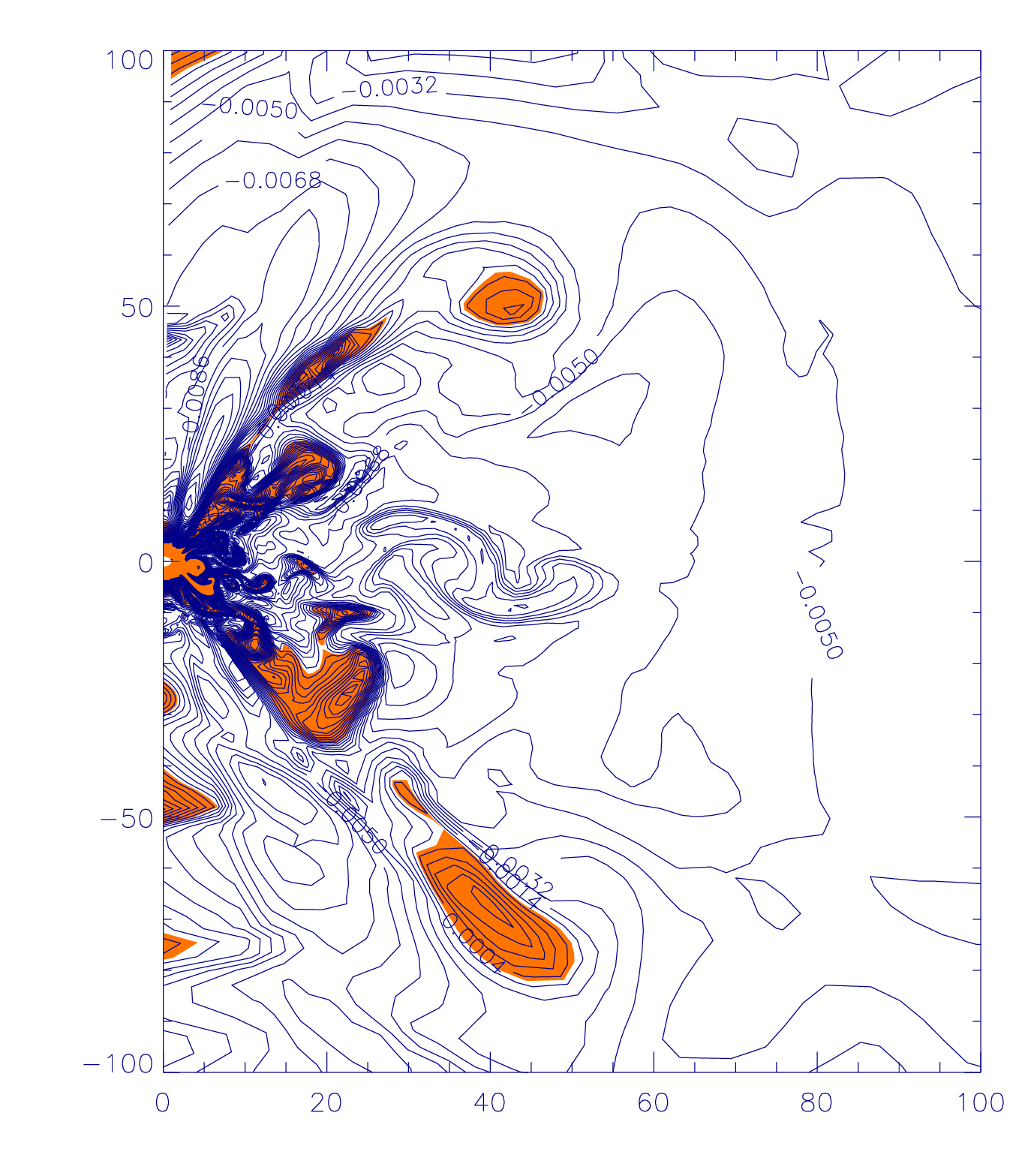}\hspace{1.cm} \epsscale{0.45}
\plotone{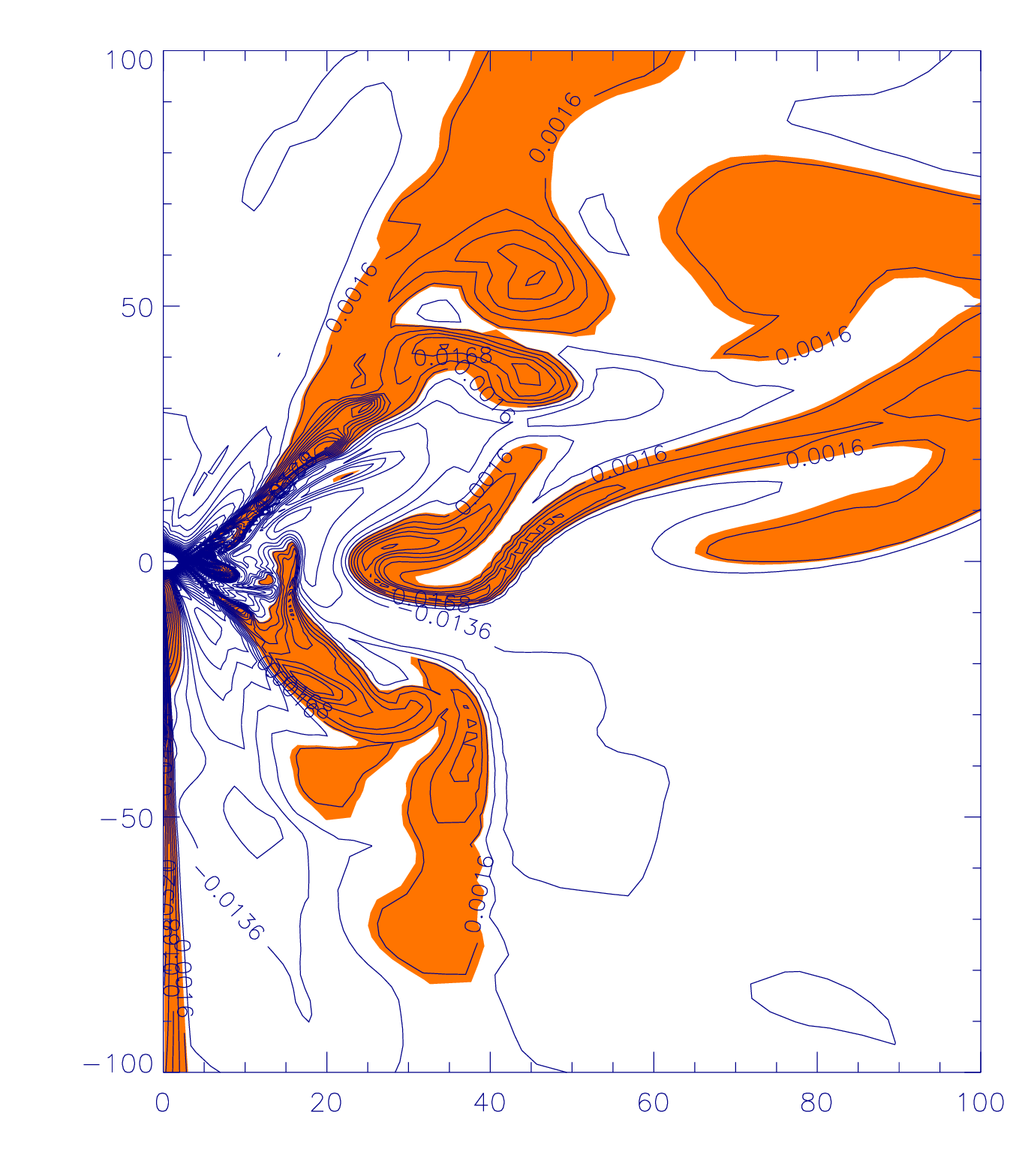} \vspace{0.2in}\epsscale{0.45} \plotone{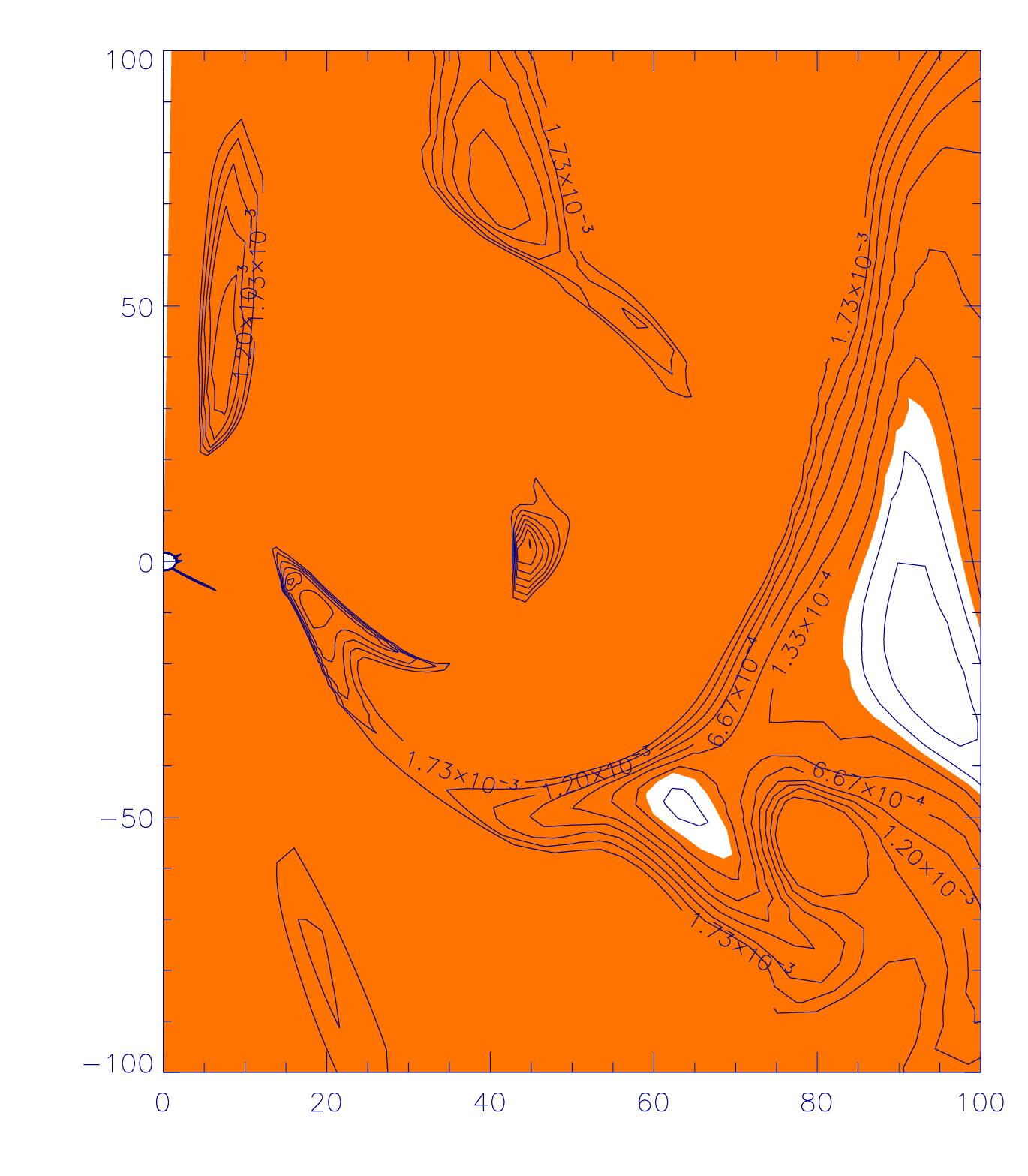}
\hspace{1cm} \epsscale{0.45} \plotone{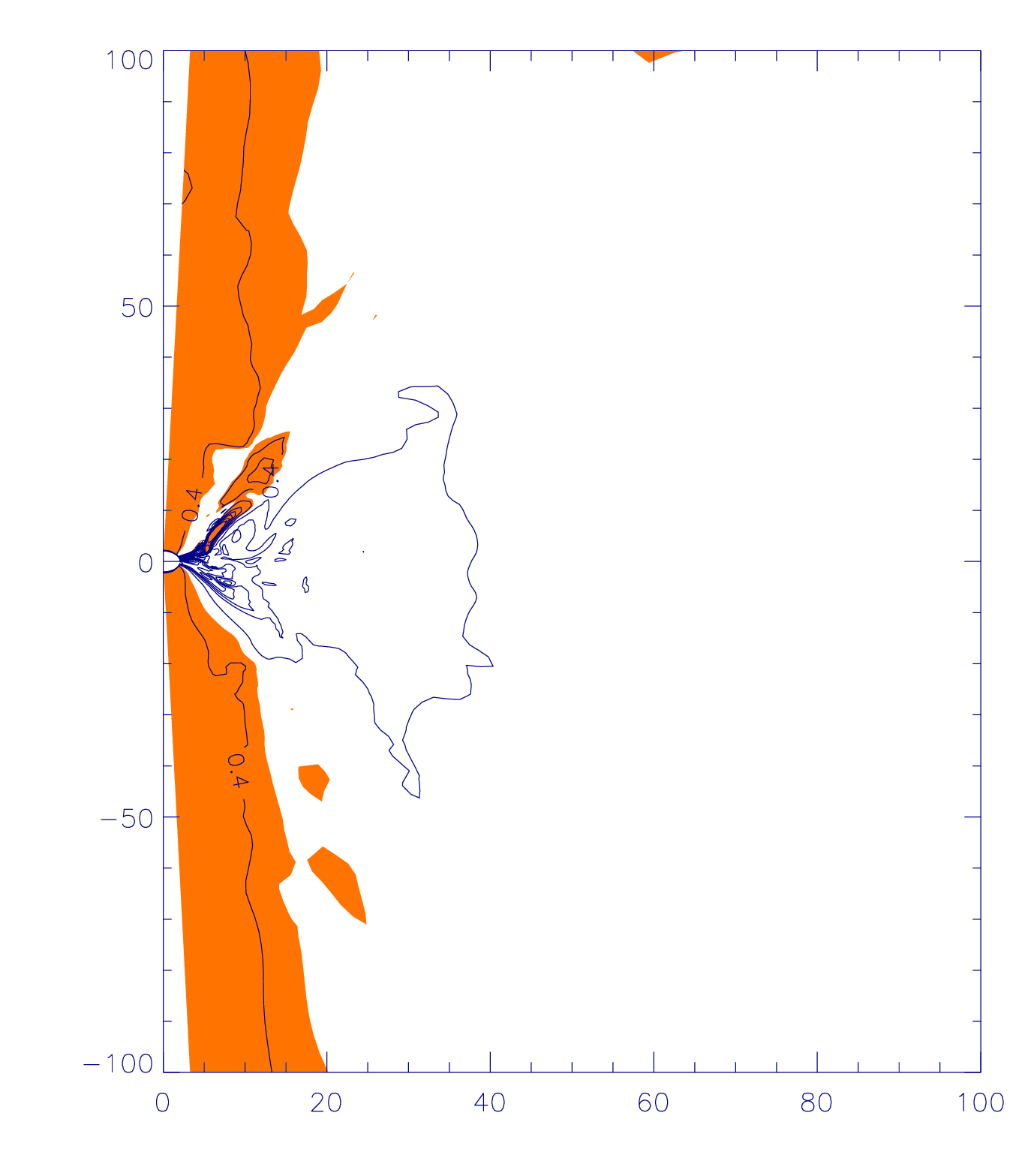}\vspace{0.3in}
\caption{The contour of Bernoulli parameters of the four models. The red-color region denotes a positive  $Be$. The top-left, top-right, bottom-left, and bottom-right plots are for Models A, B, C, and D, respectively.}
\label{Fig:bernoullicontour}
\end{figure}

Fig. \ref{Fig:bernoullicontour} presents the two-dimensional contour of $Be$ of the accretion flow (both inflow and outflow) for the four models. The red region denotes $Be>0$. We can see that the results of the four models are quite different. For Model A in most of the area $Be<0$, as stated in SPB99 and Yuan \& Bu (2010). The latter found that at the outer boundary only about 1\% of the outflow has a positive $Be$.  For Model B, the region with a positive $Be$ becomes larger. For Model C, $Be$ is positive in almost all of the area. The main reason for such differences is that the initial conditions are different, i.e, different models have different initial $Be$. The value of $Be$ along the equatorial plane of the initial condition in the four models is shown by the dotted lines in Fig. \ref{Fig:radialbernoulli}.
For Models A \& D, $Be$ is the smallest and the gas is strongly bound, i.e., $Be<0$. In fact, we find that in the initial torus the temperature $T$ of gas is about 10 times lower than the virial value. For Model B, the value of $Be$ of the initial condition is larger, but still less than zero. For Model C, $Be>0$ for all the injected gas. Since radiation is neglected in our simulation, the total energy of the gas in the computational domain should be roughly conserved (if we neglect the energy loss through the outer boundary). Therefore it is natural to expect that the final value of $Be$ is a function of $Be$ of the initial condition.

Although their initial condition is identical, the contour of Model D and Model A have one significant difference. The value of $Be$ close to the rotation axis in Model D is positive; while it is negative in Model A. Such a difference should be because of the magnetic effect. In Model D, we find that large-scale ordered poloidal magnetic field is present in the small $\theta$ region (and only in that region). Therefore compared to Model A, additional energy flux, coming from the rotational energy of the accretion flow, is transferred from small radius outward along the field line, which increases the value of $Be$ there. This is the Blandford \& Payne (1982) mechanism.

\begin{figure*}
\epsscale{0.5} \plotone{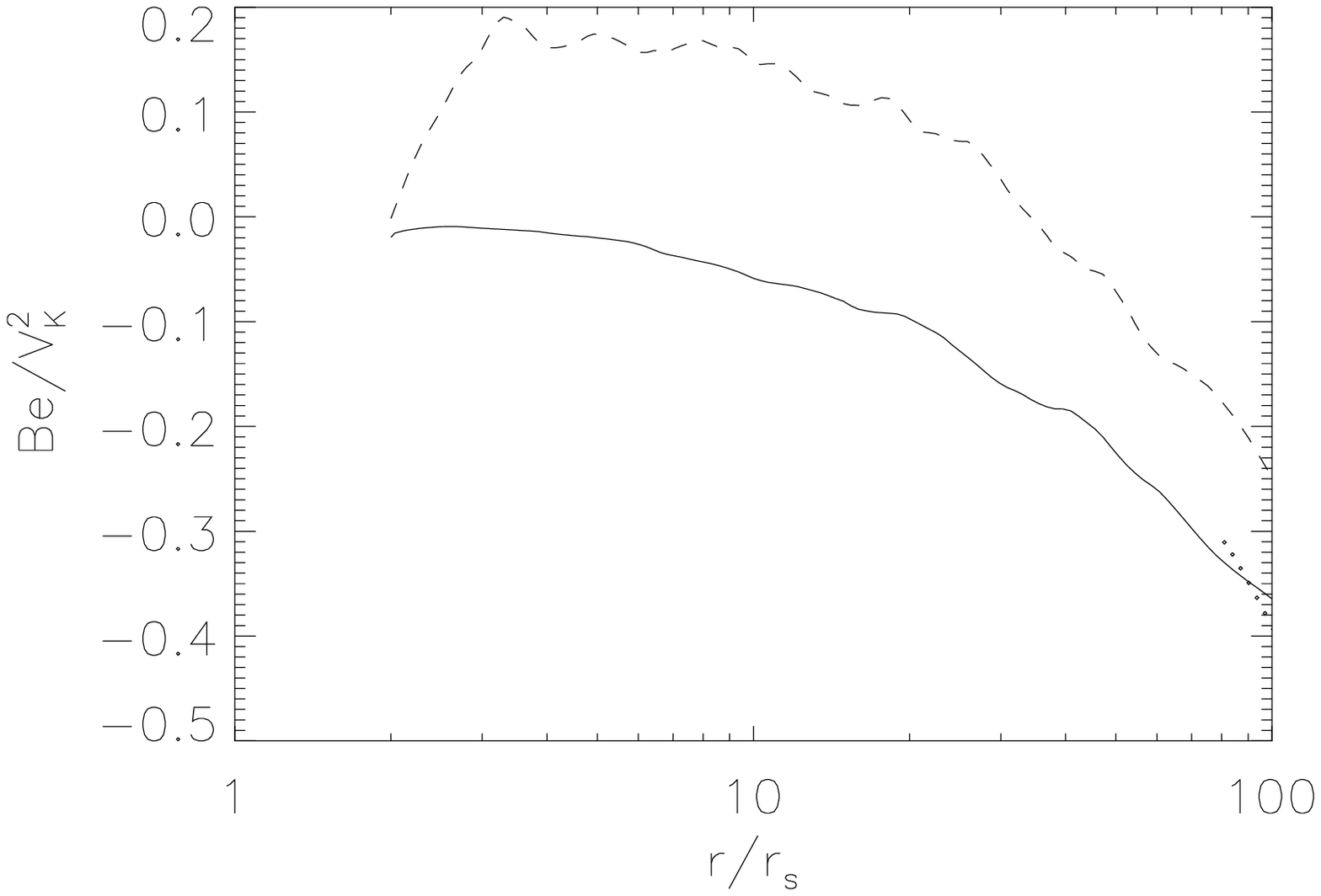}\hspace{1.cm} \epsscale{0.5}
\plotone{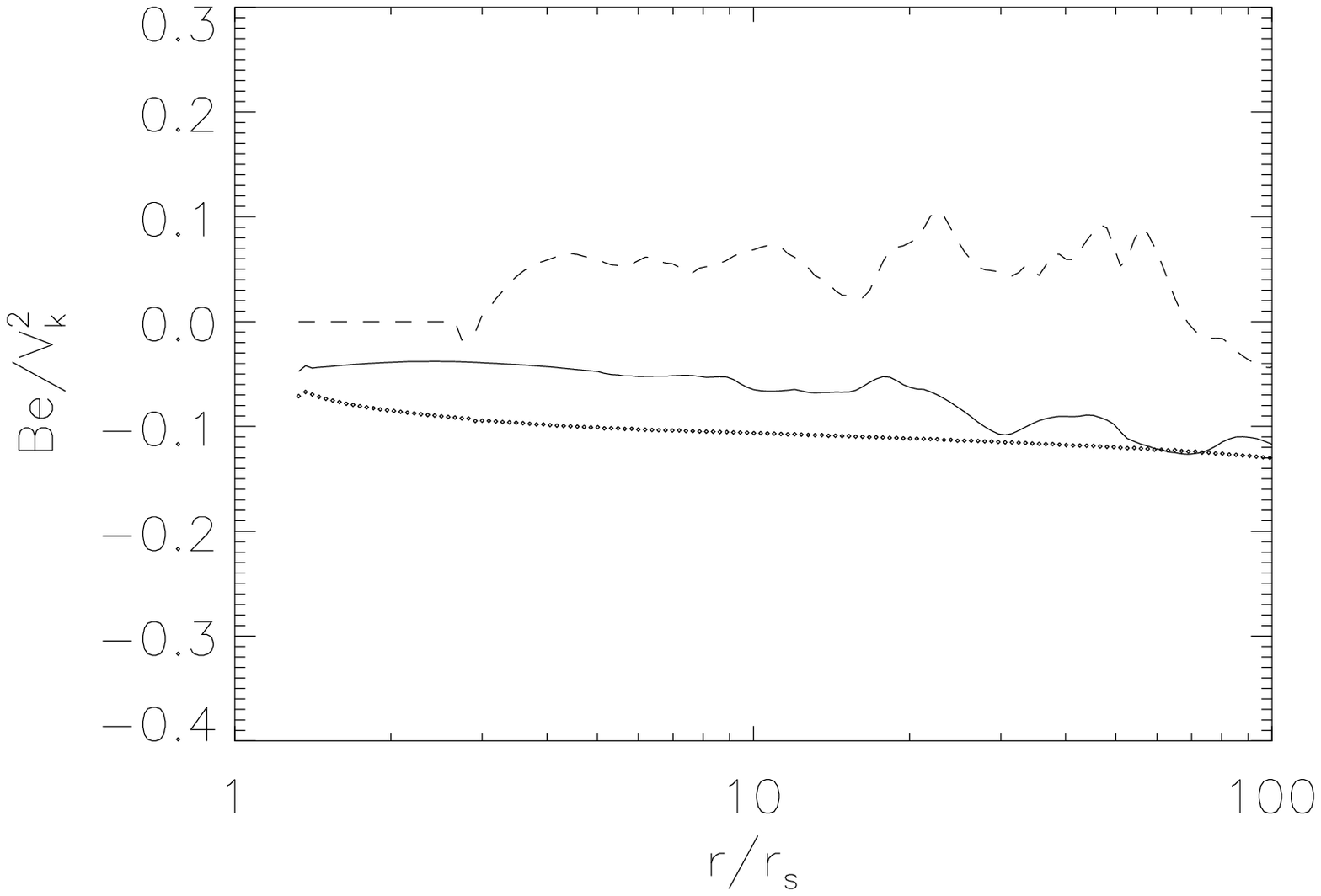} \vspace{0.2in}\epsscale{0.5} \plotone{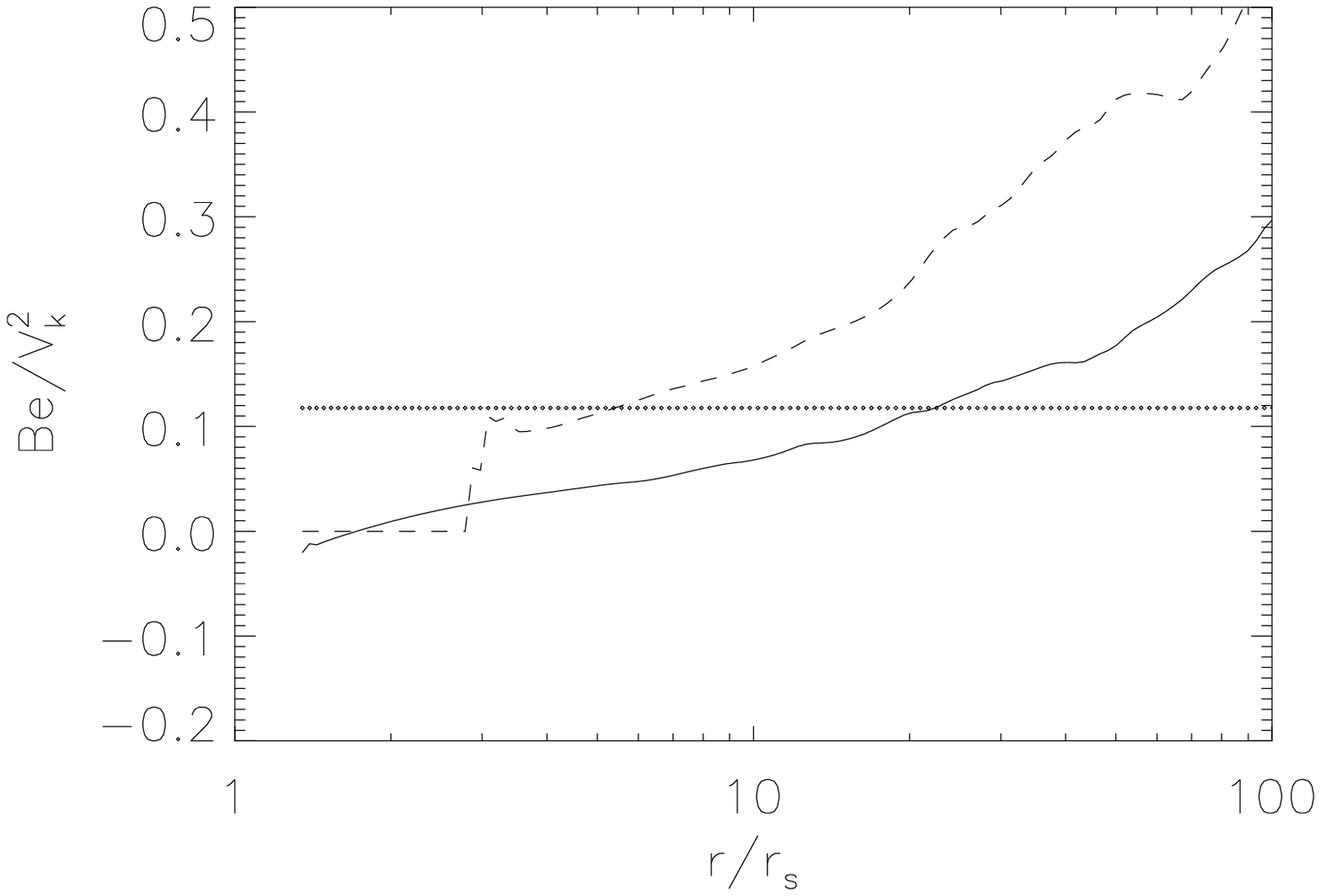}
\hspace{1cm} \epsscale{0.5} \plotone{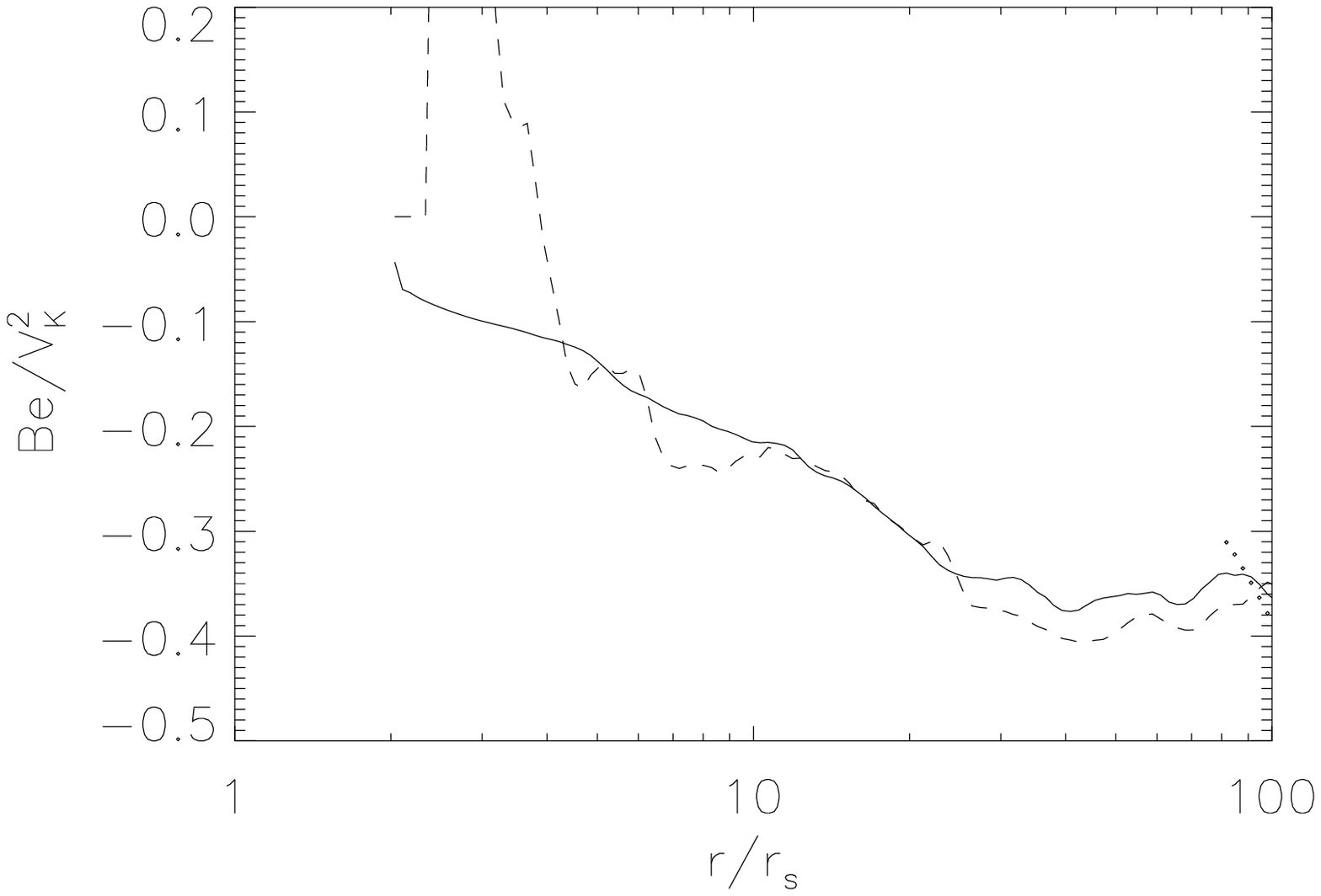}\vspace{0.2in}
\caption{The radial distribution of flux-weighted $Be$ in unit of $v_k^2$. The solid and dashed lines are for inflow and outflow, respectively. The dotted lines or segments  denote the value of $Be$ of the gas in the initial condition. The top-left, top-right, bottom-left, and bottom-right plots are for Models A, B, C, and D, respectively.}
\label{Fig:radialbernoulli}
\end{figure*}

\begin{figure}
\plotone{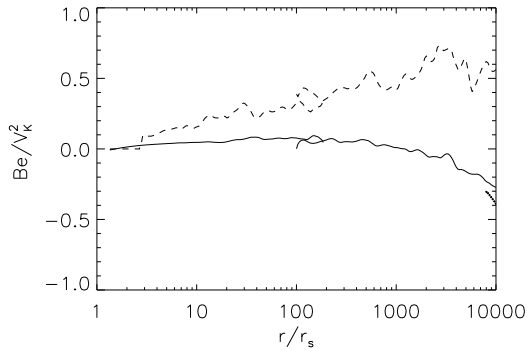}
\caption{The radial distribution of flux-weighted $Be$ for Model A in Paper I. The solid and dashed lines are for inflow and outflow, respectively. The dotted segment denotes the value of $Be$ of the gas in the initial torus.}
\label{Fig:largetorusbe}
\end{figure}

\begin{figure*}
\epsscale{0.5} \plotone{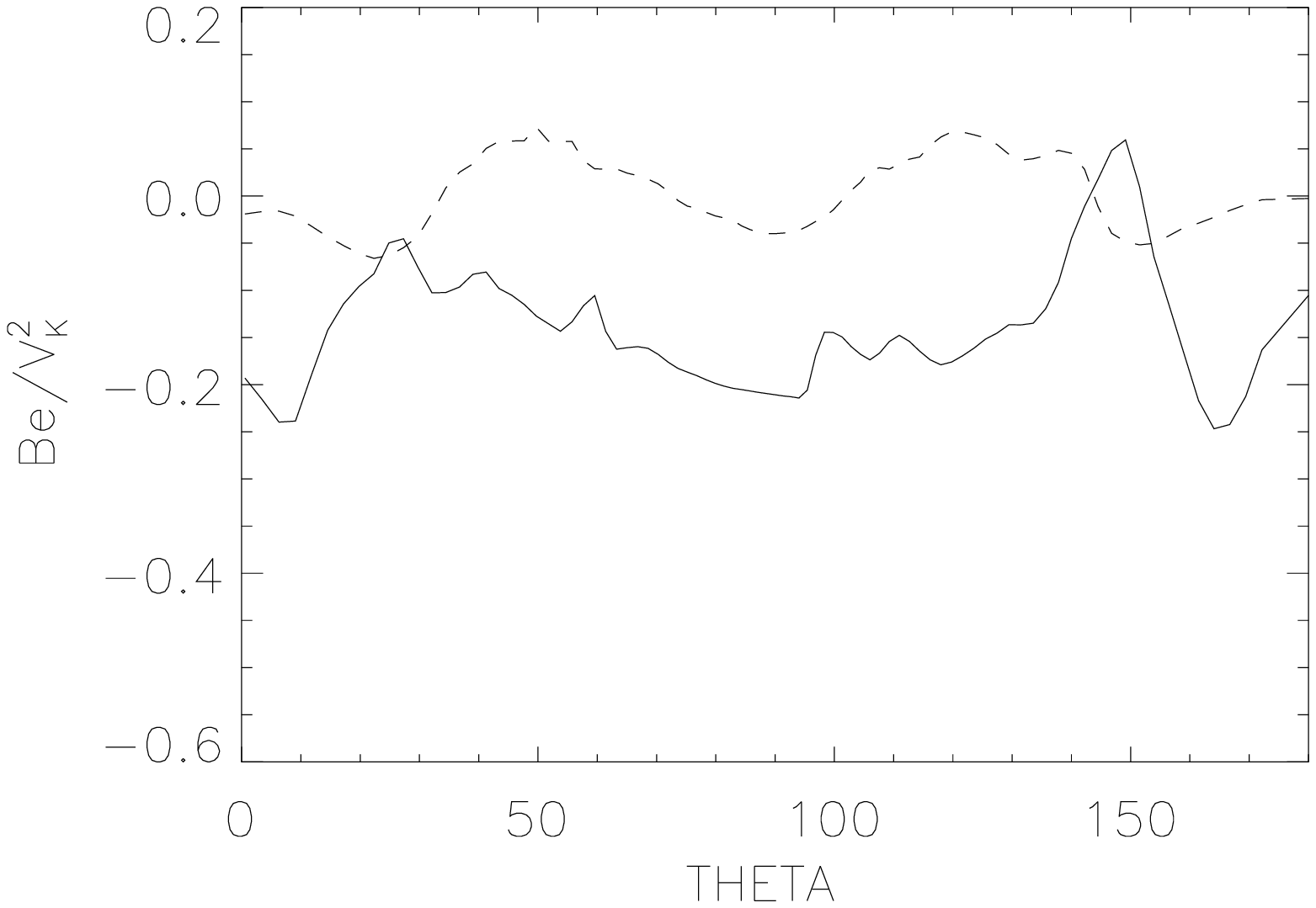}\hspace{1.cm} \epsscale{0.5}
\plotone{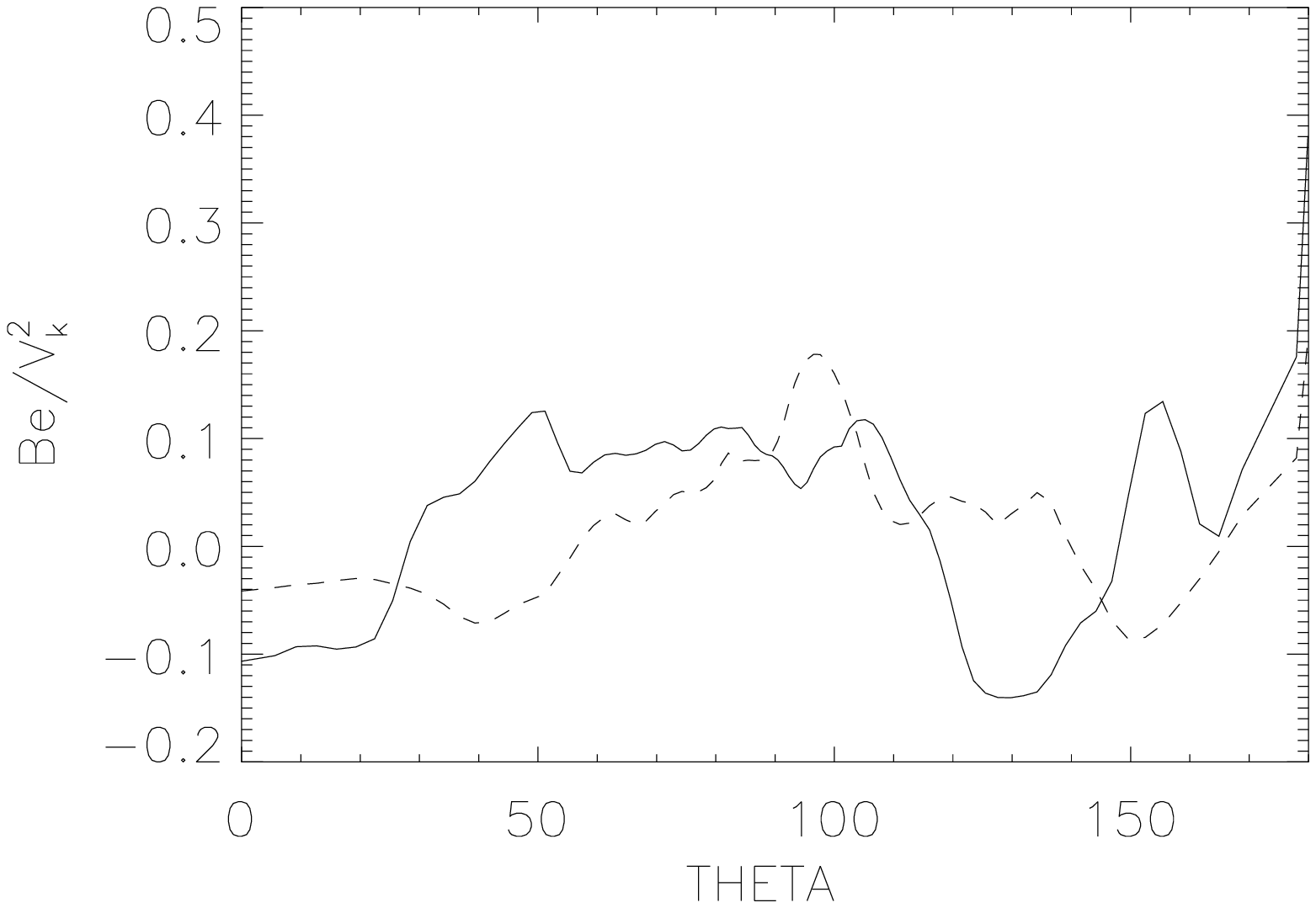} \vspace{0.2in}\epsscale{0.5} \plotone{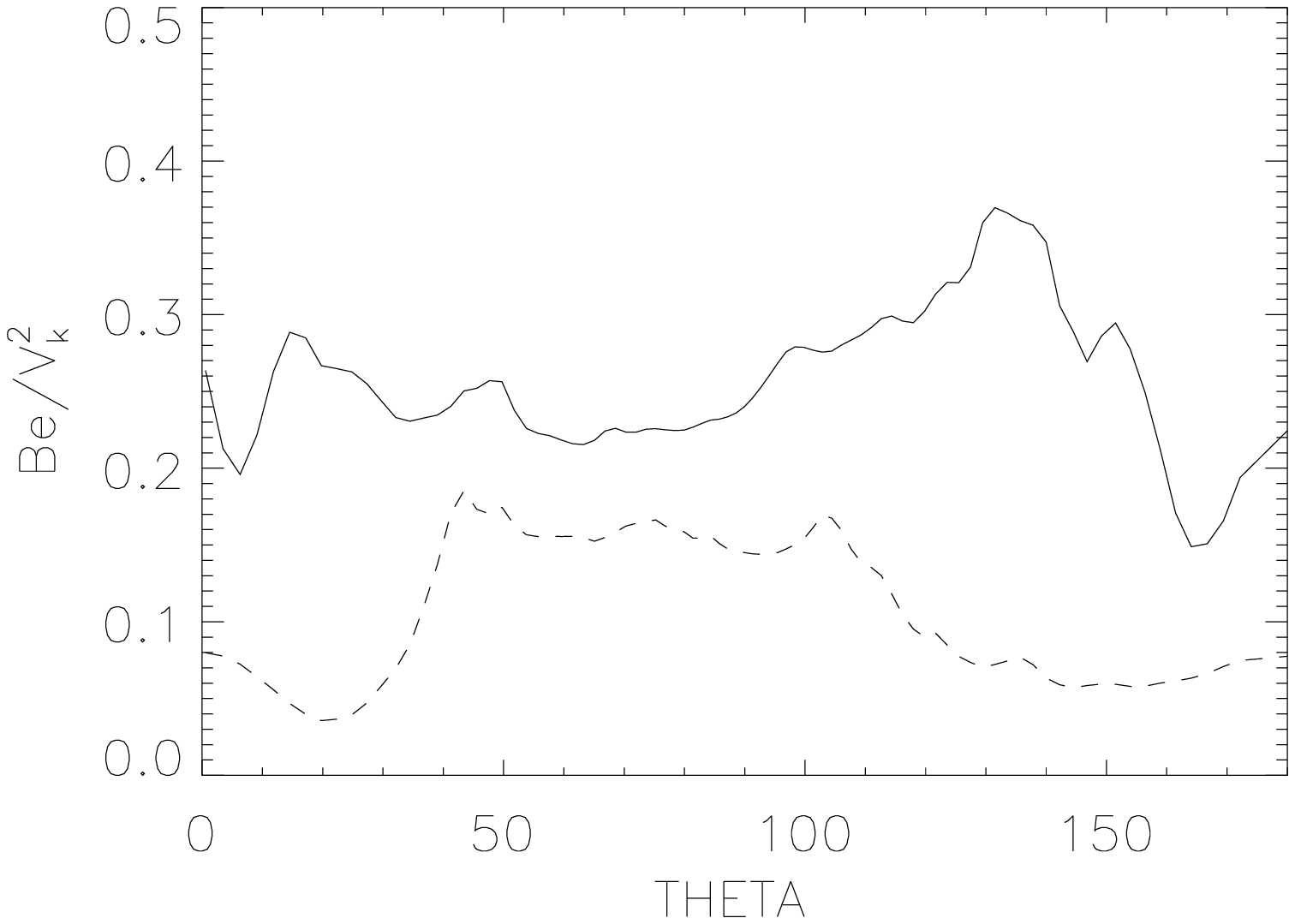}
\hspace{1cm} \epsscale{0.5} \plotone{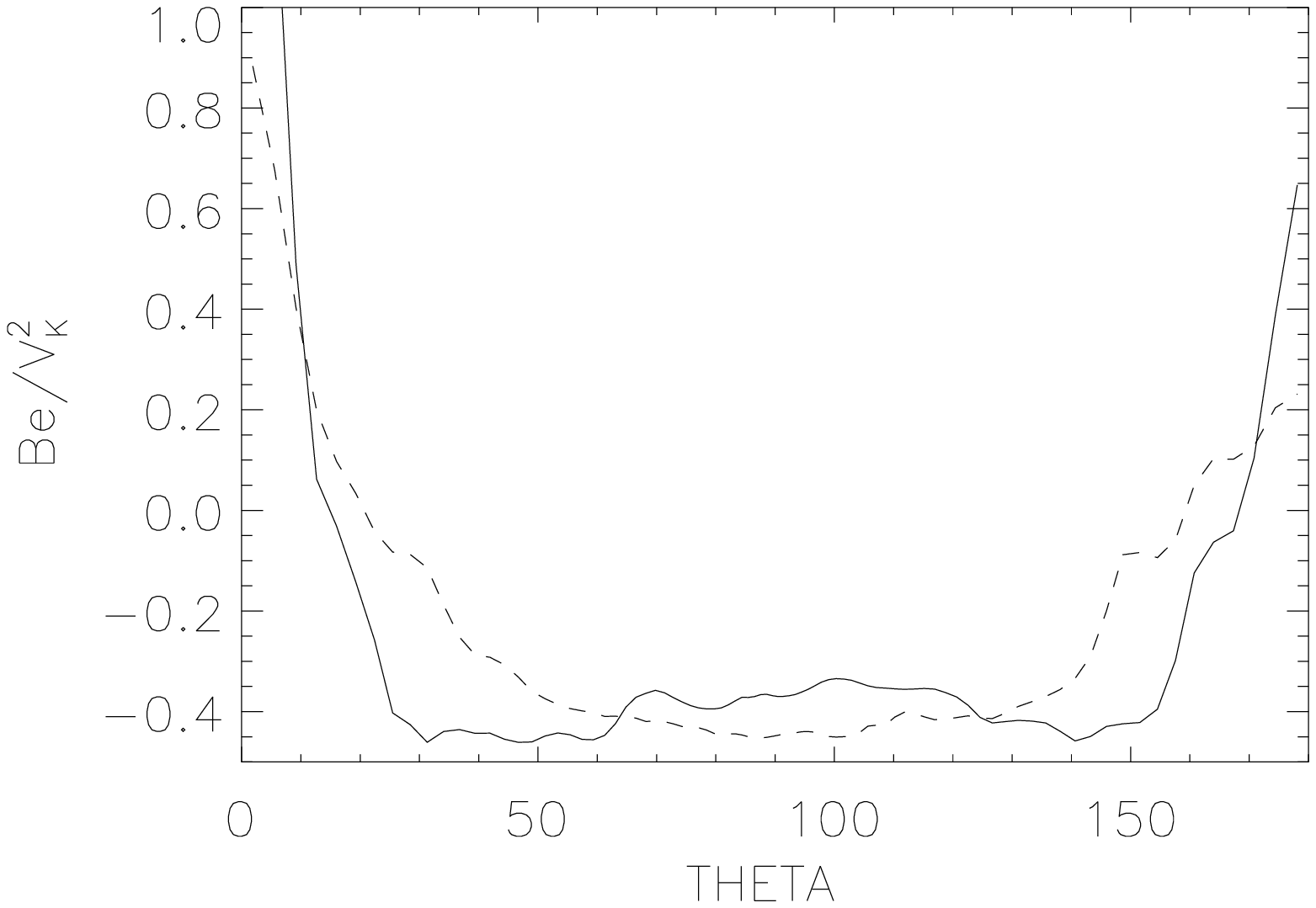}\vspace{0.2in}
\caption{The angular distribution of flux-weighted $Be$ in unit of $v_k^2$ of outflow at different radius. The solid and dashed lines are for $r=50r_s$ and $r=10r_s$ respectively. The top-left, top-right, bottom-left, and bottom-right plots are for Models A, B, C, and D, respectively.}
\label{Fig:angularbernoulli}
\end{figure*}

The thick solid lines in each plot of Fig. \ref{Fig:mdot} represent the radial profile of the outflow rate with a positive $Be$. We can see that for Models A \& B, a small fraction of the outflow has a positive $Be$. For Model C, $Be >0$ for almost all of the outflow. For Model D, a very tiny fraction of the outflow has $Be>0$. The result of Model D is consistent with Narayan et al. (2012). This is because the initial conditions in the two works are similar.

While the value of $Be$ for different models is quite different as shown by Fig. \ref{Fig:bernoullicontour}, we see from Fig. \ref{Fig:mdot} that the strength of outflow is similar for the four models, independent of the value of $Be$. This indicates that the outflow is not produced because $Be>0$. Of course, for these outflow to reach infinity, $Be>0$ is still required.

Fig. \ref{Fig:radialbernoulli} shows the radial distribution of the flux-weighted  $Be$ of outflow (dashed line) and inflow (solid line). In all models $Be$ increases inward, including the case of Model C. The apparent violation of $Be$ conservation (i.e, $Be$ is not a constant of radius) is because of the energy flux from small to large radii associated with the viscous stress. The dotted line in the figure shows the value of $Be$ in the initial condition. We can immediately find that the value of $Be$ of outflow is always larger than that of inflow for Models A, B, and C. Even, the difference between inflow and outflow seems to be similar for these three models. For Model D, however, the value of $Be$ is almost equal for inflow and outflow. This again suggests that the mechanism of the production of outflow is different for HD and MHD flows.

We are especially interested in the value of $Be$ of the outflow since it determines whether the outflow can escape to infinity and the terminal velocity of outflows if they can (\S3.5). Unfortunately from Fig. \ref{Fig:radialbernoulli} we can see that the results are quite diverse for the four models, depending on the initial conditions. For Models A \& D, the $Be$ of the initial torus is the smallest, the value of $Be$ of outflow in the two models are also roughly the smallest\footnote{The value of $Be$ in Models A \& D shown in the figure is larger than their initial value. This is compensated by the smaller $Be$ at $r>100$ which is not shown in the figure; thus the total $Be$ between initial condition and the steady state is roughly conserved.}. For Model C, $Be>0$ for the initially injected gas, subsequently $Be>0$ for both inflow and outflow. Unfortunately, as we have discussed in \S2.2, it is uncertain to determine what kind of initial condition is more realistic. But as we have argued there, in the case that the accretion flow starts from very large radius, it is likely that initial condition with $Be$ close to or larger than zero will be picked up. In this case we expect that $Be>0$ for outflow. To illustrate this point, we have calculated the value of $Be$ of the outflow of Model A in Paper I. In that model, the initial condition is a same rotating torus as in Models A \& D of the present paper, except that the torus is located at a much larger radius, $10^4r_s$. Fig. \ref{Fig:largetorusbe} shows the results. We see that $Be>0$ for all outflow. In another example, Li, Ostriker \& Sunyaev (2012) started their hot accretion flow at ten times of Bondi radius, so $Be>0$ for the initial condition. They found $Be>0$ for all their outflow (Jerry Ostriker, private communication). At last, as we will see in \S\ref{observation}, observations seem to indicate the existence of outflow from hot accretion flows. In the following discussions, we therefore assume that outflow from hot accretion flow can always escape to infinity, i.e, $Be>0$.

In this paragraph we present some speculations which further support the above assumption of $Be>0$. Note that the importance of initial condition does not mean that the final solution can be arbitrary. They must be restricted in a certain range. From Fig. \ref{Fig:radialbernoulli}, we see that the diversity of the solution at the inner region of the accretion flow away from the outer boundary seems to become smaller compared to that of the initial condition. There seems to exist a ``common'' (but not unique) solution at the inner region, which can be very roughly described by, \be Be\approx (-0.1\sim 0)v_k^2 \label{inflowbernoulli}\ee for inflow, and \be Be\approx (0.1\sim 0.2)v_k^2\label{outflowbernoulli}\ee for outflow. At large radii, the solutions strongly deviate from the above ``common'' solution, because they are still ``tied'' to the initial condition there. Based on this result, we speculate that if the outer boundary is set at a very large radius and the location of torus  is at large radius, the solution at the inner region may suffer less from the initial condition and becomes more ``converged''. In other words, the degree of the dependence of solution on the initial condition depends on the how far the gas is away from the black hole. In almost all current simulations including the present work, the outer boundary is rather small, thus it is hard for the accretion flow to fully evolve and deviate from the initial condition.  Only at small radii, the accretion flow is better evolved thus solutions there can be more approached to the ``genuine'' solution. For Model B, at both small and large radius, the simulated solution is close to the initial condition. This is likely because the initial condition of Model B is ``good'', i.e., close to the ``genuine'' solution. It is interesting to note that in this model the value of $Be/v_k^2$ is roughly a constant, consistent with the self-similar solution (Narayan \& Yi 1994). So this seems to again indicate that Model B is more realistic than the other three models.  Based on these consideration, when we calculate the terminal velocity in \S3.5, we will assume eq. (\ref{outflowbernoulli}). We emphasize again that the above consideration is rather speculative. It is crucial to explore it by simulations with a much larger outer boundary. This will be our future work. But on the other hand, we believe that the value of $Be$ determined by eq. (\ref{outflowbernoulli}) should be correct in order of magnitude; then the calculate terminal velocity (eq. [\ref{terminalvel}]) will be correct within a factor of three.

Comparing Models A \& D, we can see that the value of $Be$ in Model D is significantly smaller than in Model A, although they have the same initial condition. This is because in Model D some energy is converted into the magnetic energy, which is not include in our $Be$.

In the case that the outer boundary is small, or additional physics such as radiation is included, the value of $Be$ of the outflow should be mainly determined by the initial condition. While there is some deviation, the deviation should still be limited because of the energy conservation. If the value of $Be$ of the initial condition is far below zero, we expect that there will be no outflow with $Be>0$, i.e., no jet. This result supplies an important clue to the following long-standing observational puzzle, namely why relativistic jet can only be formed in hot accretion flow not in the standard thin disk. Black hole X-ray binaries usually come into various states (e.g., McClintock \& Remmilard 2006). The most prominent two states are hard and soft states, which are described by hot and cold accretion flows, respectively (e.g., see reviews by Zdziarski \& Gierli\'nski 2004; Done, Gierli\'nski \& Kubota 2007). Radio observations clearly show that relativistic jets exist only in hard state, not in the soft state. That is, jet can only be formed from hot accretion flow, not from the thin disk. We speculate that this may be because of the difference of the value of $Be$ of the flow in the two types of accretion flows. Because of strong radiative cooling, $Be$ in the case of thin disk is obviously much lower than that in the hot accretion flow. Following this argument, we predict that for hot accretion flows, the ratio of the mass lost rate in the jet and the accretion rate should decrease with increasing accretion rate. When the accretion rate is high, radiative cooling is strong, $Be$ should be low. This is consistent with the modeling result in Yuan \& Cui (2005).

Fig. \ref{Fig:angularbernoulli} shows the angular distribution of $Be$ at $r=10$ and $50$ for the four models. At a given radius, part of the outflow has positive $Be$ but part negative. For both Models A \& D, $Be$ is the largest away from the equatorial plane while the smallest around the equatorial plane. The difference between these two models is that in Model D the contrast of $Be$ is much larger and the region of large $Be$ is much closer to the axis. This is because of the existence of the large-scale poloidal field close to the axis, as we have explained in the previous section. The distribution of Models B \& C is  different. It is interesting to note that $Be$ is the largest at the angles of $\theta$ where the mass outflow rate is the largest (Fig. \ref{Fig:mdotoutflow}).

\subsection{Temperature}

\begin{figure*}
\epsscale{0.45} \plotone{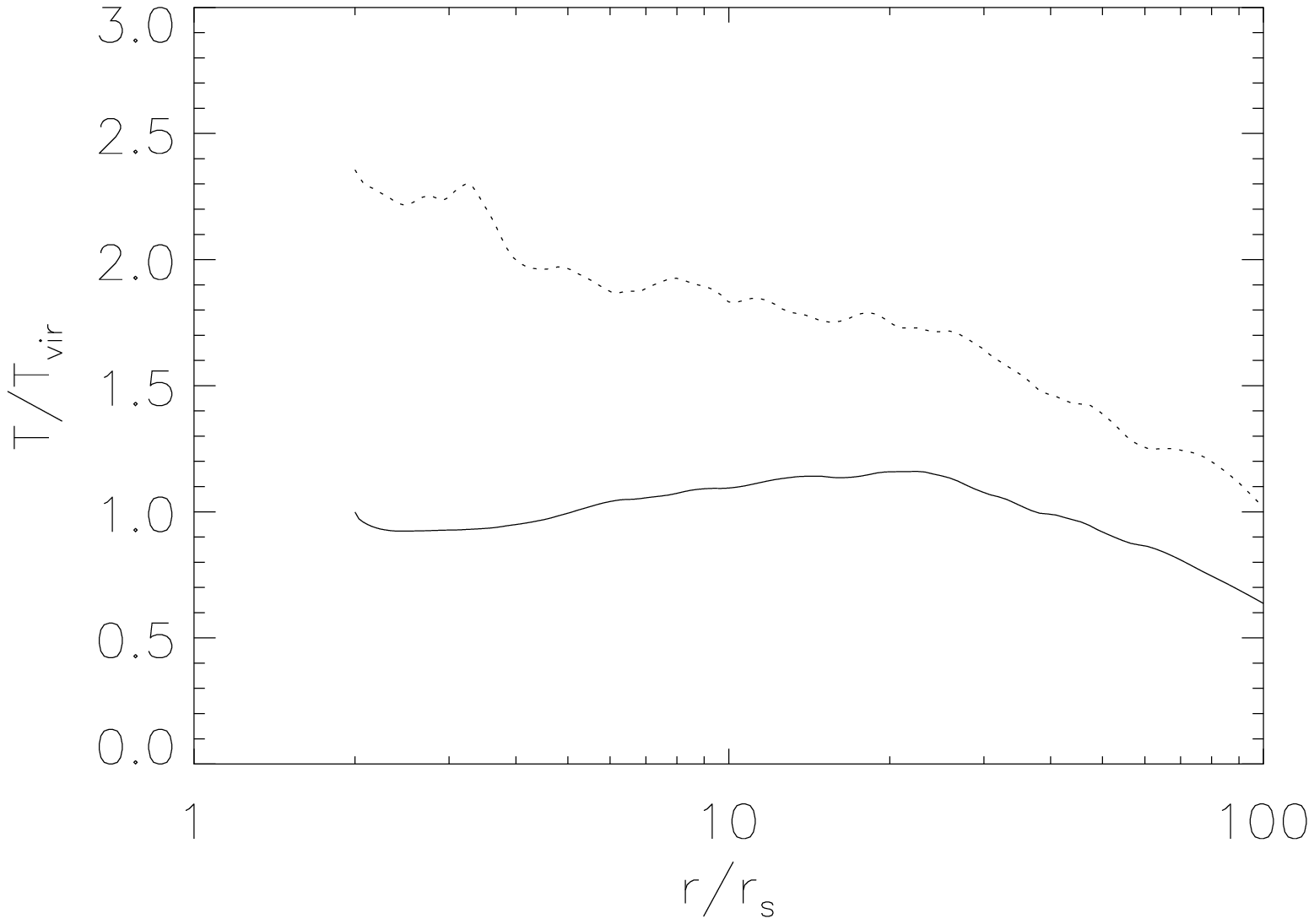}\hspace{1.cm} \epsscale{0.45}
\plotone{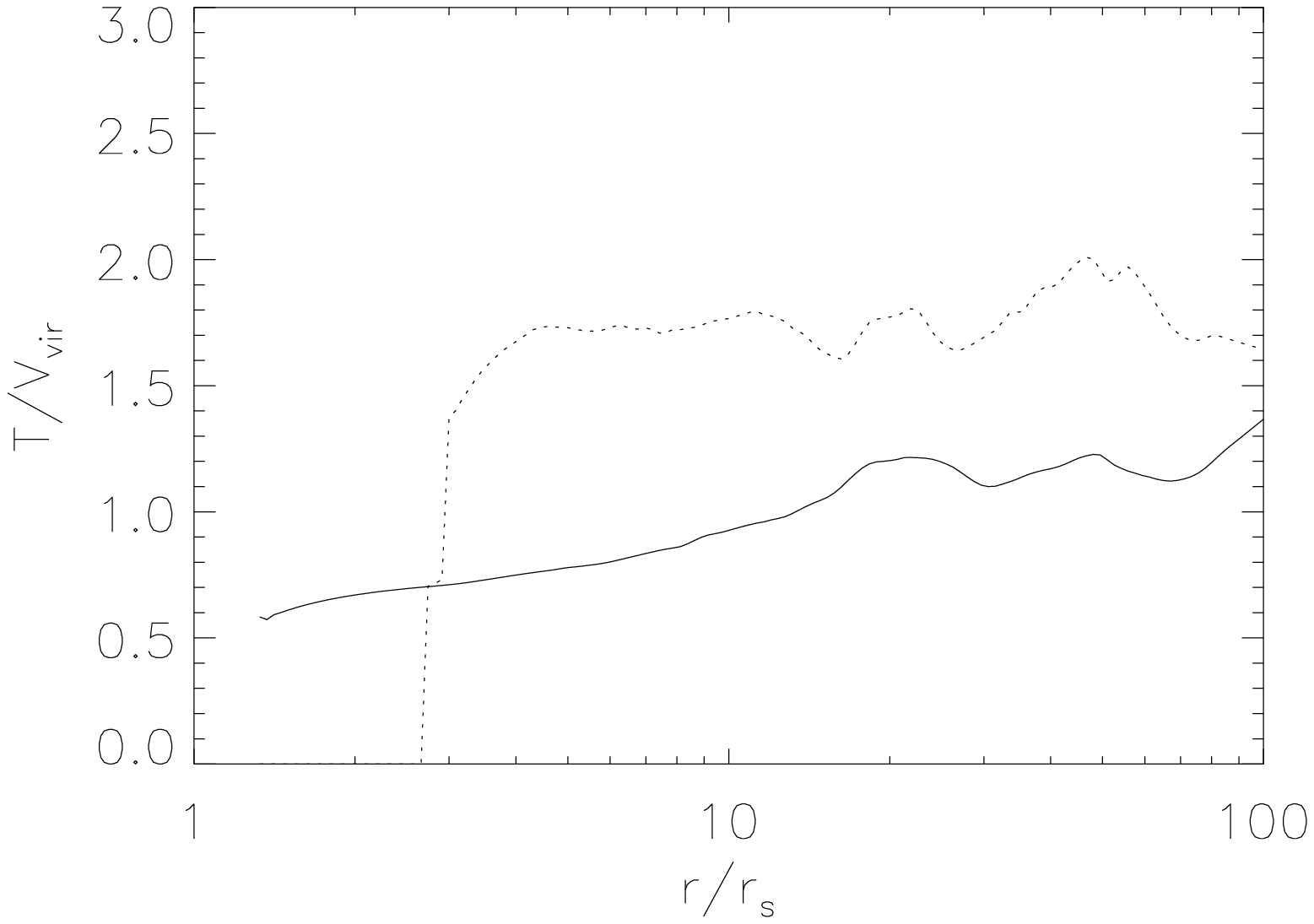} \vspace{0.2in}\epsscale{0.45} \plotone{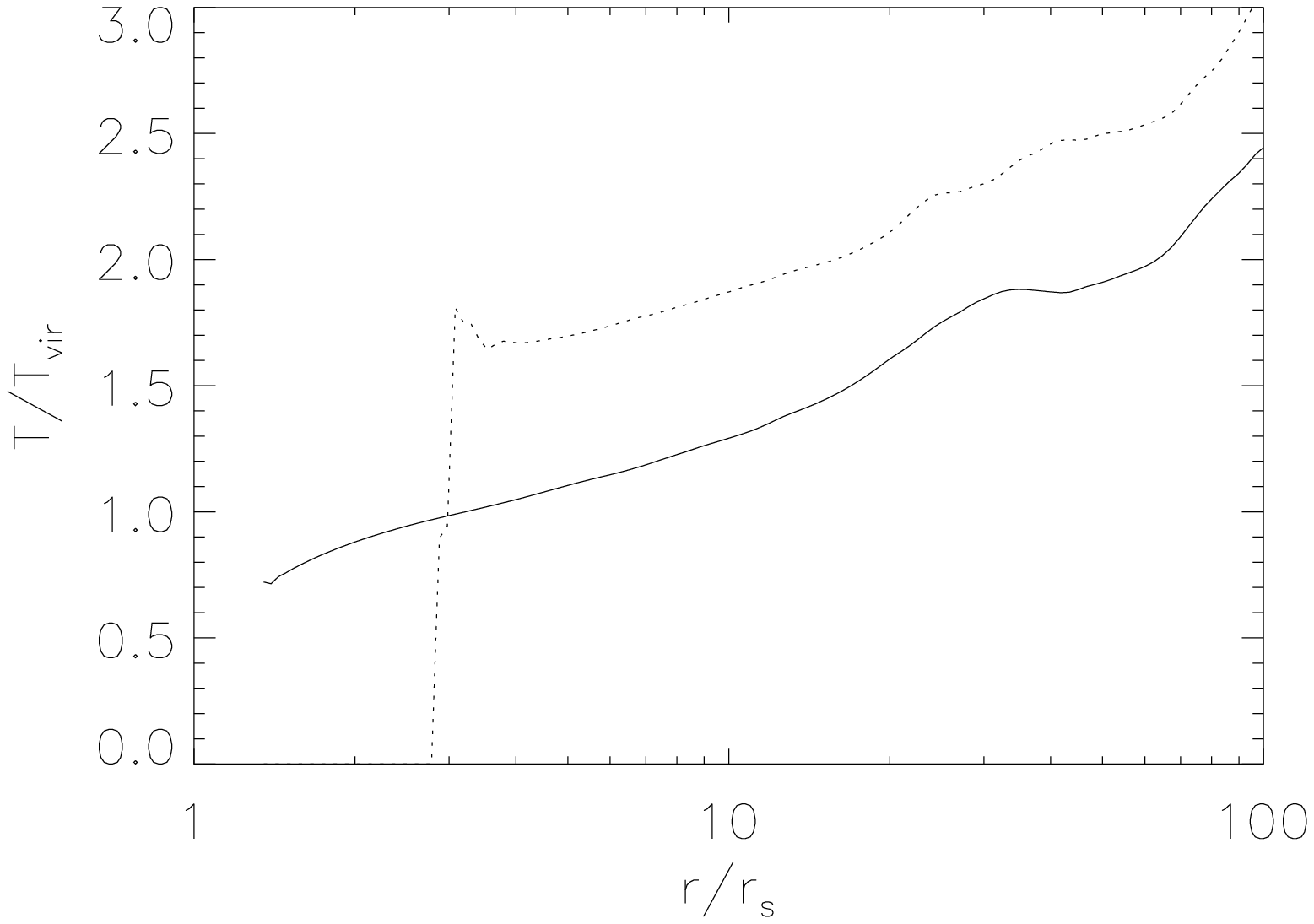}
\hspace{1cm} \epsscale{0.45} \plotone{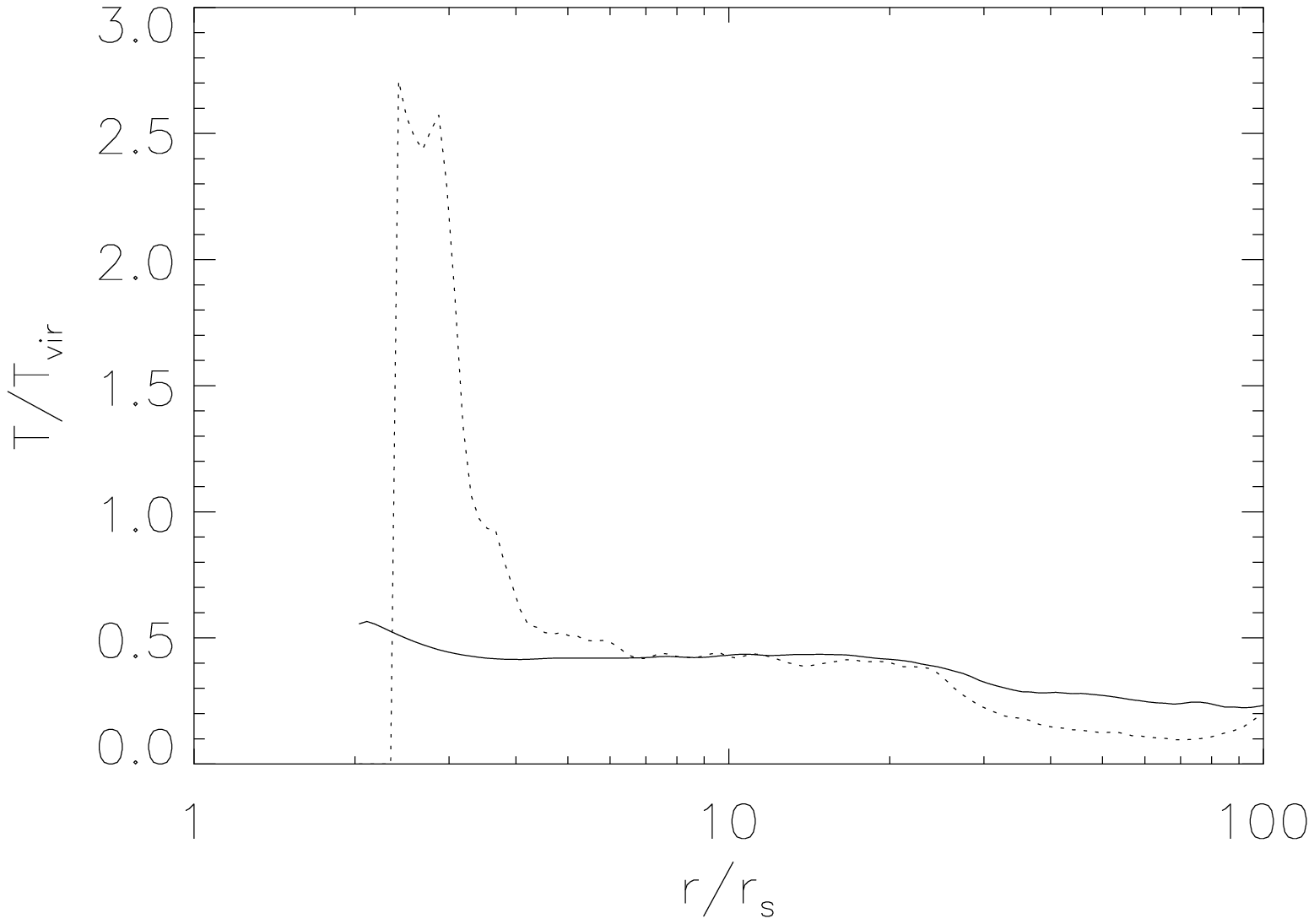}\vspace{0.3in}
\caption{The radial distribution of flux-weighted temperature. The solid and dotted lines are for inflow and outflow, respectively. The top-left, top-right, bottom-left, and bottom-right plots are for Models A, B, C, and D, respectively. }
\label{Fig:temperature}
\end{figure*}

Fig. \ref{Fig:temperature} shows the flux-weighted radial distribution of temperature of inflow and outflow for the four models.  One distinct result is that for the three HD models, the temperature of outflow is higher than that of inflow by nearly $T_{\rm vir}$. Here the virial temperature is defined as \be T_{\rm vir}\equiv \frac{GMm_p}{3kr}.\ee Since the temperature of both the inflow and outflow are nearly virial, the internal energy plays an important part in $Be$. This is why the value of $Be$ of outflow is higher than inflow (Fig. \ref{Fig:radialbernoulli}). This is a strong evidence for the convection origin of outflow in HD flows, i.e., the fluid elements with higher temperature than surrounding medium will move outward due to the buoyancy. For Model D, however, the temperature of inflow and outflow is almost identical. This suggests that an MHD flow is not convectively unstable, as we will confirm by stability analysis. Correspondingly, the mechanism of producing outflow in Model D is not convection.

\subsection{Radial Velocity}

\begin{figure*}
\epsscale{0.45} \plotone{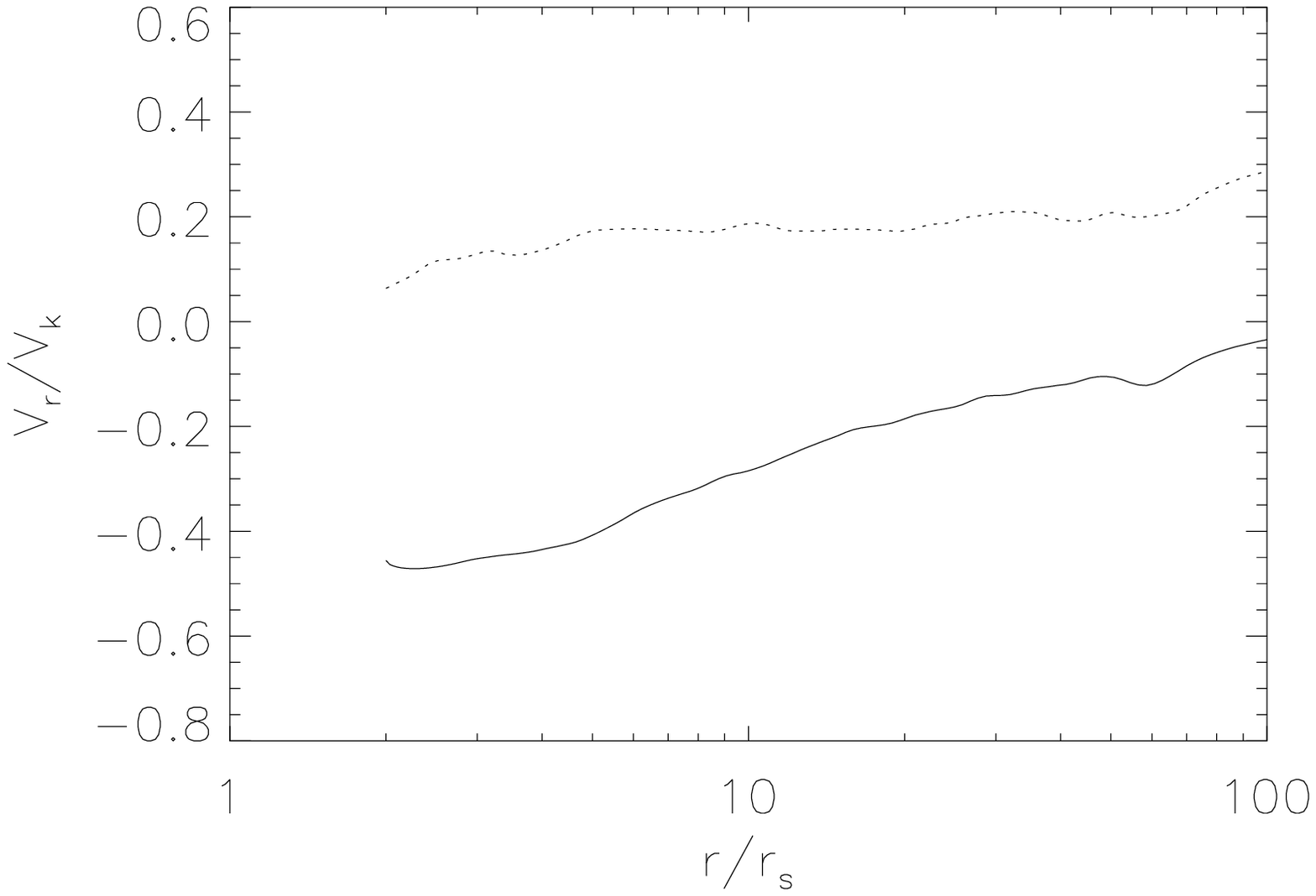}\hspace{1.cm} \epsscale{0.45}
\plotone{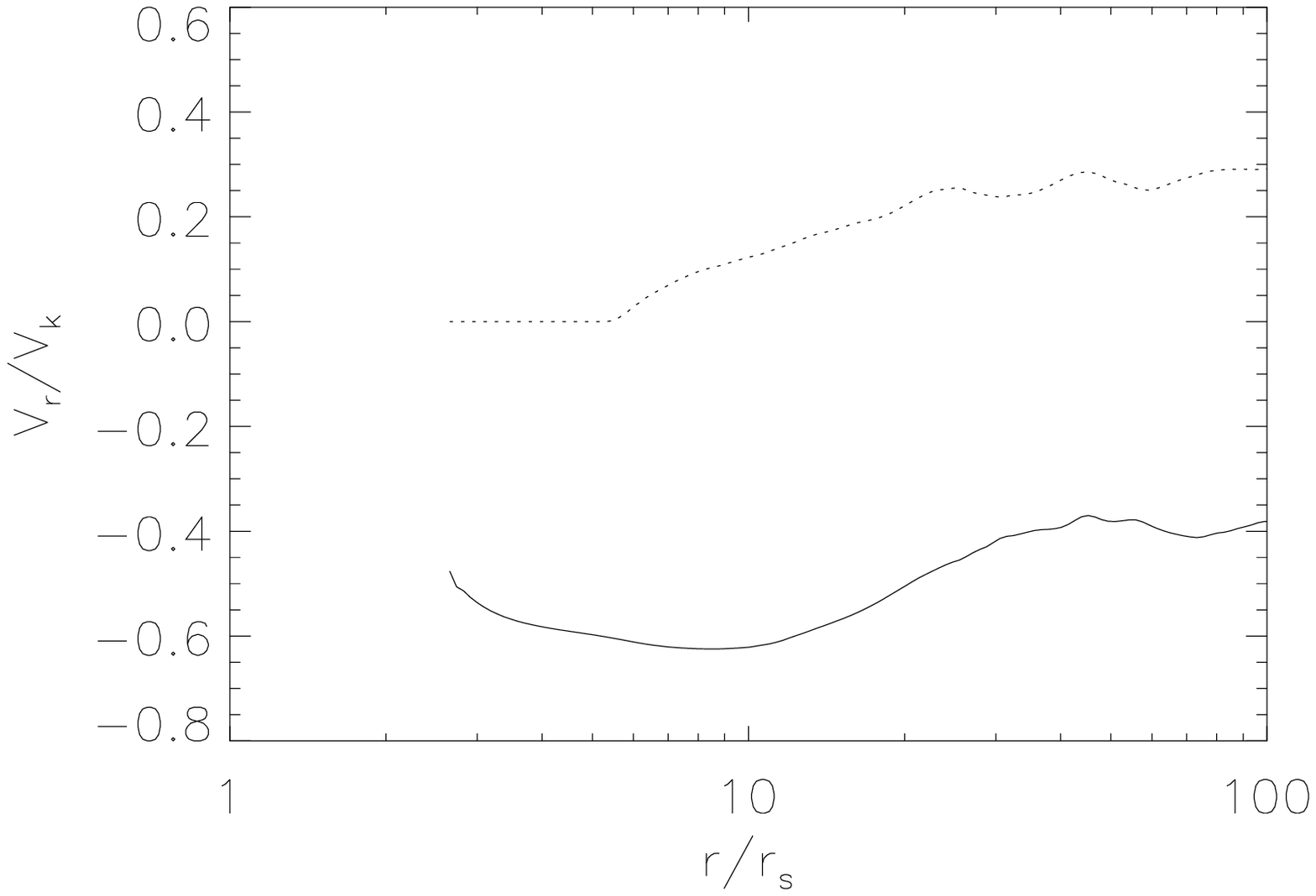} \vspace{0.2in}\epsscale{0.45} \plotone{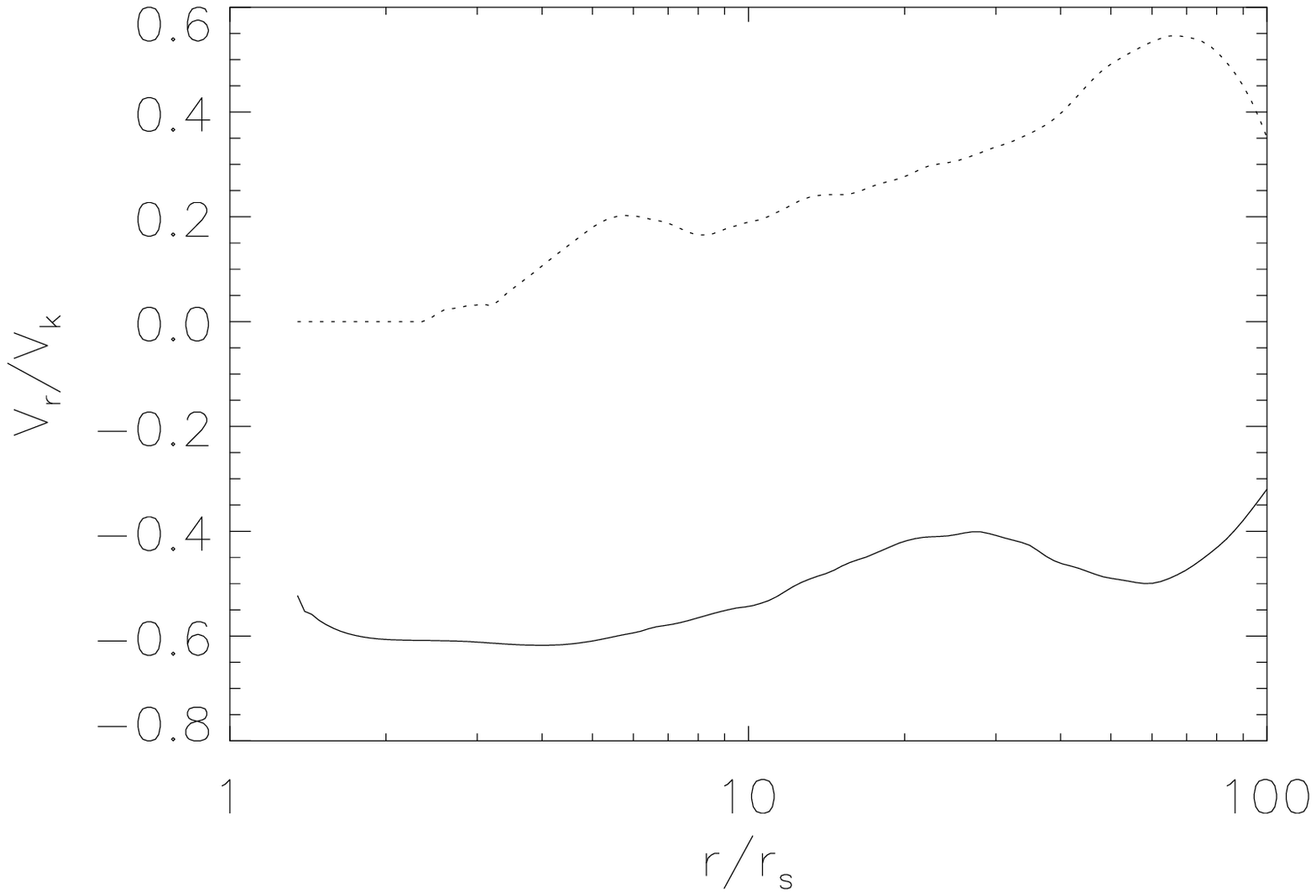}
\hspace{1cm} \epsscale{0.45} \plotone{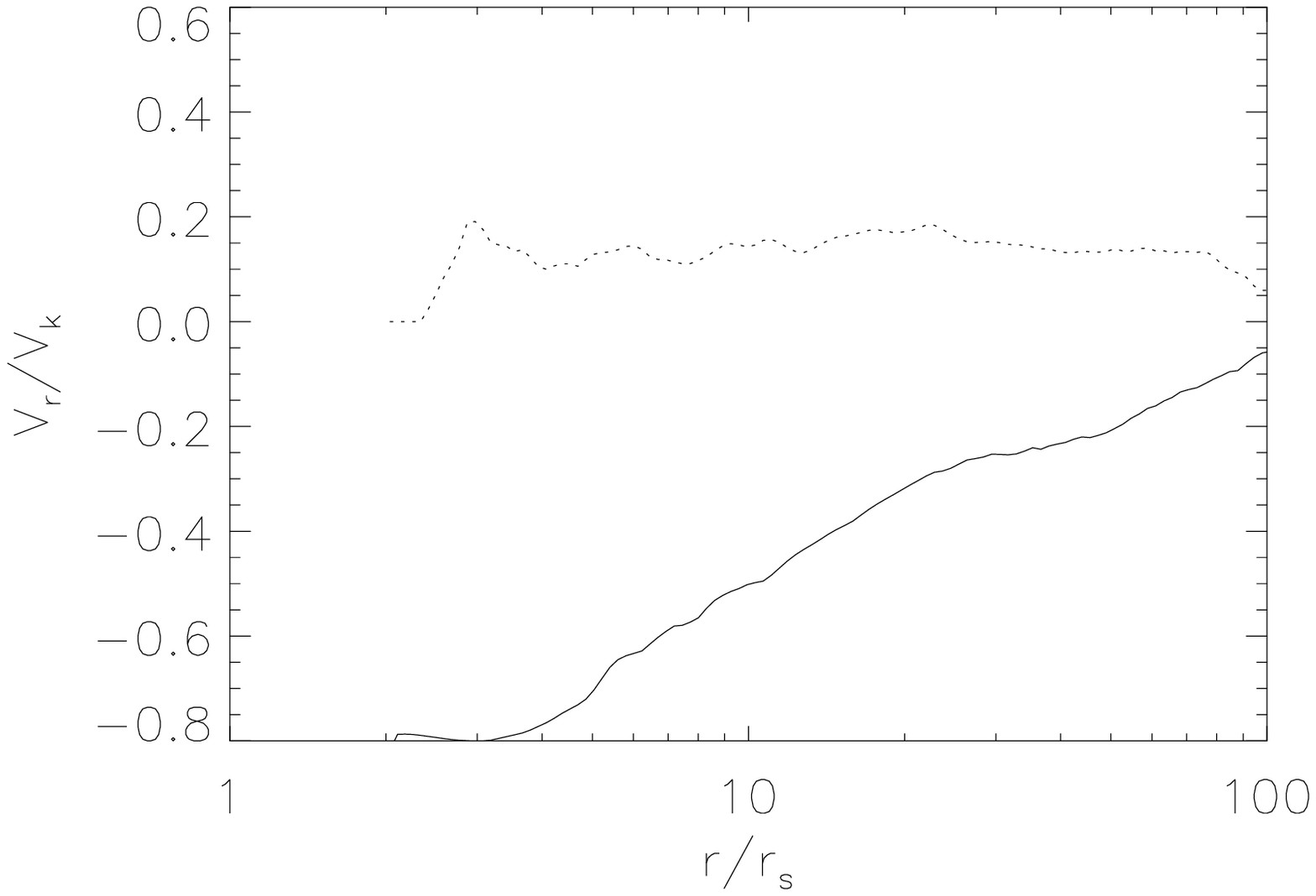}\vspace{0.3in}
\caption{The radial distribution of the flux-weighted radial velocity. The solid and dotted lines are for inflow and outflow, respectively. The top-left, top-right, bottom-left, and bottom-right plots are for Models A, B, C, and D, respectively. }
\label{Fig:radialvelocity}
\end{figure*}

Fig. \ref{Fig:radialvelocity} shows the flux-weighted radial velocity of inflow and outflow for the four models. For the profile of inflow, we see that Models B \& C are similar, while Models A \& D are similar. The former can be well described roughly by $v_r/v_k\sim const.$, or $v_r\propto r^{-0.5}$. This scaling is consistent with the self-similar solution of ADAF (Narayan \& Yi 1994). But for Models A \& D, the radial velocity scaling with radius is much steeper, i.e., the radial velocity increases faster inward compared to the Keplerian velocity. The discrepancy between the two groups of model is caused by the difference of initial condition. In Models A\&D, the radial velocity in the torus is initially zero, thus it is difficult for the gas at large radii ($\ga 100r_s$) to achieve a large velocity because of the boundary effect. On the other hand, because of the very strong gravity, the radial velocity has to rapidly increase inward at $r\la 10r_s$. These two effects make the steeper profile. If the initial torus in Models A \& D is located at a much larger radius, i.e., the dynamical range is much larger, the results of these two models will be similar to those in Models B \& C. This actually has been confirmed by our simulation in Paper I where the torus is located at $10^4 r_s$.

\begin{figure*}
\epsscale{0.45} \plotone{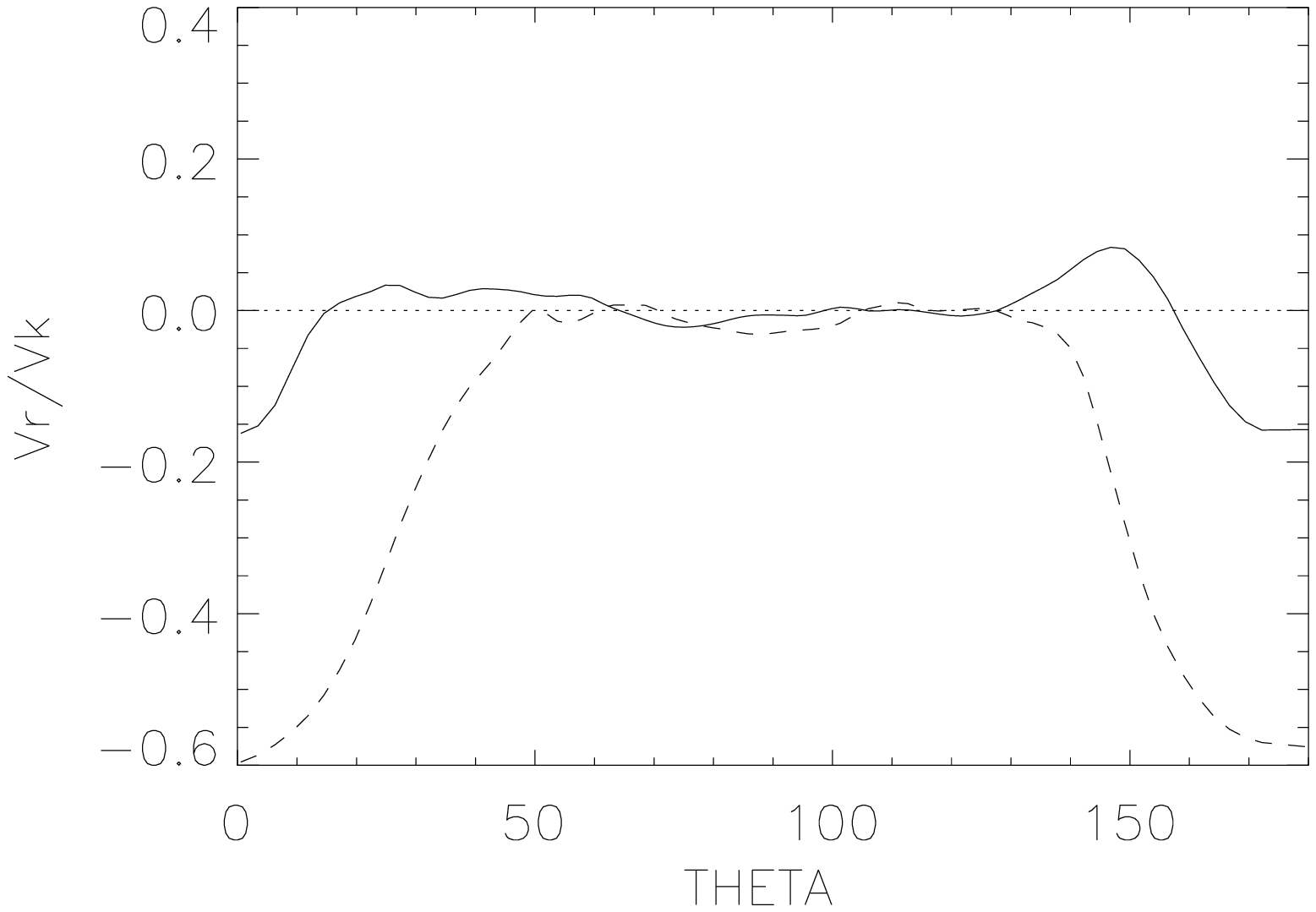}\hspace{1.cm} \epsscale{0.45}
\plotone{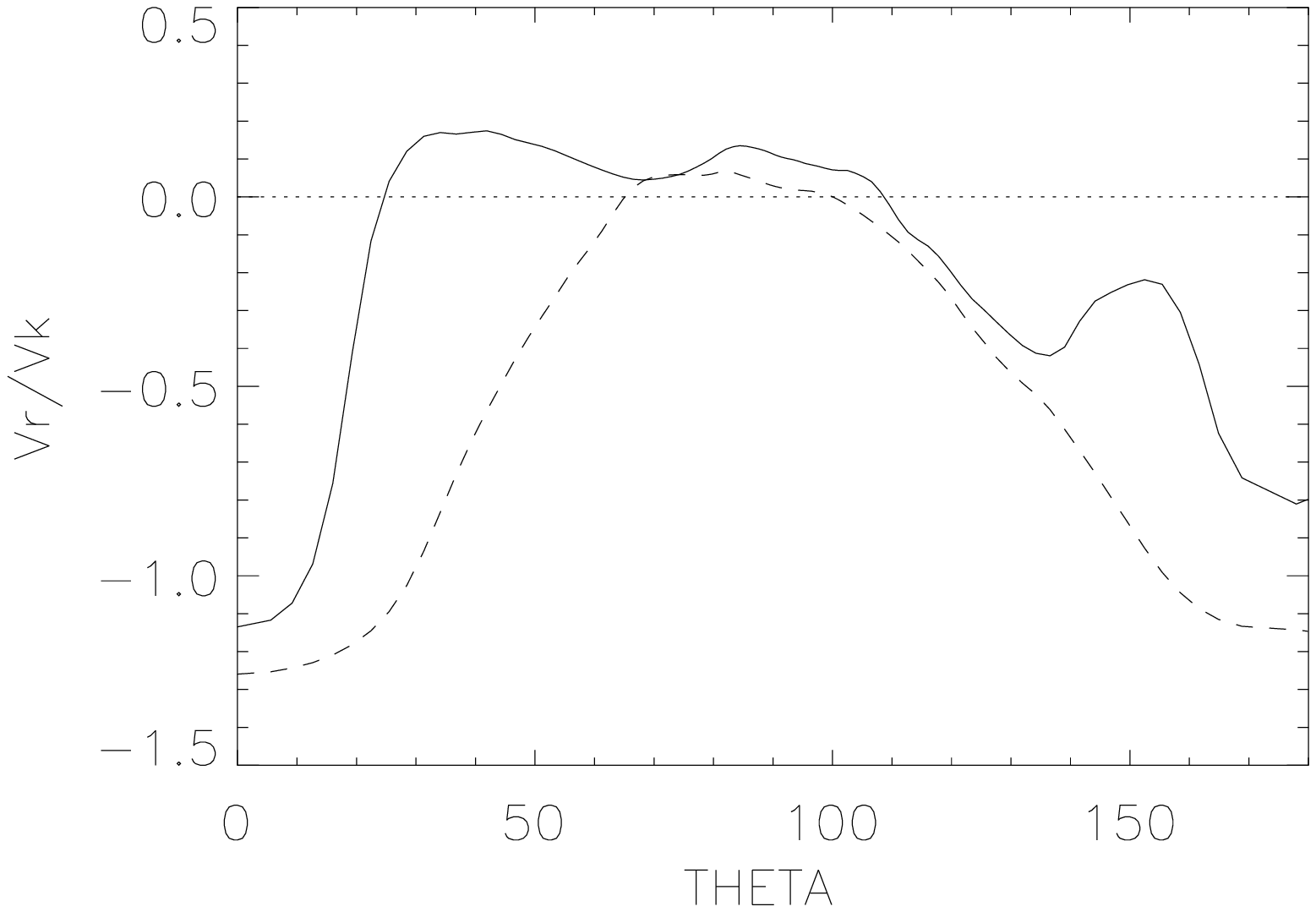} \vspace{0.2in}\epsscale{0.45} \plotone{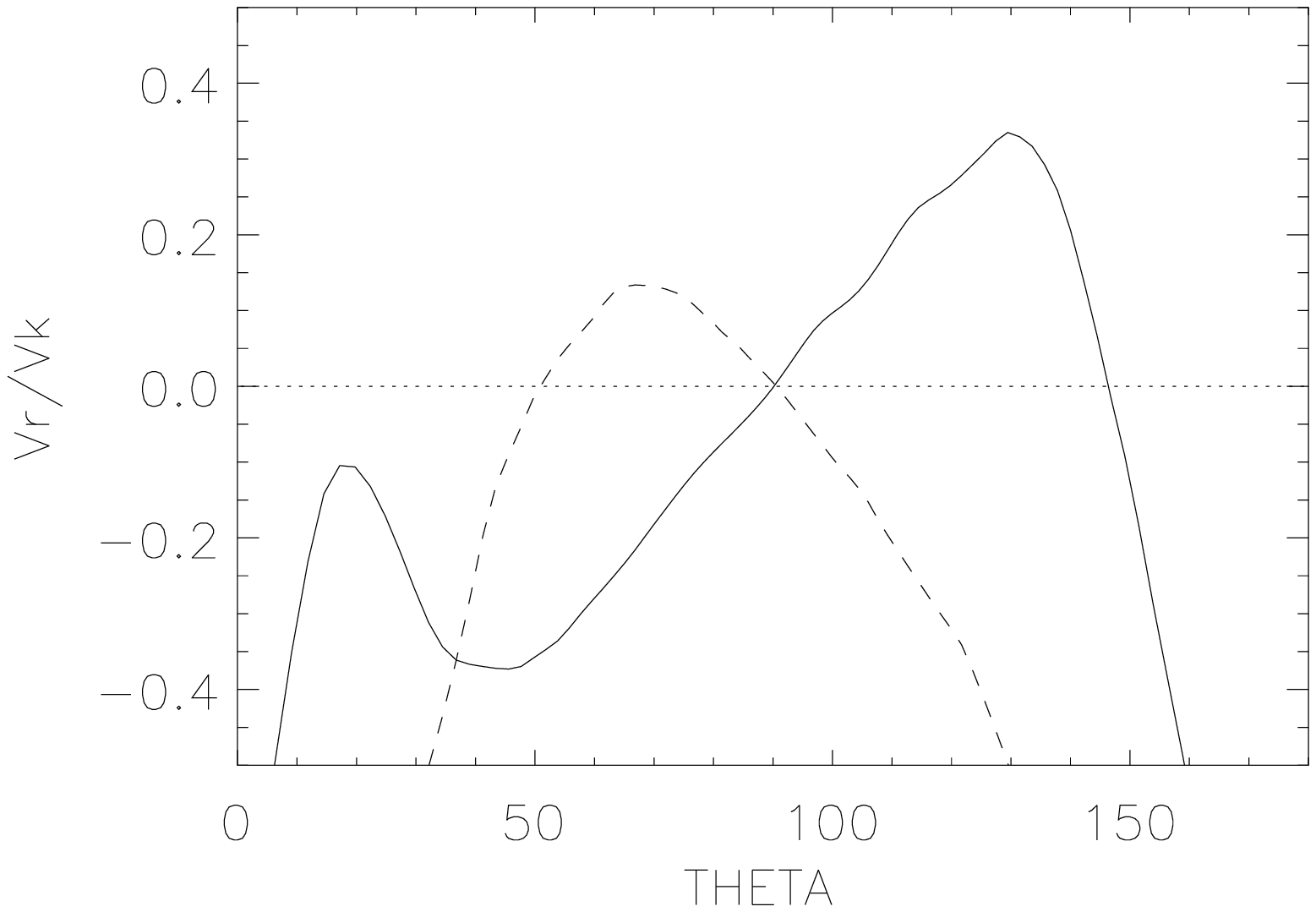}
\hspace{1cm} \epsscale{0.45} \plotone{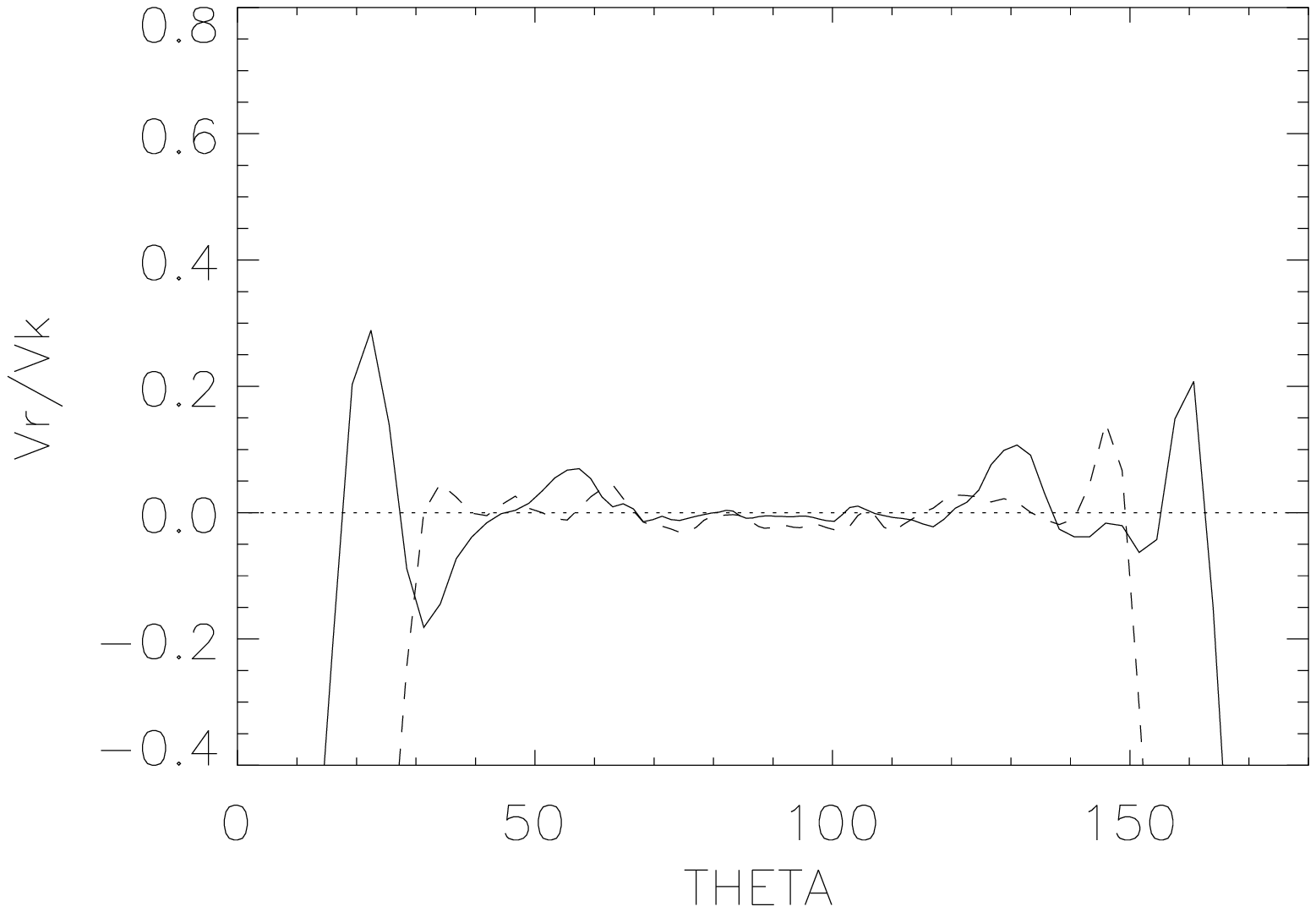}\vspace{0.3in}
\caption{The angular distribution of radial velocity at $r=50 r_s$ (solid) and $10 r_s$ (dashed). The top-left, top-right, bottom-left, and bottom-right plots are for Models A, B, C, and D, respectively.}
\label{Fig:radialvelocitytheta}
\end{figure*}

However, the radial profile of the radial velocity of outflow is completely different.  It is not so sensitive to the initial condition. The ratio of radial velocity and Keplerian velocity remains roughly a constant of radius. This ratio is even similar for all the four models, i.e., $v_r/v_k\approx (0.2-0.4)$, regardless of the difference of the initial condition and whether magnetic field is included or not. Given that the mechanism of producing outflow in HD and MHD flows are different, as we will discuss later, the reason of the same outflow radial velocity remains to be explored. We would like to emphasize that the systematic difference of the profile of radial velocity between inflow and outflow again strongly indicates that the outflow is not simply turbulent fluctuation or convective motion. They are systematic outward flux of mass.

Fig. \ref{Fig:radialvelocitytheta} shows the angular distribution of the radial velocity, averaged for all accretion flow including both inflow and outflow. It is clear that on time-average sense, the angular distributions of the radial velocity for Models A \& D are quite symmetric to the equatorial plane. Taking $r=50$ (i.e, the solid lines in each plot) as an example, the inflow concentrates around the equatorial plane, $60^{\circ} \la \theta \la 120^{\circ}$; while the outflow occurs at $15^{\circ} \la \theta \la 60^{\circ}$ and $120^{\circ} \la \theta \la 165^{\circ}$. For Models B \& C, the distribution is not symmetric to the equatorial plane. When inflow concentrates on above (below) the plane, i.e., $\theta\la 90^{\circ}$ ($\theta \ga 90^{\circ}$), the outflow then concentrates on below (above) the plane. We note that the angular distributions of radial velocity of inflow and outflow shown in this figure are consistent with the angular distributions of inflow and outflow rates, shown by Figs. \ref{Fig:mdotinflow} and \ref{Fig:mdotoutflow}. Another notable feature is that for all the four models, close to the axis, averagely the radial velocity is negative, i.e, it is inflow.

One important question is the terminal radial velocity of outflow if they reach infinity. We have interest to this question since outflow is potentially very important for AGNs feedback. We suppose that once launched, the viscous stress in the outflow and the mixing of mass, energy and angular momentum between streamlines can be ignored. For such an adiabatic expanding inviscid flow, $Be$ is conserved. When $r$ is large enough, the last two terms in eq. (\ref{bernoulli}) vanish. If the angular momentum is conserved, $v_{\phi}\approx 0$ so the velocity is mainly in the radial direction. Therefore the terminal velocity is \be v_{\rm term}\sim v_r \sim \sqrt{2Be(r_{\rm out})}. \label{terminalvel}\ee Combining with eq. (\ref{outflowbernoulli}), we have \be v_{\rm term}\sim 0.5 v_k (r_{\rm out}).\ee That is to say, the radial terminal velocity of outflow is roughly half of the Keplerian velocity at the outer radius of the hot accretion flow where most of the outflow originates.

A positive sign of $Be$ is only a condition for {\em inviscid HD} outflow to escape to infinity. When viscosity and magnetic field are present, energy may be transferred from small to large radii and magnetic force may also do work to the outflow. In this case, $Be$ defined by eq. (\ref{bernoulli}) is not conserved, and it may be not necessary for outflow to have $Be>0$ to escape to infinity. The terminal velocity given by eq. (\ref{terminalvel}) will be a lower limit.  We do not consider these complications in the present work. The exact solution of the terminal velocity as a function of its initial properties in the presence of magnetic field is an interesting question to study.

\subsection{Angular Momentum}

It has long been speculated that the outflow can carry away energy and angular momentum (e.g., Shakura \& Sunyaev 1973; Blandford \& Begelman 1999). This requires that the specific energy and specific angular momentum of outflow should be larger than that of inflow. We have discussed the Bernoulli parameter in the previous section and found that the value of $Be$ of outflow is systematically larger than that of inflow for Models A, B, and C. But for Model D, the values of $Be$ of inflow and outflow are the same. In this section we examine the angular momentum transfer by outflow.

\begin{figure*}
\epsscale{0.45} \plotone{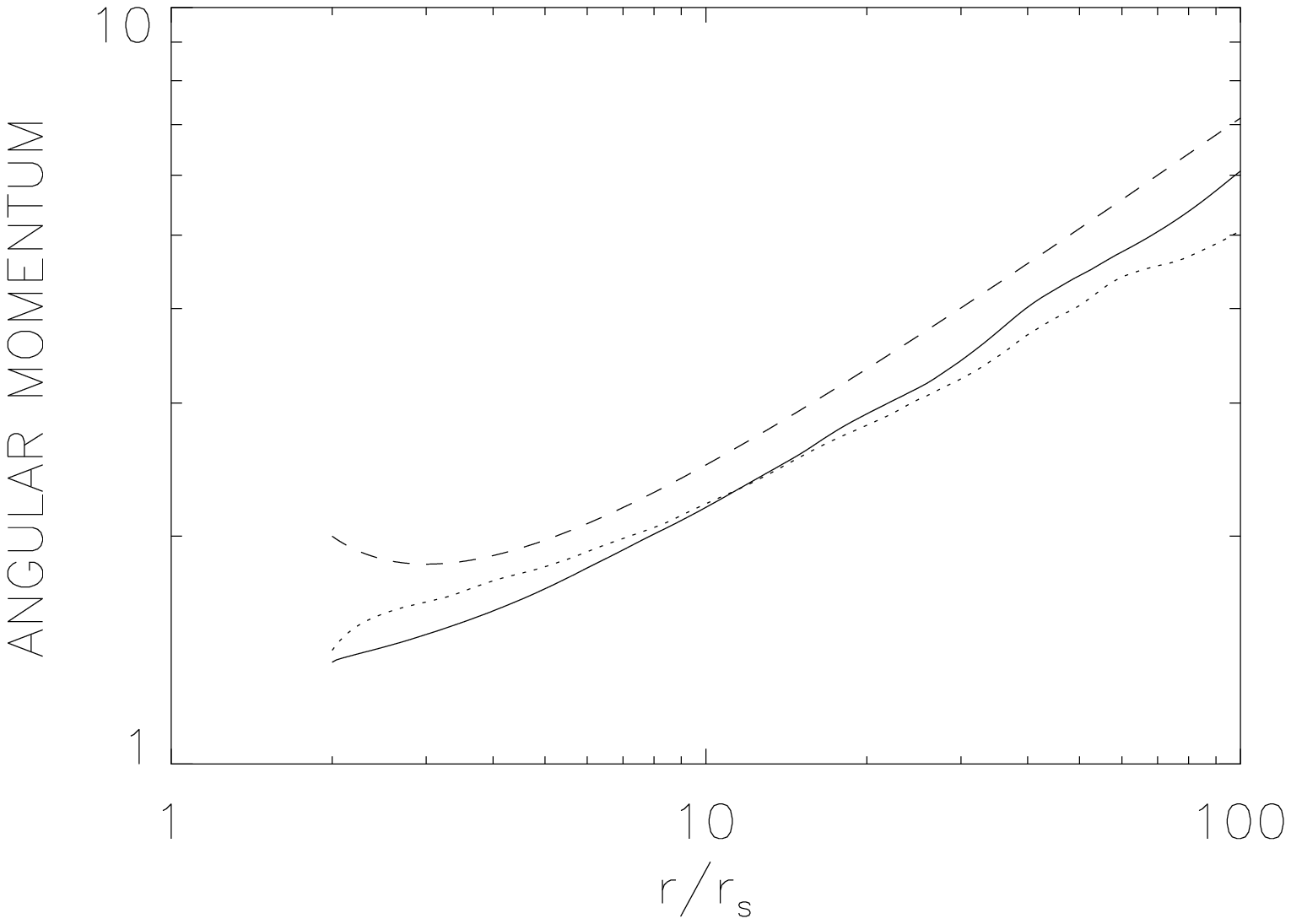}\hspace{1.cm} \epsscale{0.45}
\plotone{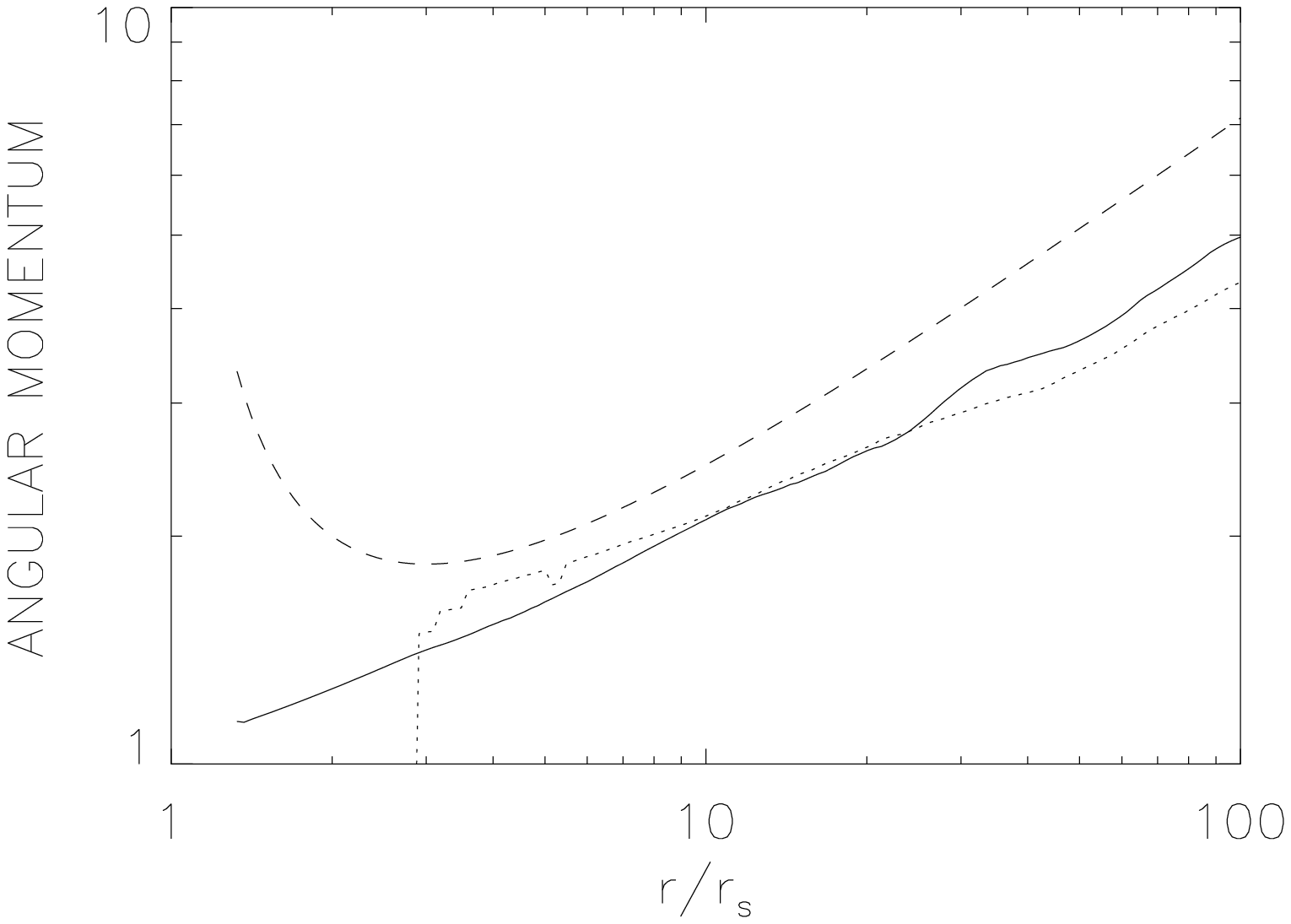} \vspace{0.2in}\epsscale{0.45} \plotone{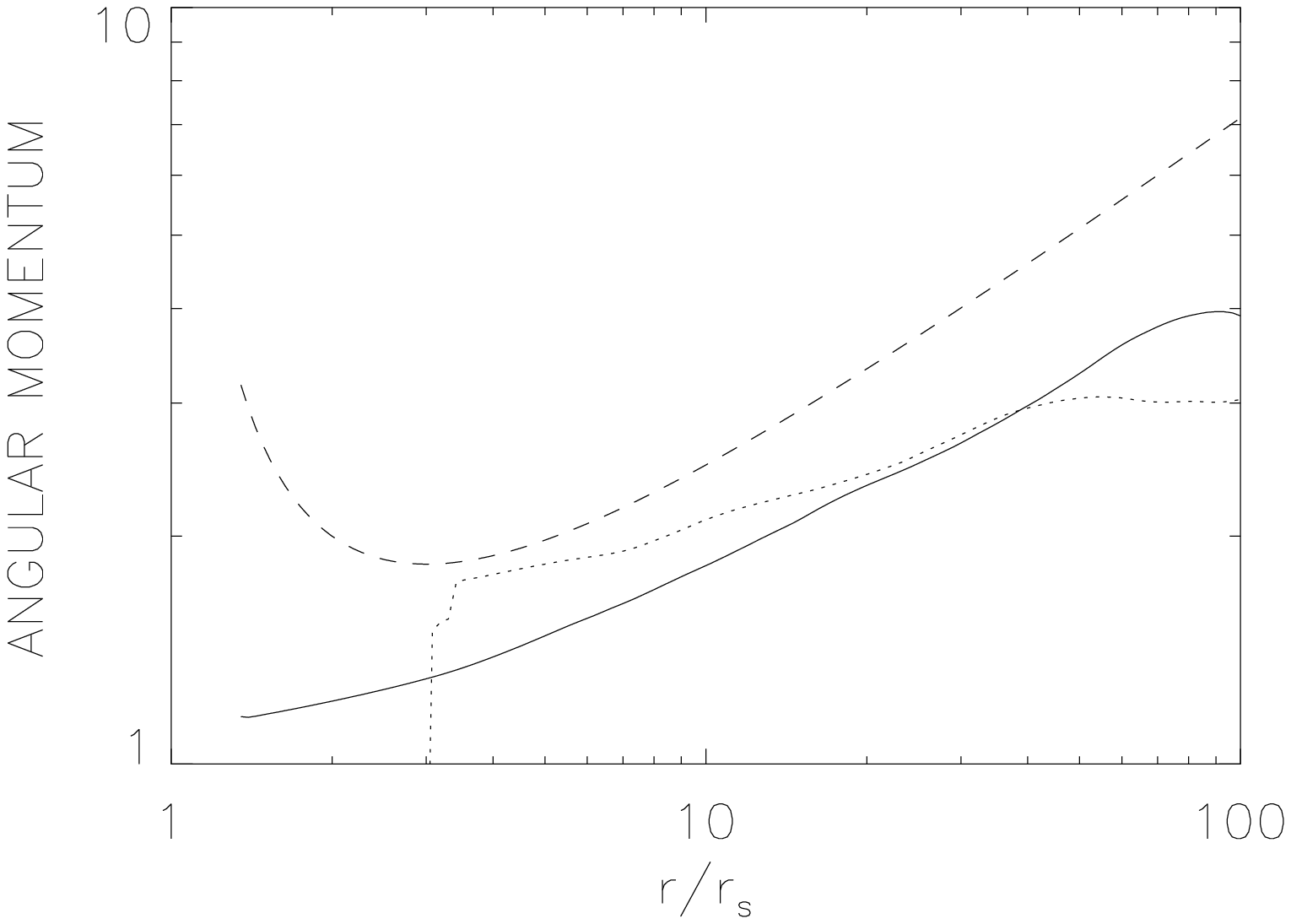}
\hspace{1cm} \epsscale{0.45} \plotone{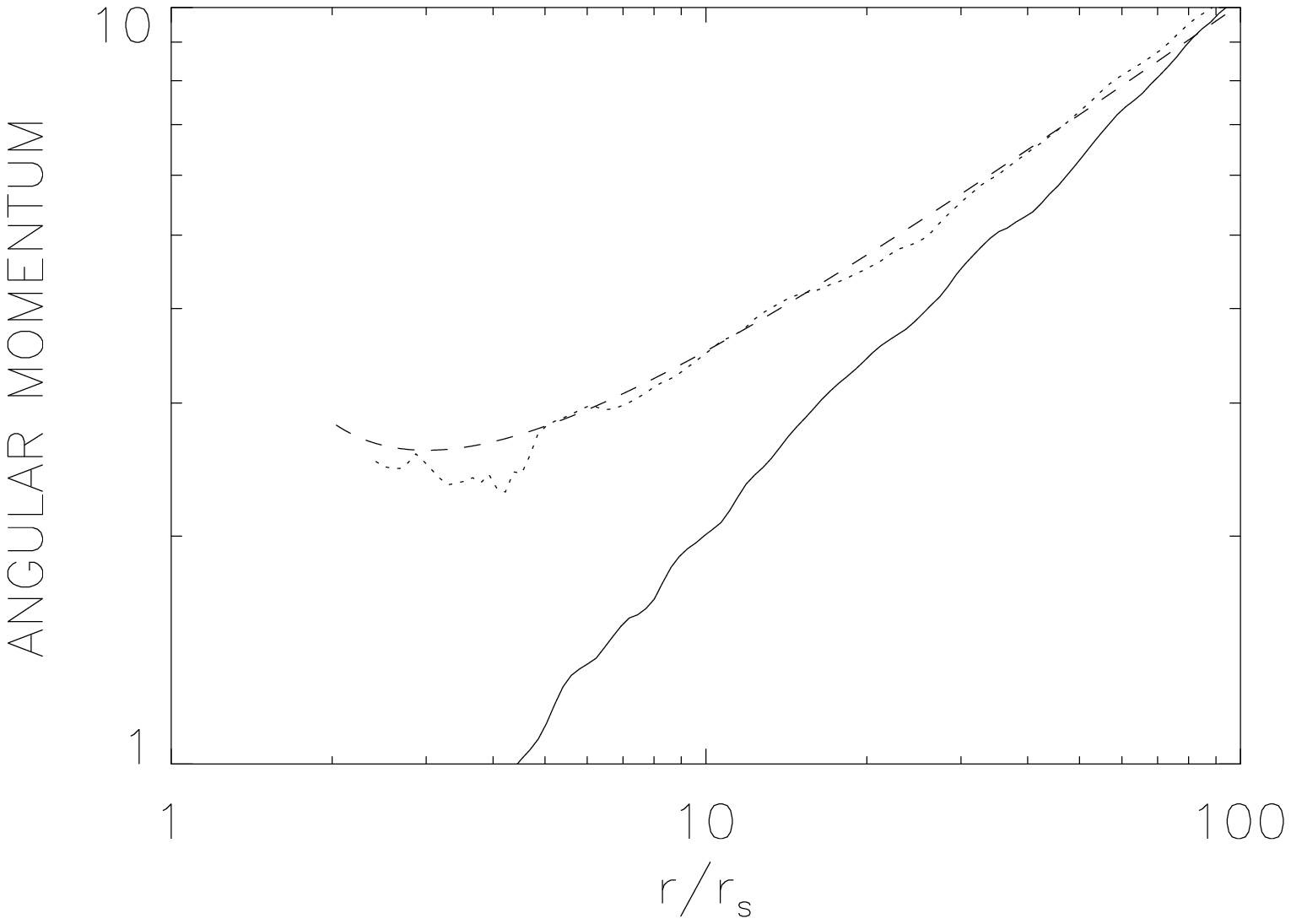}\vspace{0.2in}
\caption{The radial distribution of the specific angular momentum $l=v_{\phi}r \sin\theta$ for both inflow (solid line) and outflow (dotted line). The dashed line denotes the Keplerian angular momentum {\em at the equatorial plane}.  The top-left, top-right, bottom-left, and bottom-right plots are for Models A, B, C, and D, respectively. Note the sharp contrast between the HD and  MHD models.}
\label{Fig:rotationvelocity}
\end{figure*}

Fig. \ref{Fig:rotationvelocity} shows the distribution of specific angular momentum of both inflow and outflow. The results of the three HD models are similar, which is different from Model D. For the three HD models, the angular momentum of outflow is larger than that of inflow at small radii, consistent with Stone, Pringle \& Begelman (1999). This result indicates that outflow can transfer angular momentum at small radii. At large radii, however, the angular momentum of outflow is smaller than that of inflow. This means that the radial profile of angular momentum of the outflow is flatter than that of inflow, and the angular momentum of outflow inclines to be conserved during their outward propagation. For both inflow and outflow, the angular momentum is significantly lower than the Keplerian one at the equatorial plane.

Model D is completely different. We see from the figure that throughout the accretion flow the specific angular momentum of outflow is very close to and even in some region slightly larger than the Keplerian value at the equatorial plane. This is significantly larger than that of inflow whose value is substantially sub-Keplerian\footnote{Note that the angular momentum is calculated as the flux-weighted average over $\theta$ from $0^{\circ}$ to $180^{\circ}$, not just close to the equatorial plane.}. This indicates that in an MHD flow, outflow can transfer angular momentum much more efficiently compared to the hydrodynamical flows. This is related to the mechanism of producing outflow. We will be back to this question in \S\ref{outfloworigin}.

\subsection{Momentum Flux and Kinetic Power of Outflow}

Based on the above results, we can estimate the radial momentum flux and kinetic power of outflow. The terminal radial momentum flux is
\be \dot{p}_w=\dot{M}_{\rm out}(r_{\rm out}) v_{\rm term}\sim 0.5\dot{M}_{\rm out}(r_{\rm out})v_k(r_{\rm out}).\label{momentumflux}\ee Here $\dot{M}_{\rm out}(r_{\rm out})$ is the outflow rate at the outer boundary of the hot accretion flow $r_{\rm out}$.

It is interesting to compare this momentum flux with the radiation flux $L/c$. Adopting the result from Paper I, the inflow rate at $r= 10r_s$ is $\dot{M}_{\rm in}(10r_s)=\dot{M}_{\rm in}(r_{\rm out})(10r_s/r_{\rm out})^{0.5}=\dot{M}_{\rm out}(r_{\rm out}) (10r_s/r_{\rm out})^{0.5}$. The radiative flux is then \be L/c\sim \eta \dot{M}_{\rm out}(r_{\rm out})(10r_s/r_{\rm out})^{0.5}c. \ee Here $\eta$ is the radiative efficiency of hot accretion flow defined as $\eta\equiv L_{\rm bol}/\dot{M}_{\rm in}(10r_s)c^2$. So the ratio of momentum flux of outflow and  radiation is \be \frac{\dot{p}_w}{L/c}\approx \frac{0.1}{\eta}.\label{momentumratio}\ee
The efficiency $\eta$ is mainly a function of accretion rate, parameters $\delta$ and $\alpha$ (see Xie \& Yuan 2012 for details). For $\delta=0.5$, as indicated from the detailed modeling to Sgr A* (Yuan, Quataert \& Narayan 2003), and $\alpha=0.1$, we have (Xie \& Yuan 2012) \be \eta=0.05 (\dot{M}_{\rm in}(10r_s)/4\times 10^{-3}\dot{M}_{\rm Edd})^{0.2}\ee for $\dot{M}_{\rm in}(10r_s)=(4 \times 10^{-3}-10^{-5})\dot{M}_{\rm Edd}$; and \be \eta=0.05(\dot{M}_{\rm in}(10r_s)/4\times 10^{-3}\dot{M}_{\rm Edd})^{0.6} \ee for $\dot{M}_{\rm in}(10r_s)=(4\times 10^{-3}- 10^{-2})\dot{M}_{\rm Edd}$. Above $10^{-2}\dot{M}_{\rm Edd}$, up to $\sim 0.1 \dot{M}_{\rm Edd}$, the accretion flow is two-phase. The detailed study to the dynamics and radiation of this kind of accretion flow remains to be explored. The approximate calculation presented in Xie \& Yuan (2012) gives $\eta \approx 0.08$. So in general we expect \be \dot{p}_w\ga L/c.\label{momentumratio}\ee

The ingredient of jet is still controversial. Assuming it is dominated by normal plasma, we estimate the momentum flux of the jet: \be \dot{p}_{\rm
jet}=\Gamma_{\rm jet}^2\eta_{\rm jet} \left(\frac{10r_s}{r_{\rm
out}}\right)^{0.5}\dot{M}_{\rm in}(r_{\rm out})c=9\Gamma_{\rm
jet}^2\eta_{\rm jet}\dot{p}_w,\ee with $\eta_{\rm jet}$ is the ratio of the mass loss rate in the jet and the mass accretion rate at $10r_s$, and $\Gamma_{\rm jet}$ is the Lorentz factor of jet.

The value of $\eta_{\rm jet}$ is highly
uncertain. From the detailed modeling to a black hole X-ray binary XTE
J1118+480 (Yuan, Cui \& Narayan 2005), we get $\eta_{\rm jet}\sim
1\%$. For comparison, Fender et al. (2001) estimated the jet power in the same source based on the observed radio luminosity.  The obtained range of jet power is quite large, but the ``geometric average'' of their upper and lower limits is consistent with Yuan, Cui \& Narayan (2005) value within a factor of 3. The Lorentz factor of jets is different for AGNs and black hole X-ray binaries, and reasonable values are $\Gamma_{\rm jet}\sim  $ a few and $\Gamma_{\rm jet}\sim 1$, respectively (Gallo et al. 2003). For the jet in M~87, the estimated jet power (eq. [\ref{jetpower}]) based on $\eta_{\rm jet}\approx 1\%$, combined with $r_{\rm out}=5\times 10^5$ and $\dot{M}_{\rm in }(r_{\rm out})=0.1\msun ~{\rm {yr}^{-1}}$ from Di Matteo et al. (2003) and $\Gamma_{\rm jet}\approx 6$ (Biretta, Sparks \& Macchetto 1999), is $L_{\rm jet}\sim 7\times 10^{42}\ergs$. This value is completely consistent with the independent estimation of Reynolds et al. (1996). So the momentum flux of outflow is comparable to that of jet for AGNs; while it is a factor of 10 larger than that of jet for black hole X-ray binaries.

The kinetic energy power of the outflow is estimated to be \be \dot{E}_w=\frac{1}{2}\dot{M}_{\rm out}(r_{\rm out})v_{\rm term}^2\approx 0.1\dot{M}_{\rm out}(r_{\rm out})v_k^2(r_{\rm
out}).\ee  Here we approximately use the flux-weighted radial velocity which is not precise but correct in  order of magnitude. Since $v_k^2(r_{\rm out})=GM/r_{\rm
out}=c^2/(2r_{\rm out}/r_s)$, we have \be \dot{E}_w=\frac{\dot{M}_{\rm out}(r_{\rm out})c^2}{20r_{\rm
out}/r_s}.\label{outflowpower}\ee

We now compare $\dot{E}_w$ with several other power. The accretion power at $10r_s$ is
\be P_{\rm acc}=\left(\frac{10r_s}{r_{\rm
out}}\right)^{0.5}\dot{M}_{\rm in}(r_{\rm out})c^2\approx 60\dot{E}_w\left(\frac{r_{\rm out}}{r_s}\right)^{0.5}\gg\dot{E}_w. \ee The jet power is \be \dot{E}_{\rm
jet}=\Gamma_{\rm jet}^2\eta_{\rm jet} \left(\frac{10r_s}{r_{\rm
out}}\right)^{0.5}\dot{M}_{\rm in}(r_{\rm out})c^2=60\Gamma_{\rm
jet}^2\eta_{\rm jet}\dot{E}_w\left(\frac{r_{\rm
out}}{r_s}\right)^{0.5}.\label{jetpower}\ee  The radiation power of the accretion flow is \be \dot{E}_{\rm rad}=\eta\left(\frac{10r_s}{r_{\rm
out}}\right)^{0.5}\dot{M}_{\rm in}(r_{\rm out})c^2=60\eta\dot{E}_w\left(\frac{r_{\rm out}}{r_s}\right)^{0.5}. \label{radiationpower}\ee
The outer boundary of the hot accretion flow $r_{\rm out}$ in both
the cases of AGNs and black hole X-ray binaries is uncertain. In the case of AGNs, one possibility is that the outer boundary is the Bondi radius. This may be the case if the accretion flow starts out hot, as likely in elliptical galaxies. Another possibility is that the accretion flow starts out cool so it is described at large radii by a standard
thin disk. This may be the case of spiral galaxies. If the accretion rate is not too high, this thin disk will be truncated and replaced by a hot
accretion flow at a transition radius (Yuan 2007). This transition
radius is then the outer boundary of hot accretion flow $r_{\rm out}$. Detailed modeling show that it is a function of mass accretion rate or equivalently the
luminosity of the system (Yuan \& Narayan 2004). When the luminosity
is very low, say $L\la 10^{-6}L_{\rm Edd}$, as in the case of Sgr A*, $r_{\rm out}\sim r_{\rm Bondi}\sim 10^5 r_s$. But if luminosity is higher, $L\ga
10^{-4}L_{\rm Edd}$, $r_{\rm out}$ may be in the range from several tens to several thousands of $r_s$.

From the above assumptions on parameters, we
conclude that the power of jet and radiation is usually much stronger than that of the outflow, $\dot{E}_{\rm jet}\gg \dot{E}_{\rm rad} \gg \dot{E}_w$. But note that this does not mean the mechanical feedback of outflow is not important compared to jet. Since the jet is well collimated, it just drills through the surrounding gas with little deposition of energy and momentum within the galaxy. Rather, jet may act as the dominant feedback mechanism for galaxy clusters. For the evolution of a single galaxy, since the solid angle of outflow is very large so they can fully interact with the ISM. It has been shown that the momentum feedback of outflow dominates the growth of the central black hole (Ostriker et al. 2010).  In \S\ref{observation}, we will discuss two additional examples of such interaction.

\section{Comparison with Previous Works}

Igumenshchev \& Abramowicz (2000) performed two-dimensional HD simulation. Similar to our Model B, they adopted the ``injection-type'' initial condition. The gas was injected at the outer boundary, with rotation velocity of $0.9v_k$. The radial velocity was not given but obtains their value ``automatically'' after accretion begins, which  means that the initial radial kinetic energy is zero. The temperature are not specified either, thus the value of $Be$ of the initial gas is unknown. Interestingly, they also found nonsymmetric large-scale circular-like (but not closed) bulk motion in their low-$\alpha$ model (Model M; refer to the right plot of their Fig. 15), which is similar to our Models B \& C (refer to the discussions to Figs. \ref{Fig:mdotinflow} \& \ref{Fig:mdotoutflow} in \S3.2). Regarding outflow, they found that the value of $Be$ of most of the outflow is negative, except in the narrow region (in the $\theta$ direction) where outflow has $Be>0$. This is consistent with our result of Model A. This suggests that the initial $Be$ of their model is similar to that of our Model A, but significantly smaller than our Models B \& C.

Compared to the HD simulations, there are more MHD simulation works on hot accretion flow. When outflow is concerned, most of these works focus mainly on the jet, i.e., the outflow close to the axis and with a positive $Be$. Only some of them discussed outflow away from the axis (e.g., Stone \& Pringle 2001; Hawley \& Balbus 2002; De Villiers \& Hawley 2003; Igumenshchev, Narayan \& Abramowicz 2003; De Villers et al. 2005). These simulations often adopt a rotating torus as the initial condition, similar to our Models A \& D, and their results are also similar to our Model D (e.g., Hawley \& Balbus 2002; Narayan et al. 2012), as expected. For example, in Hawley \& Balbus (2002) the torus is centered at $200M$. Similar to us, they found that most of the outflow is bound; while close to the axis they are unbound.

De Villers et al. (2005) carried out three-dimensional MHD simulations with initial toroidal and poloidal magnetic field configuration. The center of the initial torus is located at $25M$.  In the poloidal case, jets are found in the regions of the axial funnel and funnel wall. This is consistent with our Model D, as shown by, e.g., the right-bottom plots of Fig. \ref{Fig:bernoullicontour}. They did not find jets in the toroidal case. They argued that this is perhaps because poloidal magnetic field is not formed in the funnel region. In Hawley \& Krolik (2002) the initial configuration of the magnetic field is very similar to De Villiers et al. (2005). The initial condition is also similar, i.e., a hydrostatic torus with its pressure maximum located at $20M$. Hawley \& Krolik (2002), however, did not find any jets since they found all gas is bound, even in the case of a poloidal initial magnetic field. De Villers et al. (2005) speculated that the reason is the choice of cylindrical grid in Hawley \& Krolik (2002), which prevents from the build up of the poloidal field close to the funnel. However, Hawley \& Balbus (2002) and Kato, Mineshige \& Shibata (2004) also adopted cylinder grid, similar to Hawley \& Krolik (2002). But different from Hawley \& Krolik (2002), both Hawley \& Balbus (2002) and Kato, Mineshige \& Shibata (2004) found unbound jet in their simulations. What is the reason for the discrepancy?

Whether jets can be formed or not is ultimately determined by the sign of $Be$ of the outflow. Our simulations show that the value of $Be$ of outflow is crucially determined by the initial condition of simulation. If $Be$ is large in the initial condition, the value of $Be$ of outflow will be large thus jet can be formed easily. Of course, the existence of poloidal magnetic field can also help the formation of jets, because the value of $Be$ of outflow can be increased by the Lorentz force. In Hawley \& Balbus (2002) and Kato, Mineshige \& Shibata (2004), the initial torus is located at a relatively large radius,  $\sim 200M$ and $80M$ respectively. Therefore the value of $Be$ of the initial torus is likely large enough to produce unbound jets. Following this line of thought, the temperature of the initial tours also matters. This is perhaps the case of De  Villers \& Hawley (2003). In that work, they considered hotter torus compared to De Villiers et al. (2005). Consequently, although the torus is closer to the hole, located at $\sim 15M$, a higher fractional jet rate was still found.

\section{ADIOS or CDAF?}
\label{adioscdaf}

As we have mentioned in \S1, two different models have been proposed to explain the inward decrease of the accretion rate, namely ADIOS and  CDAF. In the former, the inward decrease of accretion rate is thought to be because of the systematic mass loss in the outflow which is launched continuously from almost all radii. The rates of inflow and outflow (eqs. [\ref{inflowrate}] and [\ref{outflowrate}]) are dominated by the systematic fluxes of mass. In the CDAF model, however, they think the inward decrease of accretion rate is because that with accretion, more and more fluid circulates in convective eddies and locked in them. The inflow and outflow rates are assumed to be dominated by convective turbulence. In this section we will try to argue, based on our simulation results presented in \S3, that the ADIOS scenario is more likely.

Firstly, as pointed out by Blandford \& Begelman (2004) and Begelman (2012), the CDAF model faces the secular difficulties related to the global mass supply. The inward decrease of mass accretion rate is explained as that the fluid is locked in the convective eddies. In this scenario, most of the outflowing fluid (which contributes to the outflow rate) will eventually turn around and feed the inflow. Thus when we have continuous mass supply from outer boundary, we would expect mass accumulation; thus a steady state can't be achieved. In most previous numerical simulation works, the initial condition is a torus with limited mass. In this case the simulation can't run for too long time and the available mass is very limited, thus it is hard to examine this point. In Model C of the present paper, we continuously inject mass from an outer boundary which enable us to run the code long enough time. We do not observe any mass accumulation, and the profiles of accretion rate and density have no difference from the other three models.  The above argument also applies if the turbulence is caused by other instability such as MRI rather than convection. If there were no real systematic outflow, but the outflow rate we calculate were just the turbulent fluctuation, it would be hard to understand why the inflow rate decreases inward. Of course, as we will state below, turbulent fluctuations must contribute to the outflow rate although they are not dominant.

The second issue with the CDAF model is its applicability to MHD accretion flow.
Stone \& Pringle (2001), Hawley, Balbus \& Stone (2001) and Balbus \& Hawley (2002) argued that the dynamics of the MHD accretion flow should be dominated by MRI rather than convection, since the dominant mode of angular momentum transport in an MHD accretion flow is MRI, no matter whether or not destabilizing entropy gradients are present. Narayan et al. (2002) performed a linear MHD stability analysis of a rotating MHD accretion flow. They found that both ``convective'' and ``MRI'' modes exist in general. The stability criteria for the convective mode is the standard H{\o}iland criteria of hydrodynamics (see eq. [\ref{hoiland}] below; but see Shcherbakov 2008). When the accretion flow is convectively stable according to this criteria, MRI is the only unstable mode. When the accretion flow is convectively unstable, however,  the result is complicated, depending on the wavelength of the perturbation. The short-wavelength modes are still MRI. But the long-wavelength modes are found to be independent of the magnetic field and intrinsically a convective mode, because the buoyancy force which drives convection is much stronger than the magnetic force which drives MRI. Since the saturated value of magnetic field in accretion flow is not very strong, $\beta\gg1$, the long-wavelength fluctuation with $\lambda/H\gg \beta^{-1/2}$ can usually fit inside the disk.

\begin{figure}
\epsscale{0.9} \plotone{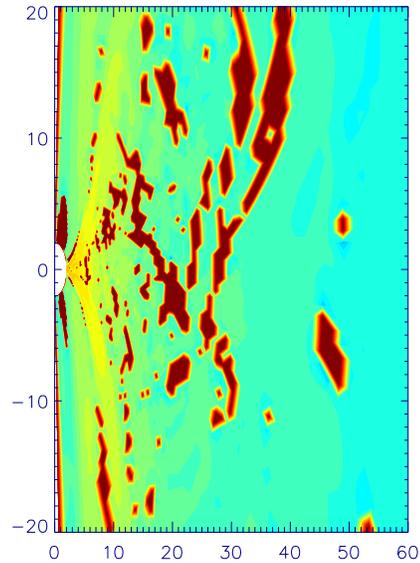}\hspace{1.cm} \epsscale{0.8}
\caption{Convective stability analysis of the MHD accretion flow (Model D) according to  eq. (\ref{hoiland}) based on the simulation data at t=3.5 orbits at $R=100r_s$. The red region denotes the unstable region. The result indicates that most region of MHD flow is convectively stable. }
\label{Fig:instability}
\end{figure}

\begin{figure}
\epsscale{0.98} \plotone{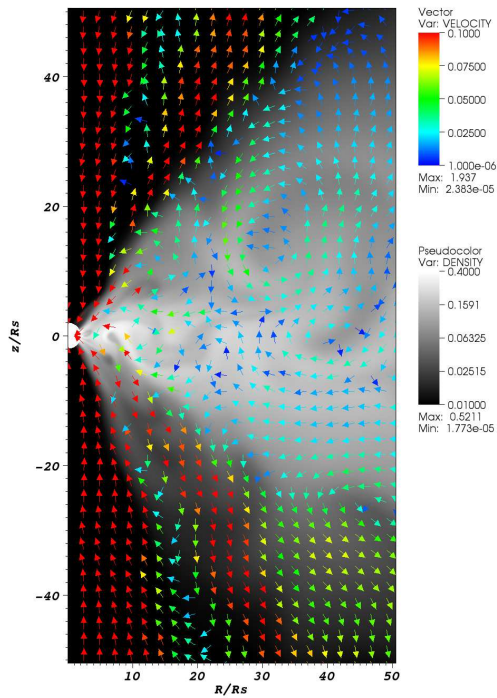}\hspace{1.cm} \epsscale{0.9}
\caption{Snapshot of the velocity vectors (arrows) of Model A in the steady state, overlaid  with density. Both systematic inflowing and outflowing motion and convective turbulence are evident. }
\label{Fig:streamline}
\end{figure}

\begin{figure*}
\epsscale{0.45} \plotone{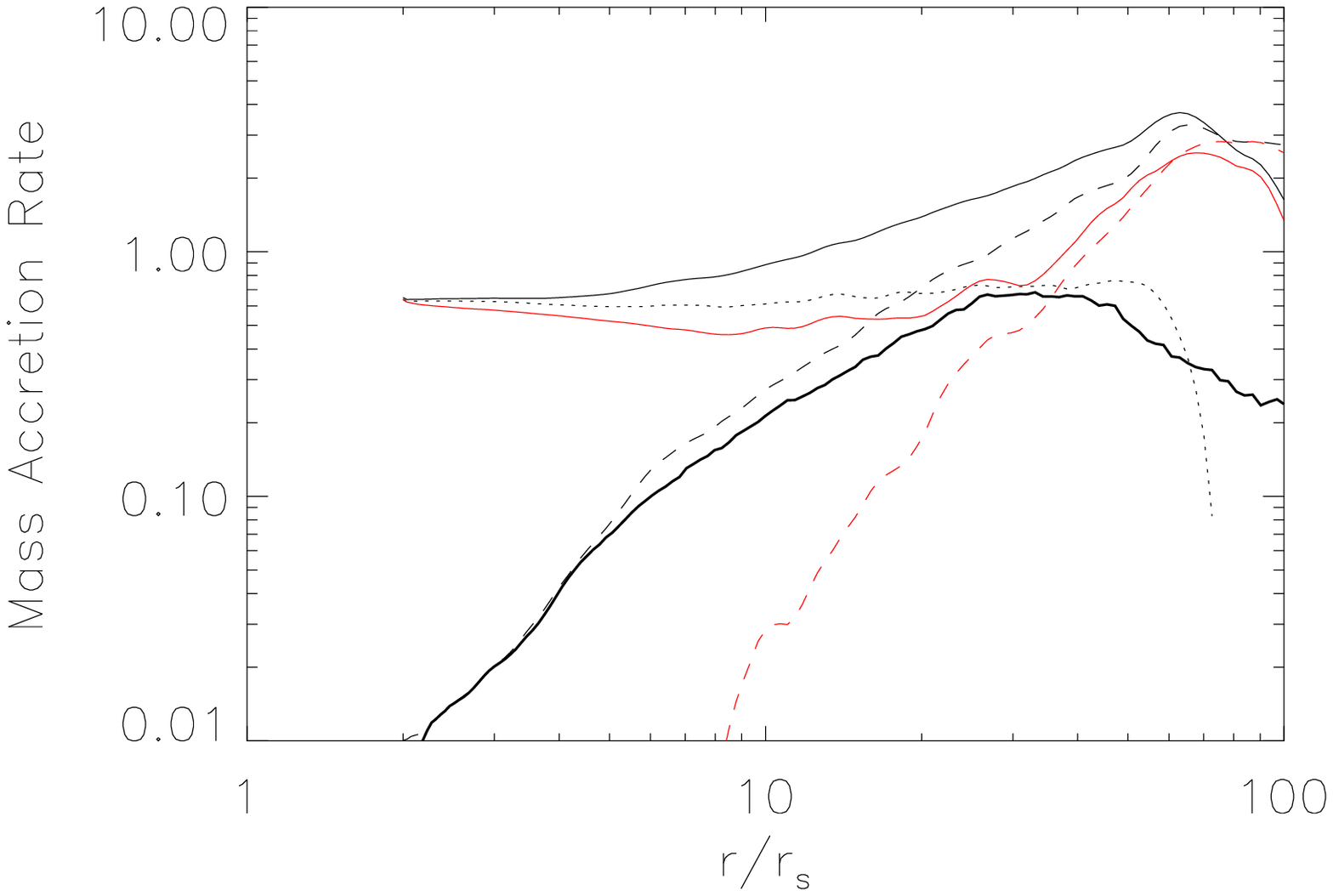}\hspace{1.cm} \epsscale{0.45}
\plotone{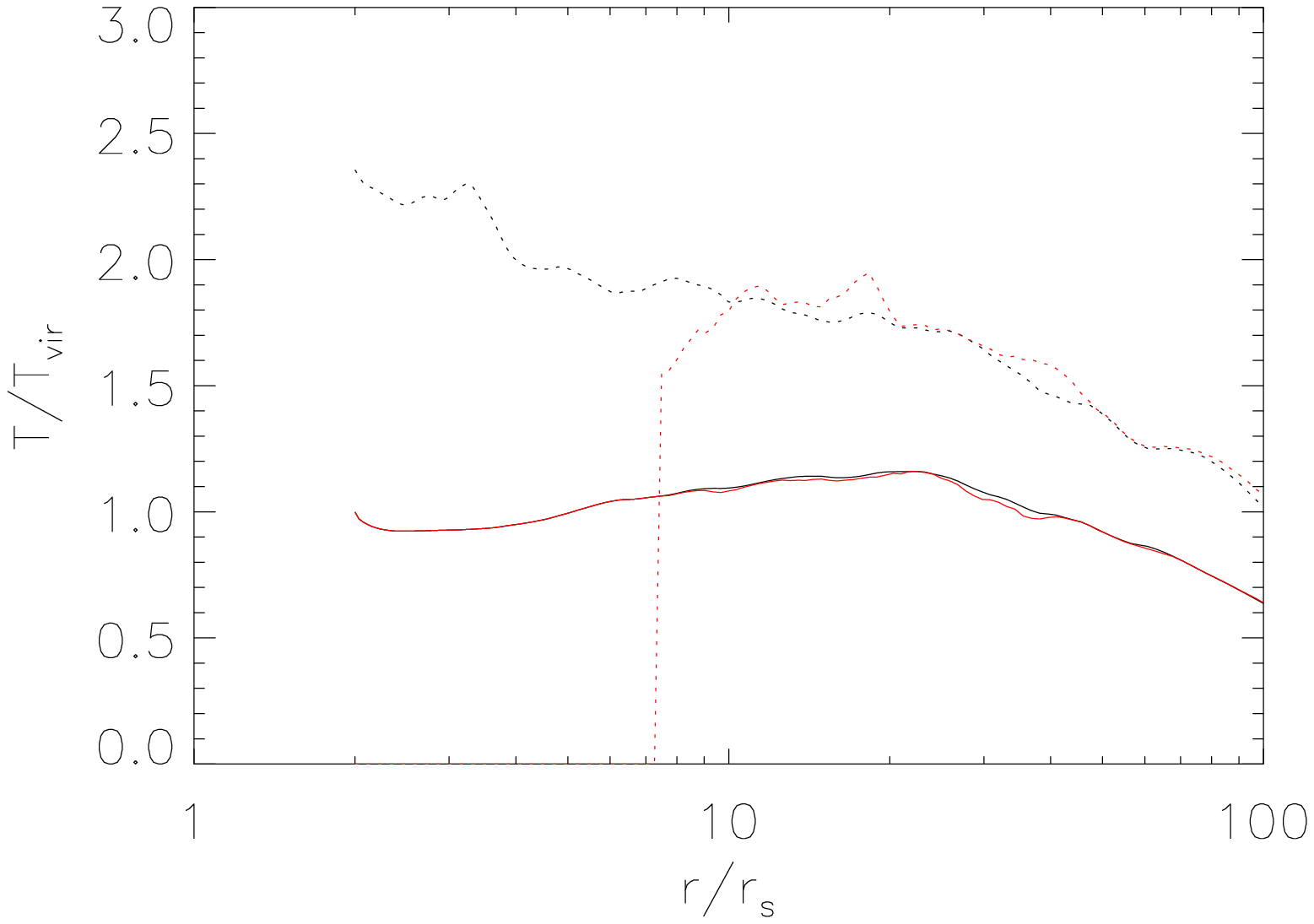} \vspace{0.1in}\epsscale{0.45} \plotone{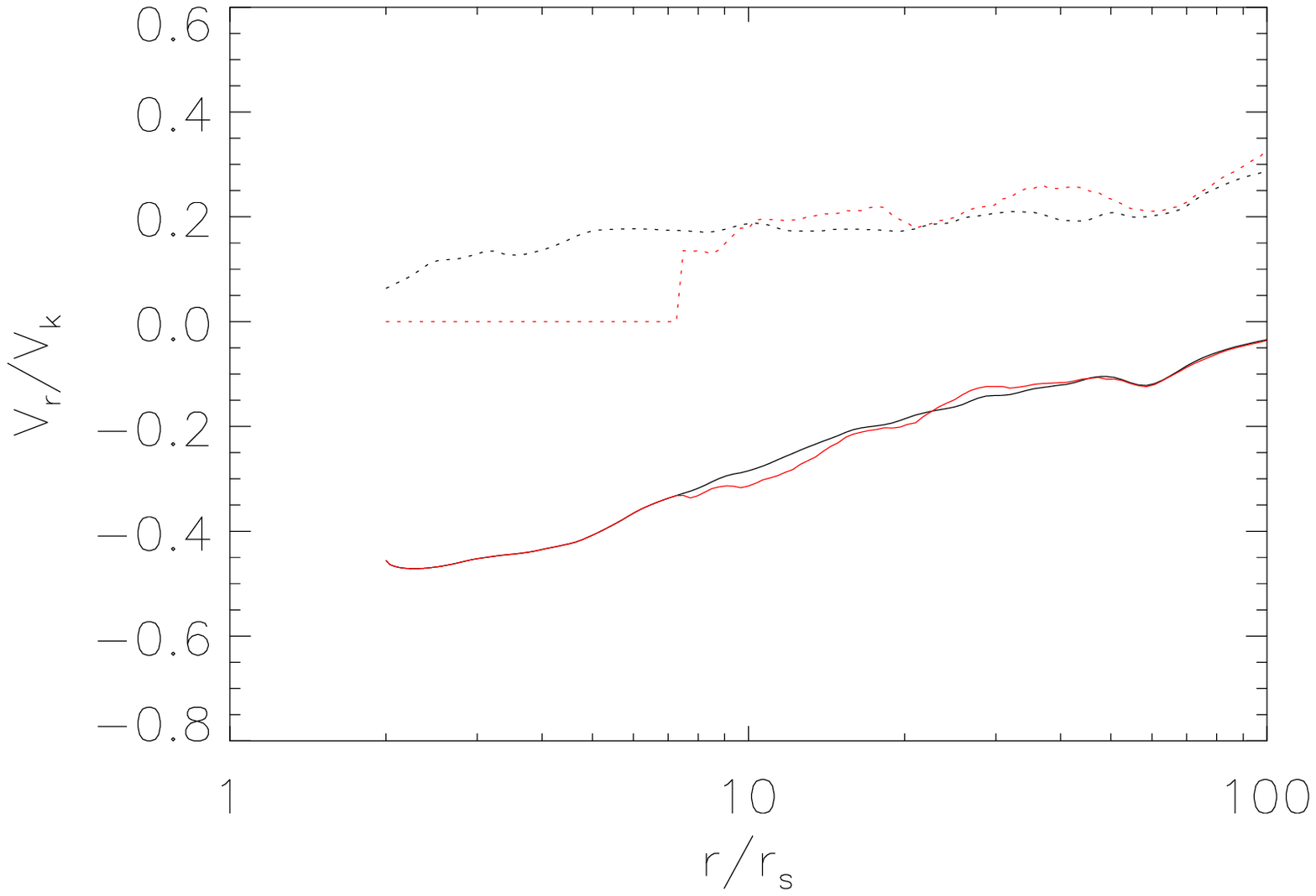}
\hspace{1cm} \epsscale{0.45} \plotone{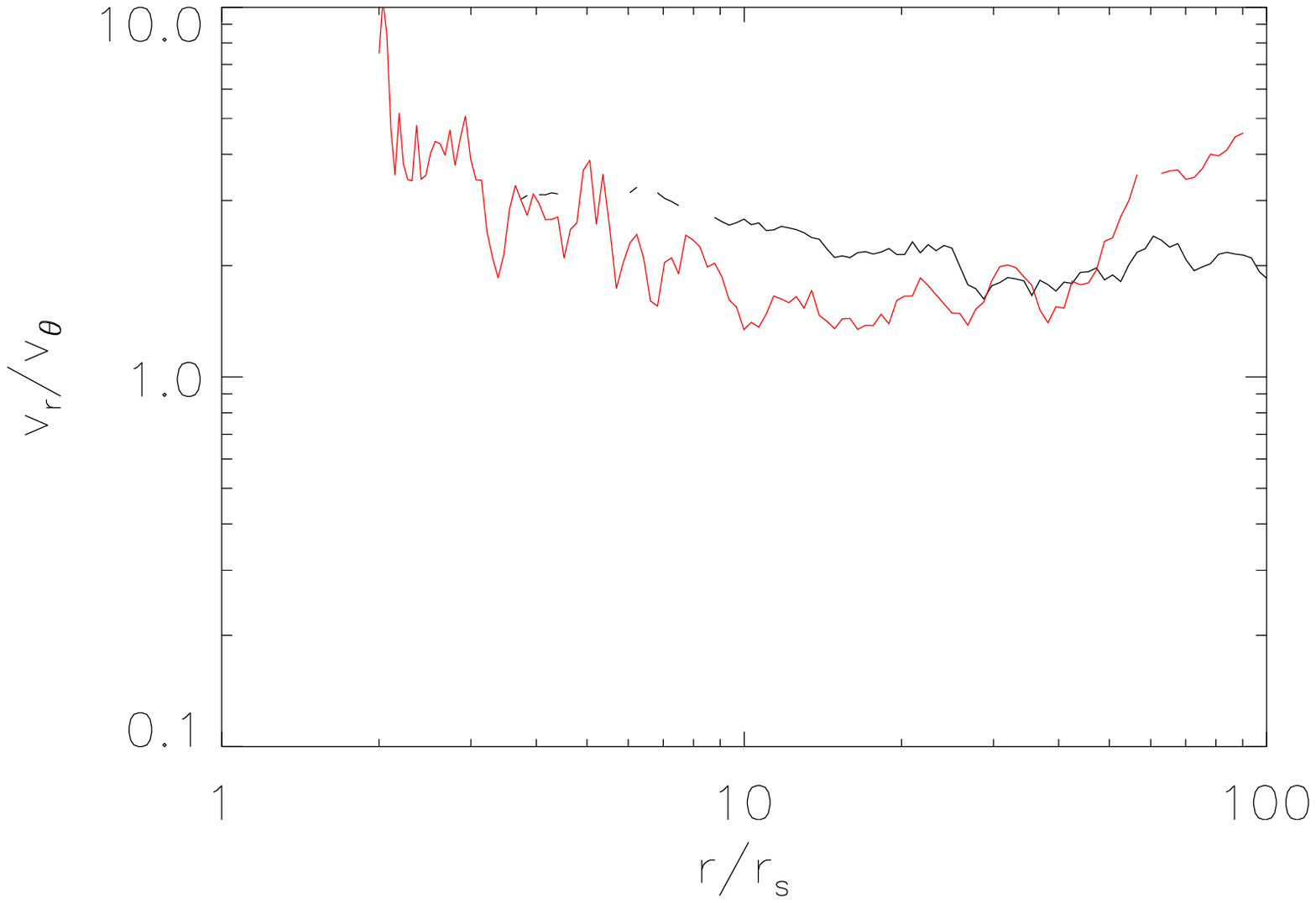}\vspace{0.1in}
\caption{Studying the contribution of turbulent fluctuation in the inflow and outflow rates (eqs. [\ref{inflowrate}] and [\ref{outflowrate}]). All data is based on Model A. {\em Top-left}: The black lines are exactly same with those in the top-left plot of Fig. \ref{Fig:mdotinflow}. The two red lines show the inflow (solid) and outflow (dashed) rates calculated by the ``second'' definition (i.e., integrating the time-averages of fluxes over $\theta$). {\em Top-right}: The black lines are exactly same with those in the top-left plot of Fig. \ref{Fig:temperature}. The red lines show the temperature of inflow (solid) and outflow (dotted) based on their ``second'' definitions. {\em Bottom-left}: The black lines are exactly same with those in the top-left plot of Fig. \ref{Fig:radialvelocity}. The red lines show the radial velocity of inflow and outflow based on their ``second'' definitions. {\em Bottom-right}: The black and red lines show the ratio $v_r/v_{\theta}$ for inflow and outflow based on our (``first'') definitions, respectively. See text for details.}
\label{Fig:turbulence}
\end{figure*}

It is therefore crucial to examine whether the MHD accretion flow is convectively stable or not. The H{\o}iland criteria for the stability is (e.g., Tassoul 1978; Begelman \& Meier 1982):
\be  (\nabla s \cdot \mathbf{dr})(\mathbf{g} \cdot \mathbf{dr})-\frac{2\gamma v_{\phi}}{R^2}[\nabla(v_{\phi}R)\cdot \mathbf{dr}]dR <0.\label{hoiland}\ee  In eq. (\ref{hoiland}), $R=r\sin \theta$ is the cylindrical
radius, ${\bf dr} = dr \hat r + r d \theta \hat \theta$ is the
displacement vector, $s = \ln(p) - \gamma \ln(\rho)$ is $(\gamma - 1)$
times the entropy, ${\bf g} = - \hat r v_K^2/r + \hat R v_{\phi}^2/R$ is
the effective gravity, and $v_K $ and $v_{\phi}$ are the
Keplerian and rotational velocities, respectively. For non-rotating flow, this condition is equivalent to an inward increase of entropy, which is the well-known Schwarzschild criteria. We study the convective stability based on the simulation data of Model D.  The results are shown in  Fig. \ref{Fig:instability}. The red region denotes it is convectively unstable. We see from the figure that most of the region the flow is not in red color, which indicates that the MHD accretion flow is convectively stable. This implies that the CDAF model can't be applied to MHD accretion flows. Given that the mass accretion rate in both MHD and HD accretion flows decreases inward, CDAF is unlikely the feasible model.

The third problem of the CDAF model is its difficulty to explaining the systematic differences of the properties of inflow and outflow. In the CDAF model, the inflow and outflow rates (eqs. [\ref{inflowrate}] and [\ref{outflowrate}]) are assumed to be dominated by the turbulent fluctuation. If this were true, we should expect that the properties of inflow and outflow should be roughly same. To check this point, in the present paper, we have done numerical simulations and calculated the Bernoulli parameter $Be$, temperature, radial and rotational velocities of inflow and outflow. Our results indicate that the properties of outflow and inflow are systematically and significantly different (Figs. \ref{Fig:radialbernoulli} -- \ref{Fig:rotationvelocity} and \S3 for details).  Such a difference is hard to be explained if the inflow and outflow rates are dominated by turbulent fluctuations. Rather, they are strong evidence that the systematic fluxes of inflowing and outflowing mass must constitute at least significantly, if not dominantly, to the inflow and outflow rates. Such systematic inflow and outflow motion can be observed in the movie made based on the HD or MHD simulations\footnote{http://www.astro.princeton.edu/~jstone/disks.html}, and also in Fig. \ref{Fig:streamline}, where the snapshot of the velocity vectors of Model A are shown (see also Fig. 3 in Li, Ostriker \& Sunyaev 2012). In this regard, the nearly equal inflow and outflow rates at large radii are simply because of the steady state requirement. This is because at the steady state the net accretion rate, which measures the difference between inflow and outflow rates, and which is much smaller than these two rates, must be a constant of radius.

On the other hand, we note that turbulence must contribute at some level.  This can again be seen from Fig. \ref{Fig:streamline} (see also Igumenshchev, Narayan \& Abramowicz 2003; Mckinney, Tchekhovskoy \& Blandford 2012; and Narayan et al. 2012). In this sense, the rates of inflow and outflow based on eqs. (\ref{inflowrate}) and (\ref{outflowrate}) may somehow overestimate the mass fluxes of systematic inflow and outflow.

Instead of using eqs. (\ref{inflowrate}) and (\ref{outflowrate}), which is the time-average of the instantaneous inflow and outflow fluxes, in the recent work of Narayan et al. (2012) an alternative approach of judging inflow and outflow was suggested. For a given grid they calculate the time-average of the  radial velocity. If it is positive they call it outflow and vice versa. In this ``second'' approach, the inflow and outflow rates are still calculated formally according to eqs. (\ref{inflowrate}) and (\ref{outflowrate}), but now $\rho$ and $v_r$ are their time-averaged values. The advantage of this approach is that it can average out the turbulence component. However, if the accretion flow is dominated by systematic inflow and outflow, and the flows are not fixed to a certain $\theta$ angle but move ``up and down'' in the $r-\theta$ space, which is quite natural, this approach will also average out the systematic mass flux thus underestimate the inflow and outflow rates. Therefore this approach roughly gives a lower limit to the inflow and outflow rates. The two red lines in the upper-left plot of Fig. \ref{Fig:turbulence} show the results of inflow and outflow rates calculated by this ``second'' approach. Comparing with the black lines which is based on our ``first'' approach, we can see that within $r\sim 30r_s$, the inflow and outflow rates now is significantly smaller, which is consistent with Narayan et al. (2012). Beyond $r\sim 40r_s$, the inflow and outflow rates obtained by the two approaches differ by a factor of less than 2. However, at such a large radius, we are not confident to the result because of the possible boundary condition effect. The over-estimation of the contribution of turbulent fluctuation causes the following two consequences. One is that, from Fig. \ref{Fig:turbulence} we can see that the inflow rate calculated by the ``second'' approach (red solid line) is even smaller than the net rate (black dotted line) which is the accretion rate at the black hole horizon; or put it in another way, the inflow rate denoted by the red solid line decreases outward in some region. The second one is that we found the inflow and outflow rates calculated by the ``second'' approach depend on the time interval adopted in the time-average integration. The longer the interval, the smaller the rates are.

To further examine the contribution of turbulent fluctuation, we have done two other tests. One is to calculate again the temperature and radial velocity of inflow and outflow, following eqs. (\ref{fluxweight}) and (\ref{fluxweightin}), but this time based on the ``second'' definition of inflow and outflow. The results are shown by the red lines in the top-right and bottom-left plots of Fig. \ref{Fig:turbulence}. The black lines show the previous result. We can see that there are very little difference. This again indicates that turbulent fluctuation should not contribute significantly in the inflow and outflow rates defined by our ``first'' approach. At last, we calculate the ratio $v_r/v_{\theta}$ at each radius geometrically averaged\footnote{Here ``geometrical average'' means that $<q>=(q_1\times q_2...\times q_n)^{1/n}$.} over all $\theta$ for inflow and outflow defined by our (``first'') approach, respectively.  The result is shown by the bottom-right plot of Fig. \ref{Fig:turbulence}. In our calculation, at each radius we exclude the grids with $v_r/v_{\theta}>10$ to avoid numerical overflow. This is why the line is broken at some radii. So this result should be regarded as the lower limit. If the motion of fluid is dominated by turbulent fluctuation, we should expect that $v_r/v_{\theta}\sim 1$. However, we see from the figure that $v_r/v_{\theta} > 2$, i.e., the motion of both inflow and outflow is dominated by systematic radial motion. We would like to emphasize that all these tests give us only qualitative idea about the contribution of turbulent. It is necessary in the future to do more precise and quantitative calculations.

\section{Mechanisms of Producing Outflows in HD and MHD Flows: Buoyancy and ``Micro''-Blandford \& Payne Mechanism}
\label{outfloworigin}

In the early version of the ADIOS scenario (Blandford \& Begelman 1999), the production of outflow was proposed to be due to the positive $Be$ of the accretion flow, a result obtained in the self-similar solution of ADAFs (Narayan \& Yi 1994). However, previous HD and MHD numerical simulations have shown the existence of outflow when $Be<0$ (e.g., Igumenshchev \& Abramowicz 1999; SPB99; Yuan \& Bu 2010), although in this case the outflow may not be able to escape to infinity.  We propose that the mechanism of producing outflow is different for HD and MHD accretion flows. In the former it is buoyant force (i.e., convection) while in the latter case it is magnetic centrifugal force.

In the case of HD flow, it has been shown that the flow is convectively unstable. This means that if the temperature of a fluid element is perturbed to be higher than the surrounding medium, its density will become lower and the fluid element will feel the buoyancy force and escape outward. But different from the CDAF model, the flow will not turn around and circulate in convective eddies. Instead, they keep flowing outward and forms outflow. This is similar to the motion of a upward running air bubble in  water.  This picture is supported by the systematically higher temperature of outflow than  inflow, as shown by Fig. \ref{Fig:temperature}.

In the case of an MHD accretion flow, however, we have shown that it is convectively stable; thus the above scenario does not apply. This is confirmed by the roughly equal temperature between inflow and outflow as shown by the right-bottom plot of Fig. \ref{Fig:temperature}. Then what is the mechanism of producing outflow? From the right-bottom plot of Fig. \ref{Fig:rotationvelocity}, we see that the specific angular momentum of outflow is nearly Keplerian, which is much larger than that of inflow. This feature provides a sharp contrast to the case of three HD models where the angular momentum of outflow is significantly sub-Keplerian and is overall very similar to that of inflow. This strongly suggests that the outflow in an MHD accretion flow is driven by the centrifugal force. An interesting question is then what makes the difference of angular momentum between the inflow and outflow. This is explained in Fig. \ref{Fig:mbp}. Consider two fluid elements located at two different radii in a differential rotating accretion flow. They are initially moving inward, connected by a magnetic field line. Magnetic stress will transport the angular momentum from the inner fluid element to the outer one since the former rotates faster. Once the angular momentum of the outer element  reaches nearly Keplerian value, the centrifugal force, combined with the gradient of the gas pressure, will be able to make the fluid element turn around and throw it outward. This mechanism is similar to the Blandford \& Payne (1982) mechanism, in the sense that outflow is produced by the centrifugal force mediated by magnetic field. But there also exist differences. In the Blandford \& Payne (1982) model, a large scale open poloidal magnetic field is required; and the model works only in the coronal region where magnetic pressure is much larger than the gas pressure (thus the fluid elements anchored in the same field line have the same angular velocity). In our case, both requirements are not necessary. In fact, in our model the magnetic field in both the accretion flow and coronal region is tangled, thus scale is not large. In addition, from the bottom-right plot of Fig. \ref{Fig:rotationvelocity}, we see that the specific angular momentum of the outflow is not super-Keplerian, as expected if the Blandford \& Payne mechanism were operating. Therefore, this mechanism is different from the Blandford \& Payne mechanism and we call it a ``micro''- Blandford \& Payne (``MBP'' hereafter) mechanism. It will be interesting to check whether this mechanism also works for the standard thin disk.

\begin{figure}
\vspace{-0.3cm}\epsscale{1.4} \plotone{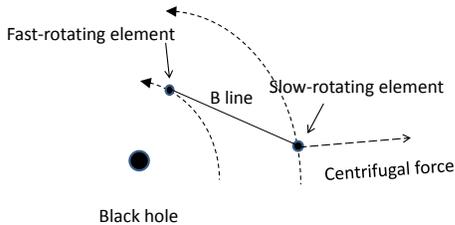}\vspace{-2.cm}
\caption{Schematic figure to illustrate the ``micro''-Blandford \& Payne mechanism of producing outflows. The fluid element at large radius obtains angular momentum via the magnetic field from the element at small radius. Consequently it can be turn around and thrown outward by the centrifugal force. See text for details. }
\label{Fig:mbp}
\end{figure}

In the literature, currently two popular models of producing outflow from accretion disks are radiation pressure driving (e.g., Murray et al. 1995) and the large scale magnetic field acceleration (e.g., Blandford \& Payne 1982). In terms of producing outflow from hot accretion flows, we propose that for hot accretion flows, the ``MBP'' mechanism may be the dominant one in nature. This is based on the fact that the radial density profile of accretion flow obtained from MHD numerical simulations, which only incorporate this mechanism, is in good consistency with that obtained from observations (see Paper I for details). Theoretically, the optical depth of the outflow from hot accretion flow may be too small for the radiation mechanism to work. The Blandford \& Payne mechanism, as we mentioned before, requires the existence of large-scale open poloidal magnetic field in a large range of radius of the accretion flow. However, the origin of such kind of field is still an open question. Two mechanisms have been considered. One is the MHD dynamo (e.g., Tout \& Pringle 1996), and another is the direct inward advection of large-scale poloidal field by the accretion flow from large radii. However, current MHD simulation shows that the dynamo mechanism seems to only produce small-scale poloidal field on the order of local disk thickness (e.g., De Villers et al. 2005). Beckwith, Hawley \& Krolik (2009) considered the second possibility. They found that large-scale poloidal field only exists close to the black hole.  However, Mckinney, Tchekhovskoy, \& Blandford (2012) show that if magnetic flux can be accumulated, large-scale poloidal field can be formed even away from the black hole.

\section{Possible Observational Applications}
\label{observation}

AGNs outflow is of great interest in recent years. One important reason is because it is believed that AGNs feedback plays an important role in galaxy formation and evolution (e.g., Di Matteo et al. 2005; Hopkins et al. 2005; Fabian 2012). In terms of the forms of the output from the central AGN, the feedback can be caused by radiative output (e.g., Ciotti, Ostriker \& Proga 2010) and mass outflow (e.g., Ostriker et al. 2010). Compared to the radiative feedback, feedback by mass outflow is more localized to the vicinity of the black hole thus more efficient in reducing the mass accretion rate and further the growth of the black hole. There are two types of mass outflow, i.e., jet and outflow (or winds). Their respective roles are different. The energy flux of the former is much larger than the latter. Perhaps because of this reason, most works on AGNs feedback concentrate only on jets. However, outflow has gradually received more and more attention because of the following reasons. Firstly, jet is well collimated thus they simply pill through the galaxy without any significant interaction with the interstellar medium. Therefore, jet can only have important role in galaxy clusters. Secondly, nearly $90\%$  AGNs are radio-quiet and don't have strong jets. Thirdly, the momentum flux of outflow is likely larger than that of jets thus outflow plays a more important role in some cases. For example, Ostriker et al. (2010) have shown that the growth of the central supermassive black hole in galactic center is mainly determined by the mass and momentum feedback of outflow.

Outflow has been directly detected in AGNs such as broad absorption lines quasars (e.g., Crenshaw et al. 2003; Tombesi et al. 2010; 2011a; 2012). But the origin and acceleration of outflow is still an unsolved problem. The widely suggested and studied mechanisms include radiation pressure (e.g., Murray et al. 1995; Murray \& Chiang 1997; Proga, Stone \& Kallman 2000), magnetic driving (e.g., Blandford \& Payne 1982; Emmering, Blandford \& Shlosman 1992; Romanova et al. 1997;  Bottorff et al. 2000), and thermal driving (e.g., Begelman, McKee \& Shields
1983; Chelouche \& Netzer 2005). While the radiation driving model is very popular, some problems have been found. For example, according to Murray \& Chiang (1997), when explaining the high ionization blueshifted lines, only
when viewed at the pole can we see the blueshifted line and the
lines must be very narrow. This is not consistent with observations
(Wang et al. 2011). The detailed modeling to some individual AGNs outflow also indicate that the radiation mechanism may not work, as we will state later in this section. The Blandford \& Payne (1982) mechanism has to face with the problem of the origin of large-scale magnetic field (refer to the discussion of the last section). In this section, we discuss the possibility that whether some AGNs outflow can be potentially explained by the model studied in the present work.

In addition to AGNs outflows detected via the absorption lines, bubbles around the black holes are also possible evidence for their existence. This is perhaps the case of the ``Fermi bubble'' detected in the center of the Galaxy. In this section, we will also discuss this possibility. To illustrate that outflow from hot accretion flow may be relevant and promising to interpreting some observed AGNs outflow, we need to introduce some background on the accretion models of AGNs. At last, outflow has also been detected in black hole X-ray binaries. Comparison between AGNs and X-ray binaries is obviously useful and important.

\subsection{Accretion Models of AGNs and Black Hole X-ray Binaries}

Outflows from AGNs are detected generally via the absorption lines. If we hope to explain the observations by the model proposed in this paper, one may ask whether the hot accretion flow model can be applied to luminous AGNs and how to produce the absorption line if the temperature of the flow is as hot as virial? Before we answer these questions we first briefly introduce the accretion model to black hole binaries. An individual black hole X-ray binary comes into several states, mainly hard, soft, and steep power-law (or very high) states. These three states are associated with continuous jets, no jets, and episodic jets, respectively (Fender 2006). The luminosity of the hard state spans a large range, from very low values up to $\sim 10\%L_{\rm Edd}$ or even $30\%L_{\rm Edd}$ (e.g., Zdziarski \& Gierli\'nski 2004). The soft state, on the other hand, only exists above $\sim 2\% L_{\rm Edd}$. Sources with $2\%L_{\rm Edd}\la L \la 10\%L_{\rm Edd}$ is not certain because of the ``hysteresis'' (Zdziarski \& Gierli\'nski 2004). The soft state is described by a standard thin disk extending to the innermost stable circular orbit; while the hard state is described by an inner hot accretion flow, which contributes most of the bolometric luminosity of the hard state, and an outer truncated thin disk (e.g., Narayan 1996; Esin, McClintock \& Narayan 1997; Yuan, Cui \& Narayan 2005; Yuan \& Zdziarski 2004; Oda et al. 2012; see Narayan 2005; McClintock \& Remillard 2006; Done, Gierli\'nski \& Kubota 2007 for reviews). The transition radius between the hot and cool accretion flows ($r_{\rm tr}$) is a function of accretion rate or luminosity (Yuan \& Narayan 2004). The model of the very high state is still unclear.

Compared to the black hole X-ray binaries, the accretion model of AGNs is far less certain. Since the physics of accretion is irrelevant to the black hole mass, by analogy to the black hole binaries, it is reasonable to assume that low-luminsoity AGNs with $L$ up to $(2-10) \%L_{\rm Edd}$ should correspond to the ``hard state'' and are powered by a hot accretion flow. This picture has received observational supports (see Yuan 2007, Ho 2008 for reviews). The model of luminous AGNs are more complicated and may not be easily associated with the standard thin disk. This is because, on one hand, some luminous AGNs may correspond to the very high state whose model is unknown; on the other hand, the observational evidence for the thin disk is not as good as in the soft state of black hole X-ray binaries. Moreover, it seems that the hard X-ray emission of luminous AGNs, which can't be explained by a standard thin disk, is much stronger than that in the soft state of black hole X-ray binaries. This indicates that hot gas must exist. A general idea is to invoke a corona sandwiching the thin disk (e.g., Galeev, Rosner \& Vaiana 1979; Haardt \& Maraschi 1991). If the dynamics of the corona can be described by a hot accretion flow, our study presented in this paper will be applicable.

Almost all works, including the present one, assume that the hot accretion flow is one-phase. Hot accretion flow is likely thermally unstable under short-wavelength perturbations (Kato 1997; Wu 1997; Yuan 2003). Yuan (2003) found that when the accretion rate of the accretion flow $\dot{M} \ga \alpha^2 \dot{M}_{\rm Edd}$ , the timescale of growth of perturbation will be shorter than the accretion timescale; consequently, cold dense clumps may be formed within the hot phase. So we expect that when $\dot{M} \ga \alpha^2 \dot{M}_{\rm Edd}$, the accretion flow and outflow should be two-phase. In this case, the presence of absorption line is expected. If $\alpha=0.1$, the corresponding $\dot{M}\sim 0.01 \dot{M}_{\rm Edd}$ and $L\sim 10^{-3}L_{\rm Edd}$.

\subsection{Bubbles Inflated by Outflow}

\subsubsection{The Fermi bubble in the Galactic center}

The first example is the two ``Fermi Bubbles'' revealed by {\it Fermi}-LAT, extending $50^{\circ}$ above and below the Galactic center (Su et al. 2010). The bubbles have approximately uniform surface brightness with sharp edges and symmetric about the Galactic plane. The bubbles are spatially correlated with the hard-spectrum microwave excess known as the {\em WMAP} haze. The initial analysis in Su et al. (2010) indicates that the bubbles are most likely created by some large episode of energy injection in the Galactic center. The age and energy of the Bubble have been  roughly estimated in Su et al. (2010) to be $10^7$ yr  and $E\sim (10^{54}-10^{55})~{\rm ergs}$, respectively.

Several models have been proposed to explain the origin of the Fermi bubble. They usually associate the bubble with the recent activity of the supermassive black hole, Sgr A* (e.g., Guo \& Mathews 2011; Zubovas \& Nayakshin 2012). Guo \& Mathews (2011) proposed that the bubble was inflated by a past jet in Sgr A*. They performed some numerical simulation and roughly re-produced the morphology of the bubble. The problems with this scenario are that the jet is more likely an outflow since it is not collimated; and the required mass flux in the  jet is super-Eddington, which may be unreasonable. In another model, Zubovas \& Nayakshin (2012) required that  the bolometric luminosity of Sgr A* in the past is very large, approaching $L_{\rm Edd}$, i.e, Sgr A* was a very luminous quasar. In this case, a powerful outflow can be produced by the radiation pressure, which then inflated the Fermi bubble. We note, however, that the required high luminosity of Sgr A* is in conflict with the study of Totani (2006) in which they argued based on other observations that Sgr A* in the past should still be within the regime of ADAF, although the mass accretion rate was 3-4 orders of magnitude higher than the present value.

Our scenario is that the bubbles are inflated by the outflow from an ADAF. Shocks are formed when the outflow interacts with the interstellar medium, and electrons are accelerated to relativistic energies in the shock front. These electrons emit synchrotron and inverse Compton radiation by scattering with CMB photons. The former explains the {\it WMAP} haze while the latter is responsible for the {\it Fermi} bubbles.  We now estimate whether the outflow can reasonably power the bubble. Taking $E\sim 10^{54}
{\rm erg}$ and $10^7~{\rm yr}$ as an example, the required bubble power is $3\times
10^{39} {\rm erg~s^{-1}}$. From eq. (\ref{outflowpower}), the
required accretion rate is \be \dot{M}_{\rm out}\sim
10^{-5}r_{\rm out}\dot{M}_{\rm Edd}.\ee Here $\dot{M}_{\rm
Edd}\equiv 10L_{\rm Edd}/c^2\sim 9\times 10^{-2}\msun~yr^{-1}$ is
the Eddington accretion rate of Sgr A*. Note that according to Yuan
\& Narayan (2004), $r_{\rm out}$ is a function of $\dot{M}_{\rm
out}$. A reasonable set of parameter would be $r_{\rm out}\sim 10^3r_s$
and $\dot{M}_{\rm out}\sim 10^{-2}\dot{M}_{\rm Edd}$.
Current accretion rate of Sgr A* at the Bondi radius ($r_{\rm
out}\sim 10^5$) is $\dot{M}\sim 10^{-6}\msun yr^{-1}\sim
10^{-5}\dot{M}_{\rm Edd}$ (Yuan, Quataert \& Narayan 2003). So we
require that the past activity of Sgr A* is about 1000 times higher in
terms of accretion rate. This is in good consistency with the estimation of
Totani (2006). Numerical simulation work is in progress (Mou et al. 2012, in preparation).

\subsubsection{M~87}

M~87 is a famous low-luminosity AGN powered by an ADAF (e.g., Di Matteo et al. 2003). The radio (90 cm) image of M~87  (Fig. 1 in Owen, Eilek \& Kassim 2000) clearly show two bubbles with radius of $\sim 20 {\rm kpc}$. The shape of the bubbles looks very similar to the Fermi bubbles in the Galactic center. The orientation of the two bubbles is in perpendicular direction of the radio jet, therefore it seems unlikely that the bubbles are inflated by the jet; instead they may be inflated by the outflow from the ADAF.

\subsection{Outflow from AGNs with High and Low Luminosities}

One popular model for producing AGNs winds is radiation driving. But detailed studies to individual sources often exhibit some problems. In the following we present several examples.

{\bf NGC~3783.} This is a Seyfert 1 galaxy, with the mass of
the black hole $\sim (3\pm1)\times 10^7 \msun$, the bolometric luminosity $L_{\rm
bol}\sim 3\times 10^{44}\ergs\sim 0.1 L_{\rm Edd}$. Chelouche \& Netzer (2005) carefully
calculated the physical properties of the highly ionized outflow and compared their theoretical calculation with the
{\em Chandra} data. The main results they obtained are: (1) The
highly ionized outflow in this source is not driven by the radiation pressure (they suggested thermal pressure gradient);
(2) The rate of outflow is $\sim 0.01-0.1 \msun {\rm yr}^{-1}\approx
0.02-0.2 \dot{M}_{\rm Edd}$ ($\dot{M}_{\rm Edd}\equiv 10L_{\rm Edd}/c^2\approx 0.5 \msun {\rm yr}^{-1}$); (3) The velocity of the outflow is $\sim 1000 ~{\rm km~s^{-1}}$; (4) The
global covering factor of the flow is $\sim 20\%$.

{\bf NGC~4151}. Kraemer et al. (2005) presented a detailed analysis to the intrinsic X-ray absorption in this source using {\em Chandra} data and found the presence of very highly ionized outflow in addition to lower
ionization gas. The authors pointed out that the outflows are so highly ionized and the
luminosity of these galaxies is only $\sim 4\% L_{\rm Edd}$ that they are unlikely to be
accelerated by radiation pressure. They pointed out that such kind of high ionization outflow is very common among Seyfert galaxies and there existence is the key to understand the origin of mass outflow.

{\bf 3C 111} Tombesi et al. (2011b) detected blueshifted absorption lines in 3C 111 based on their {\em Suzaku} observations. The mass of the black hole is $(2-30)\times 10^8 \msun$ and the bolometric luminosity is $L_{\rm bol}\approx 8\times 10^{45}\ergs\sim (2-30)\%L_{\rm Edd}$. Detailed modeling indicates that the location of absorption material is constrained at $<0.006~{\rm pc}\sim (20-300)r_s$ and the velocity and mass outflow rate of the outflow is $\sim 0.1$c and $\dot{M}_{\rm out}\sim 1\mpy$. The latter is similar to the mass accretion rate $\dot{M}_{\rm acc}\sim L_{\rm bol}/\eta c^2$ (for $\eta=0.1$).  The momentum flux of the outflow is comparable to the momentum flux of radiation, $\dot{M}_{\rm out}\sim L_{\rm bol}/c$. However, the Thompson scattering optical depth of the outflow is found to be very small, $\tau \sim 0.05$. Therefore, for the radiation mechanism to work, we have to look for other opacities. This seems to be not very easy given  the high ionization of the outflow, similar to NGC~4151. In addition, while the rather small upper limit of the location of outflow indicates their accretion flow origin and rules out other models, this location is more than one order of magnitude smaller than that predicted in the radiation driving model of Murray et al. (1995).

{\bf Low-luminosity AGNs.} Outflow can be more easily detected in luminous AGNs partly because these sources are suitable for high resolution spectroscopy. Recently, outflow has also been detected in  low-luminosity AGNs. Crenshaw \& Kraemer (2012) investigated the outflow from a sample of nearby AGNs, focusing on determining their mass outflow rate and the kinetic luminosity. Although they have to restrict their sample to apparently luminous AGNs and they do not include ultrafast outflow, among the ten nearby Seyfert 1 galaxies in their sample, six sources still have their bolometric luminosities below $5\% L_{\rm Edd}$. The detailed study to the UV and X-ray absorbers clearly show that strong outflow exists in these sources. The bolometric luminosity of one source, NGC~4395, is even as low as $10^{-3}L_{\rm Edd}$. These low-luminosity AGNs are described by an ADAF rather than a standard thin disk. Thus it is questionable whether the radiation driving model works for them.  They found that the mass outflow rate exceeds the mass accretion rate by a factor of 10 to 1000. The kinetic luminosities of outflow are approximately $0.5\%$ to $5\%$ of their bolometric luminosites for half of their moderate-luminosity AGNs in their sample.

Now we estimate whether the above examples can be possibly explained by the our outflow model. We want to emphasize that such estimations must be very crude.

{\bf NGC~3783.} During the propagation of outflow, the mass flux may increase and the velocity may decrease because of the contamination by ISM, but the momentum flux should be conserved. Assuming X-ray luminosity $L_{\rm x}\sim f_x L_{\rm bol}\sim 0.1f_x L_{\rm Edd}$ and assuming that the hard X-ray emission is produced by a hot corona which can be modeled by a hot accretion flow, the required $\dot{M}(10r_s)\sim 0.1 \eta^{-1}f_x\dot{M}_{\rm Edd}$ (here $\eta$ is the radiative efficiency of the hot accretion flow). So $\dot{M}_{\rm out}(r_{\rm out})=0.1\eta^{-1}f_x\dot{M}_{\rm Edd}(r_{\rm out}/10r_s)^{0.5}$, $v_{\rm term}\sim 0.5 v_k(r_{\rm out})\sim c/2\sqrt{2r_{\rm out}/r_s}$. The momentum flux is then $\dot{p}_w\sim 0.02f_x\eta^{-1}\dot{M}_{\rm Edd}c \sim 2\times 10^{-3}\dot{M}_{\rm Edd}c$ for $f_x=10^{-2}$ and $\eta\sim 0.1$. This is consistent with the observed $\dot{p}_w\sim (0.7-7)\times 10^{-3}\dot{M}_{\rm Edd}c$. In addition, assuming $r_{\rm out}\sim 100r_s$, the outflow rate is $\dot{M}_{\rm out}(r_{\rm out})\sim 0.03 \dot{M}_{\rm Edd}$, which also satisfies the requirement that the produced outflow rate from the hot accretion flow should be smaller than the observed one because of contamination by ISM.

{\bf 3C 111.} The location of the absorption material is very close to the black hole so we expect little contamination of the outflow. Firstly, there is obviously no problem in our model to produce outflows from radius as small as $\sim 100r_s$. Secondly, assuming $r_{\rm out}\sim 100r_s$ (since luminosity is relatively high), since $(100r_s/10r_s)^{0.5}\sim 3$, this explains why $\dot{M}_{\rm out}(100r_s)$ is not so different from the accretion rate at $10r_s$. Thirdly, the velocity of outflow is $\sim 0.5 v_k(100r_s)\sim 0.1c$, in good consistency with observations. At last, the approximate equality of the momentum flux between radiation and outflow is exactly what we expect (eq. [\ref{momentumratio}]).

{\bf Low-luminosity AGN.} Firstly, for low-luminosity AGNs, a reasonable assumption is that $r_{\rm out}\sim 10^3-10^5r_s$, depending on the exact value of accretion rate and the temperature of the fueling material. In this case, the ratio of the mass outflow rate and accretion rate (at $10r_s$) is $\dot{M}_{\rm out}(r_{\rm out})/\dot{M}_{\rm in}(10r_s)\sim (r_{\rm out}/10r_s)^{0.5}=10-100$. Given further the contamination by the ISM, this explains the observed ratio of the outflow rate and accretion rate. Secondly, from eq. (\ref{radiationpower}), for $\eta\sim 0.05$, we have $\dot{E}_w\sim 1\%\dot{E}_{\rm rad}$, again in good consistency with observations.

\subsection{Outflow from Black Hole X-ray Binaries}

A number of highly-ionized winds have been detected in black hole and neutron star X-ray binaries in recent years (see references in Neilsen \& Homan 2012). GRO J1655-40 is one of the best examples. This source was observed twice by {\it Chandra} during its 2005 outburst. The second observation was made when the source was in the soft state. A series of absorption lines from a dense and highly-ionized wind have been revealed. Detailed spectral analysis and theoretical studies have been performed by different groups and confirmed repeatedly that this outflow was driven not by the radiation or thermal pressure, but likely by magnetic mechanism (Miller et al. 2006, 2008; Kallman et al. 2009; Luketic et al. 2010). Another observation was made 20 days earlier than this one, when the source was just at the beginning of the transition from hard to soft state. This observation only detected one absorption line (Fe XXVI). Neilsen \& Homan (2012) analyzed the data and concluded that the outflow in the hard state also can't be driven by the radiation pressure because the ionization parameter is too high. Their photoionization modeling indicates that the difference of the {\it Chandra} spectrum in the two observations cannot possibly be explained by the change of the ionizing spectrum. Instead, the properties of the wind in the two states must be changed. This is reasonable since the accretion flow in the two states are different.

Most recently, King et al. (2012) investigated the kinetic power of outflows and jets across black hole mass scale. They found that both power scales with the radiative power. More importantly, they found that the scaling relations are consistent with each other within error bars, although the normalization of the jet relation is higher than the outflow. They argued that such a formal consistency suggests a common origin mechanism for the outflow and jets. Since it is generally believed that jets are magnetically originated, this would imply that outflow is also magnetically originated.

\section{Summary}
\label{summary}

Numerical simulations of hot accretion flow, both HD and MHD, have revealed that the mass accretion rate decreases with decreasing radius. Consequently the radial density profile of the accretion flow becomes much flatter compared to the self-similar solution (Narayan \& Yi 1994) which is based on assumption of a radius-independent mass accretion rate. Denoting the profiles of accretion rate and density as $\dot{M}(r)\propto r^s$ and $\rho(r)\propto r^{-p}$, the HD simulation in Paper I, which has the largest radial dynamical range so far (from $r_s$ to over $10^4r_s$), gives $s\sim 0.4-0.75$ and $p\sim 0.65-0.85$, respectively. As a comparison, the self-similar solution of ADAFs gives $s=0$ and $p=1.5$. Such a flatter density profile have obtained strong observational support in Sgr A* and NGC~3115 (see Paper I for details). Two different models have been proposed to explain such a result, namely ADIOS and CDAF. In this paper we investigate the nature of the inward decrease of accretion rate and other related questions.

For this aim, we have performed a series of HD and MHD numerical simulations, comparing the various properties of inflow (the gas with a negative radial velocity) and outflow (the gas with a positive radial velocity) such as radial and rotational velocities, temperature, and Bernoulli parameter. We found systematic and significant differences (\S3). Such differences are hard to understand if the inflow and outflow are simply the appearance of turbulent fluctuation; but strongly suggest that the motion of the accretion flow is dominated by systematic inward and outward fluxes of mass.
We have also analyzed the convective stability of MHD flows. We found that they are stable (\S5). This indicates that CDAF model at least can't be applied to MHD flows.  Based on these results, together with other arguments, we conclude that the inward decrease of accretion rate is because of mass loss via outflow (\S\ref{adioscdaf}).

An immediate question is then the origin of outflow. This is discussed in \S\ref{outfloworigin}. The detailed comparison of properties between inflow and outflow again presents important information to answer this question. In the HD case, we found that the temperature of outflow is systematically higher than that of inflow (Fig. \ref{Fig:temperature}). This suggests that the outflow is driven by the buoyancy which arises because of the convective instability of the HD accretion flow. In the case of MHD flow, we found that the specific angular momentum of outflow is very high, close to the Keplerian angular momentum at the equatorial plane; while the angular momentum of inflow is much lower (Fig. \ref{Fig:rotationvelocity}). This suggests that it is the centrifugal force that drives the production of outflow. Magnetic field in the flow efficiently transfers the angular momentum between fluid elements. Whenever one element gets enough angular momentum, it will turn around and be thrown outward. This mechanism is similar to the Blandford \& Payne (1982) mechanism, except that no large-scale magnetic field is required here. We therefore call it ``micro'' Blandford \& Payne (MBP) mechanism (Fig. \ref{Fig:mbp}).

The properties of outflow is of great interest especially because of their potential important role in AGNs feedback. We have calculated the mass flux, terminal radial velocity, momentum flux and kinetic power of outflow which should be useful in the comparison with observations of AGNs winds, and in the study of AGNs feedback (\S3). One of the most important properties is the Bernoulli parameter of outflow since its sign determines whether the outflow can escape to infinity. We have run three HD models with different initial conditions. We found that while many properties of the accretion flow are not dependent of the initial condition, the sign of $Be$ of outflow is crucially  determined by the value of $Be$ in the initial condition (Fig. \ref{Fig:radialbernoulli}). If $Be$ is large in the initial condition, $Be$ of outflow inclines to be positive. This is natural since the total energy should be conserved in the simulation. Unfortunately, it is uncertain what kind of initial condition is more realistic, and the answer may depend on circumstances. We think that the value of $Be$ of the gas is likely to be positive at least in some cases such as when the radiative energy loss of accretion flow is small.  Even in the case that $Be<0$, these outflow can still escape out of the outer boundary of the accretion flow, as shown by our simulations; thus they can still interact with the ISM and play a similar role to those outflow with $Be>0$.

We have also discussed the possible applications of the outflow from hot accretion flow in explaining observations (\S\ref{observation}). These include the formation ``Fermi bubble'' in the Galactic center, and the origin of observed winds from both AGNs and black hole X-ray binaries. These winds have been detected from sources with a variety of luminosities. Detailed analysis to some sources published in literature have shown that it is very hard for the winds to be produced by radiation or thermal driving mechanisms. Our simple estimation suggests that it is promising to explain their origin by our ``MBP'' mechanism.

\section{ACKNOWLEDGMENTS}

We are grateful to Ramesh Narayan for valuable discussions and constructive comments. We also thank  Jerry Ostriker for sending us his preprint and discussions, Jim Stone for sending us his simulation data and discussions, Jon McKinney and Francesco Tombesi for useful comments. This work was supported in part by the National Basic Research Program of China (973 Program 2009CB824800), the Natural Science Foundation of China (grants 10833002, 10825314, 11103059, 11121062, and 11133005), and the CAS/SAFEA International Partnership Program for Creative Research Teams. The simulations were carried out at Shanghai Supercomputer Center.

\clearpage

\end{document}